\documentclass[twocolumn,floats,floatfix,
               showpacs,prd,superscriptaddress,
               longbibliography,
               nofootinbib]{revtex4-2}
\usepackage[utf8]{inputenc}

\usepackage{graphicx,epsfig}
\usepackage{epstopdf}
\usepackage{amssymb,amsmath,amsthm,amsfonts}
\usepackage{bm}
\usepackage[caption=false]{subfig}
\usepackage[usenames,dvipsnames]{xcolor}
\usepackage{url}
\usepackage[inline]{enumitem}
\usepackage{dsfont}
\usepackage{float}
\usepackage{placeins}
\usepackage[T1]{fontenc}
\usepackage{mathrsfs}
\usepackage{anyfontsize}
\usepackage{xspace}
\usepackage{tensor}
\usepackage{microtype}
\usepackage{tikz}
\usetikzlibrary{positioning}
\usepackage{acronym}
\usepackage{comment}

\usepackage[linktocpage,breaklinks]{hyperref}
\hypersetup{colorlinks=true,
            citecolor=NavyBlue,
            linkcolor=NavyBlue,
            urlcolor=NavyBlue}

\usepackage[compat=1.1.0]{tikz-feynman}
\tikzfeynmanset{warn luatex=false}

\renewcommand{\vec}[1]{\boldsymbol{#1}}

\def\canuda{\textsc{Canuda}}
\def\ETK{\textsc{Einstein Toolkit}}
\def\newacronym#1#2#3{\gdef#1{\gdef#1{#2\xspace}#3 (#2)\xspace}}
\newacronym{\amr}{AMR}{adaptive mesh refinement}
\def\bh#1{black hole#1 (BH#1)\gdef\bh{BH}}
\def\bbh#1{binary black hole#1 (BBH#1)\gdef\bbh{BBH}}
\newacronym{\bssn}{BSSN}{Baumgarte-Shapiro-Shibata-Nakamura}
\newacronym{\adm}{ADM}{Arnowitt-Deser-Misner}
\newacronym{\CAH}{CAH}{common apparent horizon}

\newacronym{\gr}{GR}{General Relativity}
\def\gw#1{gravitational wave#1 (GW#1)\gdef\gw{GW}}

\newacronym{\ode}{ODE}{ordinary differential equation}
\newacronym{\pde}{PDE}{partial differential equation}

\newacronym{\eft}{EFT}{effective field theory}
\newacronym{\eob}{EOB}{effective-one-body}

\newacronym{\hpc}{HPC}{High-performance computing}
\newacronym{\sGB}{sGB}{scalar Gauss-Bonnet}
\newacronym{\nr}{NR}{numerical relativity}

\def\dif{\textrm{d}}

\def\R4D{R}
\def\rex{r_{\rm ex}}

\begin{document}

\title{Scalarization and descalarization in hyperbolic encounters of black holes}

\author{Frederick C.L. Pardoe}
\email{fpardoe2@illinois.edu}
\affiliation{The Grainger College of Engineering,
Department of Physics \& Illinois Center for Advanced Studies of the Universe, University of Illinois Urbana-Champaign, Urbana, Illinois 61801, USA}


\author{Helvi Witek}
\email{hwitek@illinois.edu}
\affiliation{The Grainger College of Engineering,
Department of Physics \& Illinois Center for Advanced Studies of the Universe, University of Illinois Urbana-Champaign, Urbana, Illinois 61801, USA}
\affiliation{Center for AstroPhysical Surveys, National Center for Supercomputing Applications, University of Illinois Urbana-Champaign, Urbana, IL, 61801, USA}

\begin{abstract}
We use numerical relativity to study the scalar field evolution sourced by hyperbolic encounters of black holes in quadratic scalar Gauss-Bonnet gravity.
In this theory, single black holes are known to acquire a scalar hair through scalarization for certain values of their mass and spin.
We work in the decoupling limit and evolve the scalar field on top of a background metric.
Seeding binary black holes with an initial scalar field, we find that configurations which initially cannot sustain a scalar hair temporarily scalarize during an encounter and thereby exhibit \textit{dynamical scalarization}. 
This is possible for both positive and negative couplings between the scalar field and curvature in black hole binaries with zero and non-zero initial spins, respectively.
Furthermore, we find that the change in the spin magnitude of black holes during certain hyperbolic encounters can lead to permanent \textit{spin-induced scalarization} (or \textit{descalarization}), which we refer to as \textit{spin-up (de)scalarization}. 
\end{abstract}

\maketitle

\section{Introduction}

Ever since the publication of \gr more than a century ago, the theory has been subject to a litany of tests. 
Historically, most tests were conducted in the weak-field regime, and only a few strong-field tests based on electromagnetic observations of pulsars were possible; see Refs.~\cite{Will:2014kxa,Berti:2015itd,Will:2018bme,Yunes:2024lzm} for a review. 
However, since the first direct detection of \gw{s} in 2015 \cite{LIGOScientific:2016aoc}, tests of \gr in the strong-field regime have become far more common. 
In the decade since, hundreds of \gw{} observations have been subject to tests~\cite{LIGOScientific:2016lio,LIGOScientific:2018dkp,LIGOScientific:2019fpa,LIGOScientific:2020tif,LIGOScientific:2021sio,LIGOScientific:2025wao,LIGOScientific:2025rid,LIGOScientific:2026qni,LIGOScientific:2026fcf,LIGOScientific:2026wpt}. 
Further strong-field tests have also been conducted using observations of supermassive \bh{s} from the Event Horizon Telescope~\cite{EventHorizonTelescope:2021dqv,EventHorizonTelescope:2022xqj}. 
To date, these tests have found observations to be consistent with \gr. However, with pulsar timing arrays \cite{Yunes:2024lzm,Liang:2023ary} and next generation \gw{} detectors such as LISA~\cite{LISA:2017pwj}, the Einstein Telescope~\cite{Punturo:2010zz}, and Cosmic Explorer~\cite{Reitze:2019iox}, opportunities to test \gr will only continue improve in the coming decades.

With these advancements in experimental capabilities has come an increased interest in theoretical alternatives to \gr.  
Numerous alternatives to \gr have been proposed, motivated in part by the seeming inevitability that quantum effects must lead to deviations from \gr in the strong-field regime \cite{Shankaranarayanan:2022wbx,Berti:2015itd,Clifton:2011jh}. One such set of theories is \sGB gravity, a subset of the broader class of Horndeski theories~\cite{Horndeski:1974wa}
that are known to lead to second order field equations.
The Horndeski theories are derived by adding a dynamical scalar field to Lovelock gravity \cite{Lovelock:1971yv,Lovelock:1970zsf,Lovelock:1972vz}, either heuristically or via Kaluza-Klein reduction from higher dimensions; see Refs.~\cite{Charmousis:2014mia,Fernandes_2022,Kobayashi:2011nu,Kobayashi:2019hrl} for a review.
Certain forms of \sGB gravity also appear as part of a dimensional reduction of heterotic string theory~\cite{Cano:2021rey,Metsaev:1987zx,Kanti:1995cp} in the low energy limit.

One of the key properties of \sGB gravity is that \bh{s} can be endowed with a scalar hair sourced by the spacetime curvature~\cite{Kanti:1995vq,Sotiriou:2013qea,Sotiriou:2014pfa}. This is due to a nonminimal coupling between the curvature and the scalar field.
In some versions of the theory, such as those with a linear or diatonic coupling between the scalar and curvature, this hair is always present. However, for quadratic or gaussian forms of the coupling, hairy solutions can only appear below a certain \bh{} mass, or equivalently, above a certain value of the coupling constant~\cite{Doneva:2017bvd,Silva:2017uqg}. 
In this case, initially hairless \bh{s} can acquire a scalar hair. This phenomenon, known as \textit{scalarization}~\cite{Damour:1993hw,Damour:1996ke}, has been studied in great detail both in \sGB gravity, as well as in other theories; see Ref.~\cite{Doneva:2022ewd} for a review. 

In practice, scalarization is often classified into a number of phenomenological subcategories. Single compact objects can experience \textit{spontaneous scalarization}, which occurs on account of the object's compactness~\cite{Damour:1993hw,Damour:1996ke,Doneva:2017bvd,Silva:2017uqg}. 
Under certain conditions, \bh{s} in \sGB gravity can also undergo \textit{spin-induced scalarization}, which occurs above a certain (dimensionless) spin determined by the coupling constant and \bh{} mass~\cite{Dima:2020yac,Herdeiro:2020wei,Hod:2020jjy}. 
Binary systems can be subject to \textit{induced scalarization}, where one scalarized compact object induces a scalar hair in another, and \textit{dynamical scalarization}, where the gravitational interaction of two compact objects causes the scalarization of both~\cite{Barausse:2012da,Shibata:2013pra,Palenzuela:2013hsa}.  Equivalent processes in which the scalar hair is lost are referred to as \textit{descalarization}~\cite{Doneva:2022ewd}. 

In recent years, numerous works have studied binary \bh{} dynamics in \sGB gravity~\cite{Silva:2020omi,Elley:2022ept,Doneva:2022byd,East:2021bqk,AresteSalo:2023mmd,Nee:2024bur,Lara:2025kzj,Julie:2023ncq}.
These studies have relied on numerical relativity, both in the decoupling limit~\cite{Witek:2018dmd,Silva:2020omi,Elley:2022ept,Doneva:2022byd,Lara:2024rwa} and with backreaction \cite{Witek:2020uzz,East:2020hgw,East:2021bqk,AresteSalo:2023mmd,Nee:2024bur,Lara:2025kzj,Kovacs:2020ywu,Corman:2024cdr}, as well as on semi-analytic methods \cite{Julie:2023ncq,Julie:2022huo,Shiralilou:2021mfl}. 
See Ref.~\cite{Ripley:2022cdh} for a review on the application of numerical relativity to various Horndeski theories.

Simulations have shown that remnants formed from the merger of two scalarized \bh{s} can descalarize if the remnant \bh{} has a larger mass~\cite{Silva:2020omi} or lower spin~\cite{Elley:2022ept} than its progenitors. 
These processes are often referred to as \textit{dynamical descalarization} and \textit{spin-induced dynamical descalarization}, respectively.
However, it is pertinent to note that the descalarization found in these cases occurs after the formation of a \CAH and is thus not dynamical in quite the same sense as defined above; originally, \textit{dynamical scalarization} was coined to describe the scalarization of inspiraling neutron stars prior to merger in other theories~\cite{Barausse:2012da,Shibata:2013pra,Palenzuela:2013hsa}.
The \textit{``dynamical'' descalarization} of remnant \bh{s} has also been studied in varieties of \sGB gravity where there is a ``jump'' between solutions with a minimum allowed scalar hair and no scalar hair at all, as opposed to a gradual transition depending on the \bh{} mass (or coupling constant)~\cite{Doneva:2022byd}. 
It is conjectured that this jump may create an abrupt qualitative difference between \gw{} emissions from systems with scalarized and unscalarized remnants.

In addition to \textit{spin-induced ``dynamical'' descalarization}, there are also examples of systems where the inspiral and merger of initially unscalarized \bh{s} leads to the \textit{spin-induced dynamical scalarization} of the remnant~\cite{Elley:2022ept}. Here, scalarization begins prior to the formation of a \CAH and is thus dynamical in the original sense of the term. 
Nonetheless, for all but exceptionally large couplings, the scalarization in these systems was found to be small prior to the formation of a \CAH.
Thus, the scalarization would likely affect merger and ringdown, but have minimal impact on the dynamics and \gw{} emission during the inspiral. 
This phenomena is known as \textit{stealth scalarization}.

In addition to the inferences drawn from studies in the decoupling limit, \gw{} emissions have also been investigated directly with backreaction.
The \gw{} emissions emitted from \textit{stealth scalarization} were directly computed by Ref.~\cite{AresteSalo:2023mmd}, which confirmed that scalarization had minimal effect on these systems prior to merger.
Backreaction has also been used to study the merger of initially scalarized \bh{s} and their \gw{} emissions~\cite{East:2021bqk}.
It was found that the \gw{} luminosity is substantially greater than in \gr when non-spinning \bh{s} are considered, but that there is negligible difference when the \bh{s} are scalarized due to spin (i.e., due to \textit{spin-induced scalarization}).

Using semi-analytic methods, it has been shown that non-spinning \bh{s} should also undergo \textit{dynamical scalarization} during the inspiral phase of a binary \bh{} merger~\cite{Julie:2023ncq}.
This is intriguing, given that such systems should descalarize if they form a \CAH. 
This phenomenon was further explored numerically through the construction of constraint satisfying initial data for binary \bh{s} in a quasicircular configuration~\cite{Nee:2024bur,Lara:2025kzj}.  
It was found that the \bh{s} had a greater value of the scalar field on their horizons than would have been the case if the \bh{s} were isolated and that such binaries could acquire a scalar hair for lower values of the coupling than would have been possible were they isolated. 

In addition to the above, scalarization has been studied in \bh{} binaries with unequal masses and scalar hairs of opposite sign~\cite{Nee:2024bur}. 
In unequal mass binaries, it was found that the larger \bh{} could support a scalar hair despite the fact that it would be unable to do so individually, in effect demonstrating \textit{induced scalarization}. It was also found that the scalar hair is smaller in systems with opposite sign than in systems with equal sign. 
Furthermore, a study considering oppositely signed \bh{s} in a backreacted inspiral found a sign flip in one of the \bh{s} leading to the introduction of eccentricity in the orbit~\cite{Lara:2025kzj}.

The studies above largely considered \bh{} binaries in head-on collisions or quasicircular inspirals. 
However, \bh{} binaries are known to display a variety of additional morphologies~\cite{Pretorius:2007jn,Shibata:2008rq,Sperhake:2009jz,Healy:2009zm,Gold:2009hr,Gold:2012tk}.
Namely, hyperbolic encounters between \bh{s} can result in scattering or zoom-whirls, where the \bh{s} perform a series of small close orbits punctuated by larger more eccentric orbits before eventually merging.\footnote{Neutron stars also exhibit these morphologies; see Refs.~\cite{Fontbute:2025vdv,Neuweiler:2025lte,Gold:2011df,East:2012ww,Radice:2016dwd,Papenfort:2018bjk,Chaurasia:2018zhg}.}
See Refs.~\cite{Palenzuela:2013hsa,Swain:2026dfr,Bernard:2026old} for eccentric orbits and scattering in other theories beyond \gr.
These temporary close encounters between \bh{s} provide a fascinating environment in which to explore the dynamical processes discussed above. 
Furthermore, it is known that scattering \bh{s} can “spin up” in the direction of their orbital angular momentum~\cite{Sperhake:2012me,Nelson:2019czq,Jaraba:2021ces,Rodriguez-Monteverde:2024tnt,Chiaramello:2024unv,Kogan:2025vml}.
This makes such encounters interesting environments in which to consider spin-induced phenomena.  

In this work, we study the scalarization and descalarization of equal-mass binary \bh{s} as they undergo hyperbolic encounters in quadratic \sGB gravity.
Working in the decoupling limit, we consider a variety of systems with different couplings, spins, and morphologies.
We observe regular and spin-induced \textit{dynamical scalarization} and \textit{descalarization} of the \bh{s}. Furthermore, we observe \textit{spin-induced scalarization} and \textit{descalarization} resulting from the spin-up of the \bh{s}. We call these phenomenon \textit{spin-up scalarization} and \textit{spin-up descalarization}.

We structure this work as follows. 
In Sec.~\ref{sec:sGB_Theory}, we introduce the theory behind \sGB gravity and discuss more precisely the conditions for the \textit{spontaneous scalarization} and \textit{spin-induced scalarization} of individual \bh{s}. 
In Sec.~\ref{sec:NRFramework}, we discuss the computational framework of our study. In Sec.~\ref{sec:Setup}, we discuss the physical setup of our simulations and summarize the simulation suite considered.
The results are presented in Sec.~\ref{sec:Results}. In Sec.~\ref{sec:Conclusion}, we discuss our conclusions and directions for further work. In the Appendix, we describe a suite of convergence tests. Throughout the following, we use geometric units where $G=c=1$.\\

\section{Scalar Gauss-Bonnet Gravity} \label{sec:sGB_Theory}

\subsection{Action and equations of motion}
\label{sec:sGB_Theory_action&eom}

In \sGB gravity, a real (dynamical) scalar field is non-minimally coupled to the spacetime curvature according to the action, 
\begin{equation}
\label{eq:SGB_action}
S = \frac{1}{16 \pi} \int d^4x \sqrt{-g} \left[ R  - \frac{1}{2} \left( \nabla\phi \right)^2 + \frac{\alpha_{\rm GB}}{4}f(\phi) \mathcal{G} \right] \, ,
\end{equation}
where $g$ is the determinant of the spacetime metric, $g_{\mu \nu}$, $R$ is the Ricci scalar associated with the metric, $\phi$ is the scalar field, and $\alpha_{\rm GB}$ is a (dimensionful) coupling constant. Several theories fall under the umbrella of \sGB gravity; the precise version is set by the coupling function $f(\phi)$, which couples the scalar field to the spacetime curvature via the Gauss-Bonnet invariant,
\begin{equation}
\label{eq:GBInvariant}
    \mathcal{G} = R^2-4R_{\mu \nu}R^{\mu \nu}+R_{\mu \nu \rho \sigma}R^{\mu \nu \rho \sigma} \, ,
\end{equation}
where $R_{\mu \nu}$ is the Ricci tensor and $R_{\mu \nu \rho \sigma}$ is the Riemann tensor. The Gauss-Bonnet invariant is a topological invariant and total derivative in four dimensions; it thus only results in modification to the Einstein field equations when coupled to a dynamical scalar field.

We work in the test field approximation, where the scalar field does not backreact onto the spacetime, which evolves according to \gr. The equations of motion are,
\begin{subequations}
\begin{align}
\label{eq:f(phi)eom}
\Box \phi=-\frac{\alpha_{\rm GB}}{4}\frac{\dif f(\phi)}{\dif \phi}\mathcal{G} \, ,
\\
\label{eq:GReom}
R_{\mu \nu}-\frac{1}{2}Rg_{\mu \nu}=0 \, ,
\end{align}
\end{subequations}
where the first equation is given by the extremization of Eq.~\eqref{eq:SGB_action} with respect to the scalar field, and the second corresponds to the Einstein field equations in vacuum.

If the coupling function contains a stationary point, $\frac{df(\phi)}{d\phi}|_{\phi=\Psi}=0$, at some value of the scalar field, $\phi=\Psi$, then Eq.~\eqref{eq:f(phi)eom} is trivially solved by the constant scalar field $\phi(x^\mu)=\Psi=\text{const}$. 
In this case, even the backreacted equations of motion can admit any metric solution of \gr. However, depending on the parameters of the theory and the geometry of a given system, the \gr solution may not be stable. 
This can lead to scalarization, whereby systems initially described by \gr develop a non-trivial scalar field. 
In this model both the solutions of vacuum \gr and scalarized solutions are thus allowed.

When expanded around the \gr solution such that $\phi=\Psi+\delta \phi$, any coupling function with a stationary point takes the form $f(\phi)=f(\Psi)+f''(\Psi)(\delta\phi-\Psi)^2/2 +\mathcal{O}(\delta\phi^3)$. Due to the topological invariance of the Gauss-Bonnet invariant, the field and coupling constant can always be redefined such that $\Psi=0$ and $f(\phi)=\phi^2+\mathcal{O}(\phi^3)$ without loss of generality. The second order term determines whether a \gr solution is stable and thus the onset of scalarization. Therefore, any theory with the same coupling constant scalarizes under the same conditions, regardless of the particular coupling function used. However, the end state of scalarization is determined by higher order terms in the coupling function and thus does depend on the exact theory~\cite{Doneva:2017bvd,Silva:2017uqg,Doneva:2022ewd,Silva:2018qhn}. 
For example, systems with coupling functions of the form $f(\phi)\propto\phi^2$~\cite{Doneva:2017bvd}, $f(\phi)\propto1-e^{-\eta \phi^2}$~\cite{Silva:2017uqg}, and $f(\phi)\propto\phi^2+\zeta\phi^4$~\cite{Silva:2018qhn} (where $\eta$ and $\zeta$ are constants) scalarize under the same circumstances, but have different end states. 

In the following, we are primarily interested in the onset of scalarization and descalarization. Therefore, we choose the coupling function $f(\phi)=\phi^2$ such that $\Psi=0$. The theory corresponding to this choice of the coupling function is known as quadratic \sGB gravity. Substituting this coupling function into Eq.~\eqref{eq:f(phi)eom} gives a Klein-Gordon like equation,
\begin{subequations}
\begin{align}
\label{eq:scalareom}
\left(\Box - \mu_{\rm eff}^2\right)\phi = 0, \, \,
\\
\label{eq:mueff}
\mu_{\rm eff}^2 = -\frac{\alpha_{\rm GB}}{2} \mathcal{G}, 
\end{align}
\end{subequations}
where the curvature acts as an effective mass, $\mu_{\rm eff}$. 

We can gain some insight into how an instability develops by considering Eq.~\eqref{eq:scalareom} in flat space. If we forget for a moment that the effective mass is curvature dependent, then we might suppose that Eq.~\eqref{eq:scalareom} would admit exponentially growing plane wave solutions with wavenumber, $k$, when $\mu_{\rm eff}^2<-k^2$. 
In principle, these plane waves could have infinite wavelength and vanishingly small wavenumber, so scalarization would occur for any tachyonic effective mass, $\mu_{\rm eff}^2<0$. 

However, the effective mass is curvature dependent, so for any non-zero effective mass, spacetime cannot truely be flat.
Nevertheless, spacetime is always \textit{locally} flat within a sufficiently small region. 
This means that plane waves can only be solutions in an approximate local sense and must therefore have finite wavelength, which imposes a lower bound on the wavenumber.
Consequently, as we increase our curvature in an attempt to satisfy $\mu_{\rm eff}^2<-k^2$, we also make the constraint more restrictive. To know when the conditions for an instability are satisfied, one must consider the particular geometry of a given system.

In stationary radially symmetric systems, one can find a condition on the existence of scalarized solutions using an effective potential formalism~\cite{Doneva:2022ewd}. However, in dynamical systems, there is no simple analytic prescription to find a condition for scalarization. 

\subsection{Black hole scalarization in a nutshell}
\label{sec:sGB_Theory_coupling}

In this work, we consider the scalarization of \bh{} binary systems. Let us begin by reviewing the conditions under which individual \bh{s} scalarize as a guide. 
In \gr, \bh{s} are uniquely described by three parameters: (1) their mass, $m$, (2) their (dimensionless) spin, $\chi$, and (3) their electric charge\footnote{\bh{s} can only acquire an electric charge in Einstein-Maxwell theory, where \gr is minimally coupled to electromagnetism.}~\cite{Israel:1967wq,Robinson:1975bv,Wald:1971iw,Bekenstein:1971hc,Bekenstein:1972ky,Carter:1971zc}. 
These parameters influence the curvature of the spacetime around a \bh{} and the corresponding Gauss-Bonnet invariant. Here we consider \bh{s} with zero electric charge. The conditions under which a \bh{} scalarizes thus depends on the \bh{} mass, the \bh{} spin, and the coupling constant.

The coupling constant is dimensionful and has units of mass squared.
Therefore, it is typical to reexpress it as the dimensionless coupling,
\begin{equation} \label{eq:beta_def}
    \beta=\alpha_{\rm GB}/\mathrm{M}^2 \, ,
\end{equation}
where $\mathrm{M}$ is some characteristic mass, which we take to be the total mass of a system. We thus use the dimensionless coupling when considering the criterion for scalarization.

Since scalarization requires a tachyonic effective mass, the dimensionless coupling and Gauss-Bonnet invariant must either both be positive or both be negative; see Eq.~\eqref{eq:mueff}.
In the vicinity of slowly spinning \bh{s}, the Gauss-Bonnet invariant is entirely positive. 
However, for \bh{s} with spin magnitude $|\chi| > 0.5$, the Gauss-Bonnet invariant becomes negative around the spin axis. 
Two types of scalarization are thus possible depending on the sign of the dimensionless coupling: \textit{spontaneous scalarization}~\cite{Doneva:2017bvd,Silva:2017uqg} and \textit{spin-induced scalarization}~\cite{Dima:2020yac,Herdeiro:2020wei,Hod:2020jjy}.

\textbf{\underline{Spontaneous scalarization} ($\beta>0$):} 
Let us first consider the criterion for the \textit{spontaneous scalarization} of \bh{s} with positive dimensionless coupling. 
A single non-spinning \bh{} of mass $m=\mathrm{M}=1$ experiences the onset of scalarization for a critical coupling of $\beta_{\rm c,1}=1.451$~\cite{Silva:2017uqg}. In the test field approximation, the \bh{} scalarizes for any dimensionless coupling above this value.  
For a \bh{} with a different mass, $m'$, expressed in the same unit system, the (dimensionless) critical coupling would be the same, but the total mass would differ (i.e., $m'=\mathrm{M}\neq1$). Consequently, the dimensionful critical coupling would be $\alpha_{\rm GB,c}(m')=m'^2 \beta_{\rm c,1}$. 

If we consider a system of isolated \bh{s} with individual masses, $m_{\rm \mathfrak{n}}$, the expression for the dimensionful critical coupling still applies.
However, the mass of a single \bh{} no longer coincides with the total system mass, $\mathrm{M}=\sum m_{\rm \mathfrak{n}}$. Thus, the formula for the (dimensionless) critical coupling of the $\mathrm{\mathfrak{n}}$-th \bh{} is
\begin{equation}
\label{eq:beta_c_nonspinning}
\beta_{\rm c}(m_{\rm \mathfrak{n}}) = (m_{\rm \mathfrak{n}}/\mathrm{M})^2 \beta_{\rm c,1} \, .
\end{equation}
This formula can be used to find the critical coupling for any non-spinning \bh{}. In the test field approximation, a \bh{} scalarizes if the dimensionless coupling is greater than its critical coupling, $\beta\geq\beta_{\rm c}(m_{\rm \mathfrak{n}})$. In a similar manner, if a system has a dimensionless coupling $\beta=\beta_{\rm c}(m_{\rm c})$, then \bh{s} with mass $m_{\rm \mathfrak{n}}\leq m_{\rm c}$ scalarize and \bh{s} with mass $m_{\rm \mathfrak{n}}>m_{\rm c}$ descalarize (or remain unscalarized). For systems with a positive coupling constant, increasing spin tends to suppress scalarization~\cite{Cunha:2019dwb}. However, we do not consider this effect here.

\textbf{\underline{Spin-induced scalarization} ($\beta<0$):} Let us now consider the criterion for the \textit{spin-induced scalarization} of \bh{s} with spin magnitude $|\chi| > 0.5$ and negative dimensionless coupling.
We are unaware of an exact solution for the spin dependence of the critical coupling.
However, Ref.~\cite{Elley:2022ept} used data from Ref.~\cite{Herdeiro:2020wei} to find a best fit for the spin-dependent critical coupling of a \bh{} expressed in units of its own mass (i.e., $m=\mathrm{M}=1$), 
\begin{equation}
\label{eq:beta_c_spinning}
\beta_{\rm c,1}(\chi)=-\frac{0.422}{(|\chi|-1/2)^2}+1.487|\chi|^{7.551} \, .    
\end{equation}
The fit has an error of $\sim 5$\% for spin magnitudes $|\chi|<0.74$ and of $\sim 15$\% for spin magnitudes $0.74<|\chi|<1$ \cite{Elley:2022ept}.
The dimensional argument used to derive the mass scaling in Eq.~\eqref{eq:beta_c_nonspinning} also applies here.
Thus, in tandem with Eq.~\eqref{eq:beta_c_nonspinning}, this formula can approximate the critical coupling for the \textit{spin-induced scalarization} of a \bh{} with any mass. In this case, a \bh{} scalarizes if the dimensionless coupling is smaller (i.e., more negative) than its critical coupling, $\beta\leq\beta_{\rm c}(m_{\rm \mathfrak{n}},\chi)$. Furthermore, if a system has a dimensionless coupling $\beta=\beta_{\rm c}(m_{\rm \mathfrak{n}},\chi_{\rm c})$, then \bh{s} with mass $m_{\rm \mathfrak{n}}$ scalarize for spins $|\chi|\geq|\chi_{\rm c}|$ and descalarize (or remain unscalarized) for spins $|\chi|<|\chi_{\rm c}|$.

\textbf{\underline{Scalar charge:}} Once a \bh{} scalarizes, it is said to acquire a scalar hair. 
Asymptotically, the scalar field has a radial dependence similar to the electric potential of a point charge,
\begin{equation}
\label{eq:scalar_charge}
\phi(r)= \Psi + Q_{\rm s}/r + \mathcal{O}(r^{-2}) \, .    
\end{equation}
Therefore, the \bh{} is also said to acquire a scalar charge, which is denoted by $Q_{\rm s}$ \cite{Silva:2017uqg}.

\section{Numerical Framework}
\label{sec:NRFramework}

In this work, we simulate a series of \bh{} binaries in quadratic \sGB gravity using the decoupling approximation, where the scalar field does not backreact onto the metric. The metric thus evolves according to \gr. We follow Refs.~\cite{Silva:2020omi,Elley:2022ept} and adopt the procedures laid out in Refs.~\cite{Witek:2018dmd,Benkel:2016kcq,Benkel:2016rlz}.

\subsection{Spacetime evolution}
\label{sec:NRFramework_spacetimeevolution}
We utilize a 3+1 implementation of numerical relativity, where the four dimensional spacetime is foliated into a set of three dimensional spatial hypersurfaces parameterized by a time coordinate, $t$; see e.g. Refs.~\cite{Alcubierre:2008,Baumgarte:2021skc,shibata2016numerical}. The hypersurfaces are described by the 3-metric, $\gamma_{ij}$, and embedded into the four dimensional spacetime as prescribed by their orthogonal unit normal vector, $n^\mu$. The unit normal vector is then used to define the projection operator, $P^\alpha_\beta=\delta^\alpha_\beta + n^\alpha n_\beta$, which projects tensors onto the hypersurfaces. For example, the covariant derivative on a 3-metric is given by $D_i=P^\alpha_i \nabla_\alpha$. Furthermore, the extrinsic curvature of a hypersurface is given by 
\begin{equation}
\label{eq:extrinsic_curvature}
K_{ij}=-P^\alpha_i P^\beta_j \nabla_\alpha n_\beta=-\frac{1}{2}\mathcal{L}_{\vec{n}}\gamma_{ij} \, ,
\end{equation}
where $\mathcal{L}_{\vec{n}}$ is the Lie derivative with respect to the normal vector. Given initial data for the 3-metric of a hypersurface and its extrinsic curvature, a system can be evolved in time. Here we summarize the mathematical formulation. 

In terms of the 3-metric, the spacetime metric is
\begin{equation}
ds^2=-(\alpha^2-\beta^i\beta_i)dt^2 +2\beta_idx^idt + \gamma_{ij}dx^idx^j,
\end{equation}
where the lapse function, $\alpha$, and the shift vector, $\beta^i$, relate the hypersurfaces at different times. 
Namely, an observer that is stationary with respect to a hypersurface is separated from their position on the next hypersurface by a proper time of $\alpha dt$. 
Their position within the next hypersurface is then given by $x^i_{t+dt}=x^i_t-\beta^i dt$. The lapse and shift are ultimately coordinate dependent and thus provide a choice of gauge. 

To generate initial data for the \gr background, we use the Bowen-York method as extended by Brandt and Brügmann~\cite{Bowen:1980yu,Brandt:1997tf} to solve the constraint equations
\begin{subequations}
\label{eq:Constraint}
\begin{align}
\label{eq:Hamiltonian_Constraint}
& {}^{(3)}R + K^2-K_{ij}K^{ij} = 0\, ,
\\
\label{eq:Momentum_Constraint}
 & D_{j}\left( K^{ij}-\gamma^{ij}K \right)=0 \, ,
\end{align}
\end{subequations}
for the 3-metric and extrinsic curvature in vacuum. Here, ${}^{(3)}R$ is the three dimensional Ricci scalar describing the intrinsic curvature of a hypersurface, and $K=\gamma^{ij}K_{ij}$ is the trace of the extrinsic curvature; see e.g. Refs.~\cite{Alcubierre:2008,Baumgarte:2021skc,shibata2016numerical} for further details on the constraint equations. In Arnowitt-Deser-Misner-York form, the vacuum evolution equations are written as
\begin{subequations}
\label{eq:ADM_evolution}
\begin{align}
\label{eq:ADM_evolution_1}
 &\partial_{t}\gamma_{ij}=-2\alpha K_{ij}+\mathcal{L}_{\vec{\beta}}\gamma_{ij} \, ,
\\
\label{eq:ADM_evolution_2}
& \begin{aligned}
 \partial_{t}K_{ij}=&-D_{i}D_{j}\alpha + \mathcal{L}_{\vec{\beta}} K_{ij} \\
 & +\alpha\left[ {}^{(3)}R_{ij} + K K_{ij} -2 K_{ik} K^k_j \right] \, ,
 \end{aligned}
\end{align}
\end{subequations}
where ${}^{(3)}R_{ij}$ is the three dimensional Ricci tensor describing the intrinsic curvature of a hypersurface, and $\mathcal{L}_{\vec{\beta}}$ is the Lie derivative along the shift vector. To evolve the simulations, we use the \bssn implementation of the 3+1 formalism~\cite{Shibata:1995we,Baumgarte:1998te} together with moving puncture gauge~\cite{Campanelli:2005dd,Baker:2005vv}.

\subsection{Scalar evolution}
\label{sec:NRFramework_scalarevolution}

Let us now consider the numerical description of the scalar field. 
In analogy to the extrinsic curvature in Eq.~\eqref{eq:extrinsic_curvature}, we introduce the momentum of the scalar field as
\begin{equation}
\label{eq:scalar_momentum}
K_{\phi}=-\frac{1}{2}\mathcal{L}_{\vec{n}}\phi\, .
\end{equation}
We initialize the scalar field around each \bh{} using the bound state solution of an isolated, non-spinning \bh{} obtained by Ref.~\cite{Silva:2017uqg}.
Following Refs.~\cite{Silva:2020omi,Elley:2022ept}, we thus approximate the solution of a many-body system using the formulae,
\begin{subequations} \label{eq:initialscalar}
\begin{align}
\label{eq:initialscalar_sum}
&\phi|_{t=0}=\sum_{\rm \mathfrak{n}} \phi_{\rm \mathfrak{n}} \, , \hspace{0.7cm} K_{\phi}|_{t=0}=0 \, , 
\\
\label{eq:initialscalar_n}
&\phi_{\rm \mathfrak{n}}=c_{\rm 0} \left[ c_{\rm 1} \rho_{\rm \mathfrak{n}} + c_{\rm 2}\rho_{\rm \mathfrak{n}}^2 + c_{\rm 3} \rho_{\rm \mathfrak{n}}^3 \right] \, ,
\\
\label{eq:initialscalar_rho}
&\rho_{\rm \mathfrak{n}}=\frac{ m_{\rm \mathfrak{n}} r_{\rm \mathfrak{n}} }{(m_{\rm \mathfrak{n}} + 2r_{\rm \mathfrak{n}})^2} \, ,
\end{align}
\end{subequations}
where $\phi_{\rm \mathfrak{n}}$ is the bound state solution of the $\mathfrak{n}$-th \bh{} and $r_{\rm \mathfrak{n}}$ is the distance to the $\mathfrak{n}$-th \bh{} in quasiisotrpic radial coordinates. 
The coefficients $c_{\rm i}$ are expansion parameters used to fit the numerical solution in Ref.~\cite{Silva:2017uqg}. They are fit to be $c_{\rm 1}=3.68375$, $c_{\rm 2}=4.97242$, and $c_{\rm 3}=2.29938\times10^2$. The parameter $c_{\rm 0}$ sets the initial scalar field amplitude. 

In the following, we consider \bh{} binaries, where the \bh{s} begin widely separated.
Note that Eq.~\eqref{eq:scalareom} is linear and homogeneous with respect to the scalar field. It is therefore reasonable to approximate the initial scalar field as a sum of bound states around the individual \bh{s}. Moreover, the initial scalar field amplitude only influences the evolution up to an overall constant and is thus arbitrary. 
Note that in Eq.~\eqref{eq:initialscalar} we assume the field is momentarily at rest, i.e., its momenta is set to zero. Furthermore, Eq.~\eqref{eq:initialscalar} does not take into account a \bh{'s} spin. 
Therefore, the scalar field takes time to equilibrate. For a backreacted solution see Ref.~\cite{Nee:2024bur}.

Once the initial data for the scalar field is specified, it is evolved according to the equations, 
\begin{subequations}
\label{eq:scalar_evolution}
\begin{align}
\label{eq:scalar_evolution_1}
(\partial_{t}-\mathcal{L}_{\vec{\beta}}) \phi = & -\alpha K_{\phi}
\, ,\\
\label{eq:scalar_evolution_2}
(\partial_{t}-\mathcal{L}_{\vec{\beta}}) K_{\phi}  = &
    -D^i \alpha D_i \phi
\\ &
  - \alpha \left(D^i D_i \phi - K K_{\phi} + \frac{1}{2} \alpha_{\rm GB} \mathcal{G} \phi \right)
\, .\nonumber
\end{align}
\end{subequations}
Since we work in the decoupling approximation, the Gauss-Bonnet invariant is calculated from the curvature of the time-dependent background spacetime~\cite{Witek:2018dmd,Benkel:2016kcq,Benkel:2016rlz}.

We note that every system must be provided with some initial scalar field as otherwise there would be no scalar dynamics; see Eq.~\eqref{eq:scalareom}. Therefore, we seed every \bh{} with an initial scalar field using Eq.~\eqref{eq:initialscalar}. 
However, depending on the parameters of a system, the \bh{s} may not be able to sustain a scalar field, and thus the scalar field decays; see Sec.~\ref{sec:sGB_Theory_coupling}. 
In the following, we refer to these \bh{s} as unscalarized.

\subsection{Code}

We run simulations with the \ETK~\cite{maxwell_rizzo_2025_15520463,Loffler:2011ay, Zilhao:2013hia}, an open-source software for computational astrophysics, and the \canuda~code~\cite{witek_2023_7791842,Okawa:2014nda,Zilhao:2015tya,Witek:2018dmd,Silva:2020omi,Richards:2025ows} for fundamental physics.
The \ETK~is built upon the \textsc{Cactus} computational framework \cite{Goodale2002a,Cactuscode} and uses \textsc{Carpet} \cite{Schnetter:2003rb,Carpetbitbucket} to implement box-in-box adaptive mesh refinement as well as hybridized message passing interface and open multiprocessing parallelization.

We generate the initial data for the background metric using the \verb|TwoPunctures| spectral thorn \cite{Ansorg:2004ds}. The initial data for the scalar field is generated using \canuda's \verb|EdGB_dec_Init| thorn. We then evolve the system using \canuda's \verb|LeanBSSNMoL|\footnote{This thorn is adapted from the \texttt{Lean} code \cite{Sperhake:2006cy}.} and \verb|EdGB_dec_Evol| thorns, which handle the background and scalar evolutions, respectively. \verb|LeanBSSNMoL| and \verb|EdGB_dec_Evol| provide up to eighth order finite differences for spatial derivatives.
Here, we use fourth order finite differencing for spatial derivatives and employ the fourth order Runge-Kutta scheme for the time integration. The system is evolved via the method of lines.

\subsection{Extraction of variables}

In order to analyze the simulations, we require a variety of information about the scalar field and spacetime evolution.
The scalar field is handled by \canuda's \verb|EdGB_dec_Base| thorn, and the Gauss-Bonnet invariant is calculated from the background in \canuda's \verb|EdGB_dec_Evol| thorn. In order to visualize the behavior of these quantities, we use the \verb|CarpetIOHDF5| thorn to output the values of the quantities across two dimensional cross sections of our grid as a function of time. Furthermore, we decompose the scalar field into a basis of spherical harmonics,
\begin{equation} \label{eq:phi_mode_extract}
    \phi_{\rm \mathfrak{l}\mathfrak{m}}(t,\rex)=\int   \phi(t,\rex,\theta,\phi) Y^*_{\mathfrak{l}\mathfrak{m}}(\theta,\phi) d\Omega\, ,
\end{equation}
on spheres of constant extraction radii, $\rex$, using the \verb|Multipole| thorn~\cite{MultipoleThorn}. Here, $Y^*_{\mathfrak{l}\mathfrak{m}}(\theta,\phi)$ are the complex conjugates of the spherical harmonics, $Y_{\mathfrak{l}\mathfrak{m}}(\theta,\phi)$, with spin-weight zero. The effective scalar charge of a system is then given by $\rex \phi_{\rm 00}$ for a sufficiently large extraction radius; see Eq.~\eqref{eq:scalar_charge}. 

We compute the \bh{} apparent horizons and their properties, such as the horizon area, $A_{\rm H}$, and equatorial circumference, $C_{\rm e}$, using the \verb|AHFinderDirect| thorn \cite{Thornburg:1995cp,Thornburg:2003sf}. Furthermore, we use these quantities in postprocessing to compute the (dimensionless) spin,
\begin{equation} \label{eq:chi_extract}
    \chi=\sqrt{1-\left( \frac{2\pi A_{\rm H}}{C_{\rm e}^2}-1 \right)^2} \, ,
\end{equation}
of the \bh{s}. Note that the dimensionless spin is only well defined when the \bh{s} are relatively isolated. We find that other common methods for computing the spin, such as via a ratio of horizon circumferences or quasilocal measures, give similar results; see also Refs.~\cite{Anninos:1994pa,Kiuchi:2009jt,Nelson:2019czq}. 
We also use the \verb|AHFinderDirect| thorn to track the position of the \bh{} punctures in tandem with the \verb|PunctureTracker| thorn~\cite{PunctureTrackerThorn}. Finally, the \verb|AHFinderDirect| thorn outputs the minimum and maximum radii of the \bh{} horizons, which we use to reconstruct the approximate shape of the horizons during the evolution.

\subsection{Grid setup and refinement}

We evolve the \bh{} binaries on a three dimensional cartesian grid with outer boundary located at $x,y,z=\pm256\mathrm{M}$, where $\mathrm{M}$ is the total mass of the system, and $\mathrm{M=1}$ in code units. 
To reduce the computational cost, we impose rotation symmetry about the z-axis and reflection symmetry across the xy-plane. 
We use \textsc{Carpet} to employ seven levels of box-in-box adaptive mesh refinement centered around the \bh{s}.
The outermost refinement level has a resolution with step size $dx=1\mathrm{M}$.
Within consecutive refinement levels, we halve the step size such that the innermost refinement level has step size $dx=\frac{1}{64}\mathrm{M}$.
We set the Courant factor to \verb|dtfac| $=0.225$.

We use two different setups for the refinement levels around the \bh{s}. The first (setup A) places the refinement boundaries at radii $r/\rm{M}=\{64.0, 16.0, 6.0, 3.0, 1.5, 0.75\}$ around the \bh{} centers. The second setup (setup B) places the refinement boundaries at radii $r/\rm{M}=\{64.0, 16.0, 4.0, 2.0, 1.0, 0.6\}$.
We use setup A in the majority of cases and setup B in systems where the \bh{s} have zero initial spin. We present a series of convergence tests in the Appendix.

\section{Setup}
\label{sec:Setup}

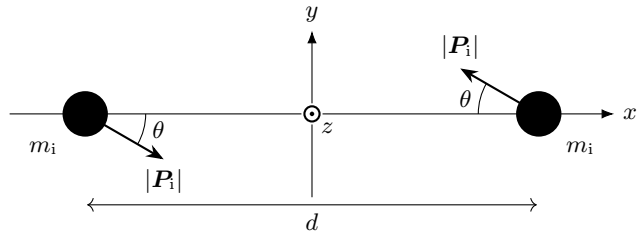
\begin{figure}[htbp!]
    \begin{center}
  \begin{tikzpicture}
    \fill[black] (-3 , 0) circle (0.3 ) node[below left=0.35] {$m_{\rm i}$};
    \draw[thick,-Stealth] (-3,0) -- (-3+0.866*1.2,-0.5*1.2) node[below] {$|\vec{P}_{\rm i}|$};
    \draw (-3+0.8,0) arc (0:-30:0.8) node[midway, right] {$\theta$};

    \fill[black] (3, 0) circle (0.3 ) node[below right=0.35] {$m_{\rm i}$};
    \draw[thick,-Stealth] (3,0) -- (3-0.866*1.2,0.5*1.2) node[above] {$|\vec{P}_{\rm i}|$};
    \draw (3-0.8,0) arc (180:150:0.8) node[midway, left] {$\theta$};

    \draw[-Latex] (-4,0) -- (4,0) node[right] {$x$};
    \draw[-Latex] (0,-1.1) -- (0,1.1) node[above] {$y$};
    \fill[white] (0 , 0) circle (0.14 );
    \draw[thick] (0 , 0) circle (0.1 ) node[below right] {$z$};
    \fill[black] (0 , 0) circle (0.03 );

    \draw[<->] (-2.975,-1.2) -- (2.975,-1.2) node[midway,below] {$d$};
  \end{tikzpicture}
    \end{center}
    \caption{ \label{fig:Setup_GR_background} Diagram illustrating the initial setup of the \gr background for the \bh{} binary systems that we consider. The system has rotational symmetry about the z-axis such that the \bh{s} have equal initial mass, $m_{\rm i}$, initial momentum, $|\vec{P}_{\rm i}|$, and incident angle, $\theta$. The \bh{s} have initial separation $d=100\mathrm{M}$.}
\end{figure}

In the following, we analyze the evolution of a scalar field in the (\gr) background of \bh{} binaries with the initial setup depicted in Fig.~\ref{fig:Setup_GR_background}.
The \bh{s} have equal initial masses $m_{\rm i}=\mathrm{M}/2$, where $\mathrm{M}$ is the total system mass. Note that the \bh{s} have equal, but opposite, initial momenta, $|\vec{P}_{\rm i}|$, inclined at an incident angle, $\theta$, with respect to the x-axis. The \bh{s} are equidistant from the z-axis with initial separation $d=100\mathrm{M}$. Consequently, the systems are rotationally symmetric about the z-axis.

In several of the simulations, we also provide the \bh{s} with an initial (dimensionless) spin given by the parameter $\chi_{\rm 0}$. 
We orient the \bh{} spins along the z-axis such that they are both aligned (+$z$) or anti-aligned (-$z$) to the orbital angular momentum. When referring to the spin, $\chi$, we therefore refer solely to the z-component and take positive (negative) values to refer to aligned (anti-aligned) spins. Note that the spins maintain rotational symmetry.
The \bh{s'} (\gr) configuration is the same as in Ref.~\cite{Kogan:2025vml}. Here, we consider \bh{s} with initial spin parameter $\chi_{\rm  0}=0.0$ and initial momentum $|\vec{P}_{\rm i}|=0.245\mathrm{M}$, as well as \bh{s} with initial spin parameters $\chi_{\rm  0}=\{-0.7, 0.7 \}$ and initial momentum $|\vec{P}_{\rm i}|=0.490\mathrm{M}$.

For sufficiently large initial momenta, such \bh{s} are initially unbound in a Newtonian sense and are thus considered to be in a hyperbolic configuration.
However, as they pass one another during a close encounter, they lose energy to gravitational radiation, which can cause them to become bound and undergo dynamical capture. This results in three distinct morphologies as the incident angle is increased: (1) direct mergers, where the \bh{s} merge during the first encounter; (2) zoom whirls, where the \bh{s} undergo a series of close encounters (i.e., small orbits) punctuated by relatively large eccentric orbits until they lose sufficient energy such that they merge; and (3) scattering, where the \bh{s} separate and escape to infinity after a single close encounter~\cite{Pretorius:2007jn,Shibata:2008rq,Sperhake:2009jz,Healy:2009zm,Gold:2009hr,Gold:2012tk}.

\subsection{Spin-up of scattering black holes}
\label{sec:Setup_spinup}

When \bh{s} undergo a close encounter, they emit and reabsorb gravitational radiation. This gravitational radiation carries energy and angular momentum, which changes the mass, angular momentum, and ultimately the (dimensionless) spin of the \bh{s}~\cite{Sperhake:2012me,Nelson:2019czq,Jaraba:2021ces,Rodriguez-Monteverde:2024tnt,Chiaramello:2024unv,Kogan:2025vml}. This change in parameters usually leads to a spin-up, where the spin increases along the direction of a system's orbital angular momentum (i.e., the z-axis), although a spin-down, where the spin decreaces with respect to the direction of the orbital angular momentum, is also possible~\cite{Sperhake:2012me,Rodriguez-Monteverde:2024tnt,Chiaramello:2024unv,Kogan:2025vml}. 

The change in spin can be quantified as the difference between the final spin and the initial spin, 
\begin{equation}
\label{eq:change_in_spin}
\Delta\chi=\chi_{\rm f} - \chi_{\rm i} \, ,
\end{equation}
where here we compute the final spin, $\chi_{\rm f}$, and initial spin, $\chi_{\rm i}$, according to Eq.~\eqref{eq:chi_extract}. We compute the initial spin at a time, $t_{\rm i}=45 \mathrm{M}$, approximately halfway between the start of the simulation and the close encounter (i.e., after the initial gauge adjustment). We compute the final spin at a time, $t_{\rm f}=270 \mathrm{M}$, long after the encounter. Note that the initial spin computed from Eq.~\eqref{eq:chi_extract} typically differs slightly ($<1\%$) from the initial spin parameter, $\chi_{\rm 0}$, due to numerical and theoretical limitations in the initial data construction of spinning \bh{s}.

\begin{figure}[htbp!]
    \centering
    \includegraphics[width=\columnwidth]{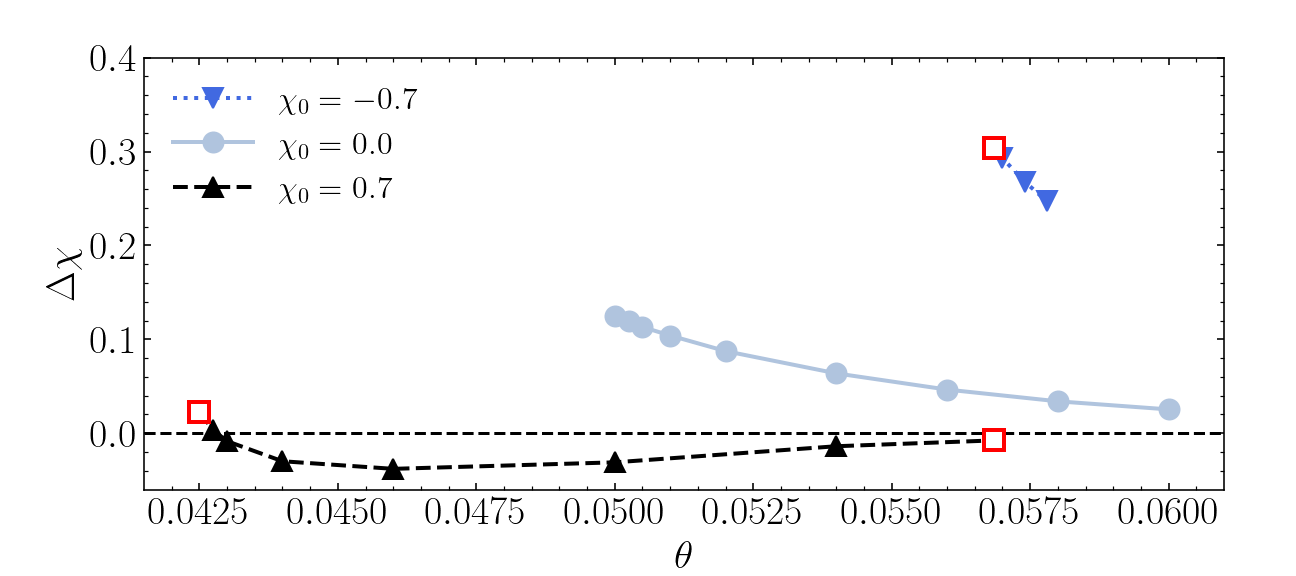}
    \caption{\label{fig:Spinup_Explanation} The change in spin, $\Delta\chi$, of scattering binary \bh{s} with initial momentum $|\vec{P}_{\rm i}|=0.490\mathrm{M}$ and initial spin parameters $\chi_{\rm0}=\{-0.7,0.0,0.7\}$ as a function of the incident angle, $\theta$. The plot is adapted from Fig.~6 in Ref.~\cite{Kogan:2025vml}. In this work, we consider systems with the same \gr parameters as those denoted by the red squares.
    }
\end{figure}

In this work, we are concerned with the role that \bh{} spins, and their change during an encounter, may play in the \textit{spin-induced scalarization} of \bh{s}, which can only occur for spin magnitudes $|\chi|>0.5$; see Eq.~\eqref{eq:beta_c_spinning}. 
Therefore, we consider systems with initial spin parameters of magnitude $|\chi_{\rm 0}|=0.7$. In Fig.~\ref{fig:Spinup_Explanation}, we plot the change in spin of scattering \bh{} binaries against their incident angle for \bh{s} with initial momenta $|\vec{P}_{\rm i}|=0.490\mathrm{M}$ and initial spin parameters $\chi_{\rm 0}=\{-0.7,0.0,0.7\}$; the plot is adapted from Fig.~6 in Ref.~\cite{Kogan:2025vml}. We include initial spin parameter $\chi_{\rm 0}=0.0$ for comparison. 
Typically, the spin-up of scattering \bh{s} increases with decreasing initial spins and incident angles.
The behavior of simulations with initial spin parameter $\chi_{\rm 0}=0.7$ differs and exhibits a spin-down at intermediate angles. At large incident angles the change in spin approaches zero. 

The red squares in Fig.~\ref{fig:Spinup_Explanation} denote systems with the same \gr parameters as those studied below. The setup with initial spin parameter $\chi_{\rm 0}=0.7$ and small incident angle yields an increase in spin magnitude. The setup with initial spin parameter $\chi_{\rm 0}=-0.7$ leads to a decrease in spin magnitude. The setup with initial spin parameter $\chi_{\rm 0}=0.7$ and large incident angle corresponds to a minute change in spin. Such systems are useful as we consider the construction of hyperbolic encounters in which different types of scalarization may occur.

\subsection{Simulation Suite}
\label{sec:Setup_SimulationSuite}

\begin{figure*}[htbp!]
    \centering
    \subfloat[ExptBp0355X0 \label{fig:Feynmann_NoSpin}]{
        \resizebox{!}{0.5\columnwidth}{
            \begin{tikzpicture}
            \begin{feynman}
            \vertex (c);
            \vertex [above left=of c] (a) {\( \overline S \)};
            \vertex [below=1 of c] (d);
            \vertex [below left=of d] (b) {\( \overline S \)};
            \vertex [right=0.7of c] (e) ;
            \vertex [right=0.7of d] (f) ;
            \vertex [below=0.5of e] (g) ;
            \vertex [right=0.7of g] (h) ;
            \vertex [right=0.7of h] (i) {\( \overline S \uparrow \)};
            \vertex [above=0.24 of h] (j) {};
            \vertex [right=0.23 of j] (k) {\( S{-}S \)};
            
            \diagram* {
            (a) -- (c),
            (d) -- [boson, edge label=\( S{-}S \)] (c),
            (b) -- (d),
            (c) -- [edge label=\( \overline S \)] (e),
            (f) -- [edge label=\( \overline S \)] (d),
            (e) -- [bend left] (h),
            (f) -- [bend right] (h),
            (h) -- (i),
            };
            \end{feynman}
            \end{tikzpicture}
        }
    }
    \hfill
    \subfloat[ExptBm300Xm07 \label{fig:Feynmann_SpinEncounter}]{
        \resizebox{!}{0.5\columnwidth}{
            \begin{tikzpicture}
            \begin{feynman}
            \vertex (c);
            \vertex [above left=of c] (a) {\( \overline S \downarrow \)};
            \vertex [below =1of c] (d);
            \vertex [below left=of d] (b) {\( \overline S \downarrow \)};
            \vertex [above right=of c] (e) {\( \overline S \downarrow \)};
            \vertex [below right=of d] (f) {\( \overline S \downarrow \)};
            \diagram* {
            (a) -- (c),
            (b) -- (d),
            (c) -- [boson, edge label=\( S{-}S \)] (d),
            (c) -- (e),
            (d) -- (f),
            };
            \end{feynman}
            \end{tikzpicture}
        }
    }
    \hfill
    \subfloat[ExptBm300Xp07\label{fig:Feynmann_SpinScalarize}]{
        \resizebox{!}{0.5\columnwidth}{
            \begin{tikzpicture}
            \begin{feynman}
            \vertex (c);
            \vertex [above left=of c] (a) {\( \overline S \uparrow \)};
            \vertex [below =1of c] (d);
            \vertex [below left=of d] (b) {\( \overline S \uparrow \)};
            \vertex [above right=of c] (e) {\( S \Uparrow \)};
            \vertex [below right=of d] (f) {\( S \Uparrow \)};
            \diagram* {
            (a) -- (c),
            (b) -- (d),
            (c) -- [boson, edge label=\( S{-}S \)] (d),
            (c) -- (e),
            (d) -- (f),
            };
            \end{feynman}
            \end{tikzpicture}
        }
    }
    \hfill
    \subfloat[ExptBm350Xm07 \label{fig:Feynmann_SpinDeScalarize}]{
        \resizebox{!}{0.5\columnwidth}{
            \begin{tikzpicture}
            \begin{feynman}
            \vertex (c);
            \vertex [above left=of c] (a) {\( S \Downarrow \)};
            \vertex [below =1of c] (d);
            \vertex [below left=of d] (b) {\( S \Downarrow \)};
            \vertex [above right=of c] (e) {\( \overline S \downarrow \)};
            \vertex [below right=of d] (f) {\( \overline S \downarrow \)};
            \diagram* {
            (a) -- (c),
            (b) -- (d),
            (c) -- [boson, edge label=\( S{-}S \)] (d),
            (c) -- (e),
            (d) -- (f),
            };
            \end{feynman}
            \end{tikzpicture}
        }
    }
    \caption{
        \label{fig:Feynmann_Diagrams}
        Schematic depiction of the phenomenology of close encounters in quadratic \sGB gravity. Time moves from left to right. Black lines denote the trajectories of the \bh{s}. Wavy lines denote the interaction of the \bh{s} during a close encounter. The symbol \(S\) (\(\overline{S}\)) denotes scalarized (unscalarized) \bh{s}. Arrows \(\uparrow\) (\(\downarrow\)) denote spins aligned (anti-aligned) with the orbital angular momentum. When the change in spin is important to the dynamics of the process, arrows \(\Uparrow\) (\(\Downarrow\)) denote aligned (anti-aligned) spins with greater magnitude. \underline{Panel~\ref{fig:Feynmann_NoSpin}:} The \textit{dynamical scalarization} of two \bh{s} during a zoom-whirl and merger. \underline{Panel~\ref{fig:Feynmann_SpinEncounter}:} The \textit{spin-induced dynamical scalarization} of two scattering \bh{s}. \underline{Panel~\ref{fig:Feynmann_SpinScalarize}:} The permanent \textit{spin-up scalarization} of two scattering \bh{s} as a consequence of their spin-up. \underline{Panel~\ref{fig:Feynmann_SpinDeScalarize}:} The permanent \textit{spin-up descalarization} of two scattering \bh{s} as a consequence of their spin-up, which leads to a decrease their spin magnitude. 
    }
\end{figure*}
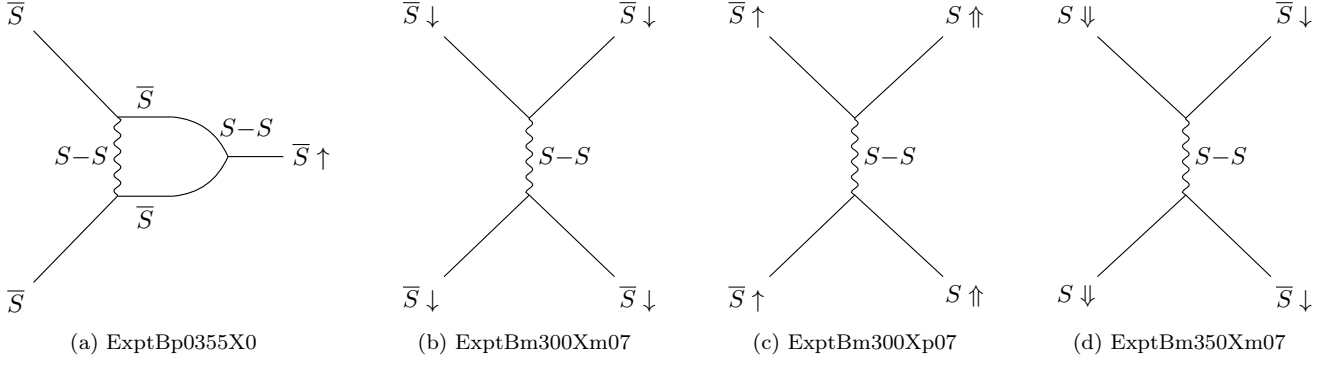

\begin{table*}[htbp!]
    \centering
    \begin{tabular}{|c|c|c|c|c|c|c|c|c|c|c|}
    \hline
     Run & $\beta$ & $c_{\rm 0}$ & $\chi_{\rm 0}$ & $\chi_{\rm i}$ & $\chi_{\rm f}$ & $\Delta \chi$ & Type & $|\vec{P}_{\rm i}|/\mathrm{M}$ & $\theta$ & Phenomenon \\
    \hline
     ExptBp0355X0 & $0.355$ & $1.0$ & $0.0$ & N/A & N/A & N/A & Zoom-Whirl & $0.245$ & $0.0580$ & Dynamical scalarization \\
     \hline
     CtrlBm300Xp07 & $-3.00$ & $1.0$ & $0.7$ & $0.695$ & $0.687$ & $-0.008$ & Scattering & $0.490$ & $0.05685$ & Control, unscalarized  \\
     ExptBm300Xm07 & $-3.00$ & $1.0$ & $-0.7$ & $-0.695$ & $-0.391$ & $0.304$ & Scattering & $0.490$ & $0.05685$ & Dynamical scalarization  \\
     ExptBm300Xp07 & $-3.00$ & $1.0$ & $0.7$ & $0.695$ & $0.718$ & $0.023$ & Scattering & $0.490$ & $0.0425$ & Spin-up scalarization \\
    \hline
     CtrlBm350Xp07 & $-3.50$ & $0.1$ & $0.7$ & $0.695$ & $0.687 $ & $-0.008$ & Scattering & $0.490$ & $0.05685$ & Control, scalarized \\
     ExptBm350Xm07 & $-3.50$ & $0.1$ & $-0.7$ & $-0.695$ & $-0.391$ & $0.304$ & Scattering & $0.490$ & $0.05685$ & Spin-up descalarization \\
     \hline
    \end{tabular}
    \caption{\label{tab:Simulation_Suite} List of simulations. For each simulation we provide a run name, the dimensionless coupling, $\beta$, and the initial scalar field amplitude, $c_{\rm 0}$. Furthermore, we provide the \bh{s'} initial spin parameter, $\chi_{\rm 0}$, as well as the initial spin, $\chi_{\rm i}$, final spin, $\chi_{\rm f}$, and change in spin, $\Delta\chi$, computed from Eq.~\eqref{eq:chi_extract}. We also provide the type of the encounter, the initial momentum of the \bh{s}, $|\vec{P}_{\rm i}|$, and the incident angle, $\theta$. Finally, we state the phenomenon each simulation is designed to display. Each system has an initial separation of $d=100\mathrm{M}$. }
\end{table*}

The fact that \bh{s} can undergo extremely close flybys without forming a \CAH while scattering and during the first encounter(s) of a zoom-whirl, makes such morphologies natural settings in which to search for \textit{dynamical scalarization}~\cite{Julie:2023ncq,Nee:2024bur,Lara:2025kzj}. 
Furthermore, the spin-up of \bh{s} during such encounters makes the scattering of initially spinning \bh{s} an interesting setting in which to look for \textit{spin-induced scalarization} (or \textit{descalarization}). 
Here we have two goals: We seek to observe instances of (1) temporary \textit{dynamical scalarization} during the course of these encounters and (2) permanent \textit{spin-induced scalarization} (or \textit{descalarization}) owing to the spin-up of the \bh{s}, which we call \textit{spin-up scalarization} (or \textit{descalarization}). 

In order to achieve these goals, we run a series of simulations depicted schematically in Fig.~\ref{fig:Feynmann_Diagrams} and described in Table~\ref{tab:Simulation_Suite}. Fig.~\ref{fig:Feynmann_Diagrams} contains “Feynmann”-like diagrams illustrating the phenomenology of different numerical “experiments.” Time moves from left to right. We denote \bh{} trajectories by solid black lines. Close encounters are denoted by wavy lines, and a merger is denoted by the intersection of two solid lines. We denote segments of the trajectories in which \bh{s} are scalarized (unscalarized) with the symbol $S$ ($\overline S$). Note that every \bh{} is seeded with some initial scalar field; see Sec.~\ref{sec:NRFramework_scalarevolution}. However, around “unscalarized” \bh{s}, the scalar field decays. We denote the combined scalarization of the \bh{s} due to a close encounter as $S{-}S$. We denote positive (negative) spins aligned (anti-aligned) to the orbital angular momentum with arrows $\uparrow$ ($\downarrow$). In systems where the change in spin magnitude is important to the scalar dynamics, we use arrows $\Uparrow$ ($\Downarrow$) to denote spins with relatively large magnitude. Table~\ref{tab:Simulation_Suite} lists the simulation parameters and other information describing the experiments. Table~\ref{tab:Simulation_Suite} also details a set of “control” simulations to be discussed in due course. 

We now summarize these simulations and their outcomes. We first consider a binary of non-spinning \bh{s} that undergoes a zoom-whirl and merger. We then discuss two control simulations for binaries with spinning \bh{s} and negative dimensionless coupling. Finally, we describe a series of scattering encounters between spinning \bh{s}, which constitute the remainder of our experiment runs.

\subsubsection{Numerical experiment without spin}
\label{sec:Setup_withoutspin}

\textbf{ExptBp0355X0}, depicted in Fig.~\ref{fig:Feynmann_NoSpin} and listed in Table~\ref{tab:Simulation_Suite}, is set up to achieve the first goal of finding \textit{dynamical scalarization} during the course of a close encounter. In this case, we consider \bh{s} that are initially non-spinning. The parameters of ExptBp0355X0 are chosen such that the \bh{s} undergo a single zoom-whirl and then merge. 
In order to see if we can observe \textit{dynamical scalarization}, we set the dimensionless coupling just below the critical value of the positive dimensionless coupling given by Eq.~\eqref{eq:beta_c_nonspinning} in Sec.~\ref{sec:sGB_Theory}, such that the \bh{s} begin unscalarized. For a \bh{} binary system with equal initial masses $m_{\rm i}=\mathrm{M}/2$, the critical coupling given by Eq.~\eqref{eq:beta_c_nonspinning} is $\beta_{\rm c}=0.36275$. 
In this simulation, we therefore set the dimensionless coupling slightly below that value at $\beta=0.355$. The results are displayed in Sec.~\ref{sec:results_NoSpin}. We ultimately find that the system does temporarily display \textit{dynamical scalarization} during the first encounter. We then find that the \bh{s} begin to descalarize during the orbit between encounters, briefly show some signs of dynamically rescalarizing during the inspiral, and then descalarize after merger. The remnant \bh{} has a spin of $\chi=0.733$.

A natural followup question to ask is whether one can observe a spin-induced version of \textit{dynamical scalarization}. However, in order to consider scalarization in systems of initially spinning \bh{}, we must take extra care in considering the initial conditions of the scalar field.

\subsubsection{Control simulations}
\label{sec:Setup_controls}

When considering setups with a positive coupling constant and non-spinning \bh{s}, determining the \bh{s'} critical coupling and selecting an appropriate dimensionless coupling for the system is straightforward. 
In this case, the formula for the critical coupling, Eq.~\eqref{eq:beta_c_nonspinning}, is exact. Moreover, the initial data reliably produces the intended initial masses with high accuracy. Therefore, we can confidently control whether the \bh{s} begin scalarized or unscalarized by setting the dimensionless coupling above or below the critical value given by Eq.~\eqref{eq:beta_c_nonspinning}. 

For spinning \bh{s} in systems with a negative coupling constant, this process is far less straightforward.
First, the formula for the (negative) critical coupling in Eq.~\eqref{eq:beta_c_spinning} is merely a fit to numerical data and thus carries an inherent uncertainty. 
For example, the critical coupling $\beta_{\rm c}=-2.61$, associated with a spin magnitude of $|\chi|=0.7$ and a mass $m_{\rm \mathfrak{n}}=\mathrm{M}/2$, has an error of $\pm 0.13$; see Ref.~\cite{Elley:2022ept}. We also note that Ref.~\cite{Elley:2022ept} estimates an error of $15\%$ if the spin magnitude is above $|\chi|=0.74$. 

Another issue is that the initial spin parameter, $\chi_{\rm 0}$, deviates from the initial spin, $\chi_{\rm i}$, found using Eq.~\eqref{eq:chi_extract} after gauge adjustments. As reported in Table~\ref{tab:Simulation_Suite}, for simulations with an initial spin parameter of $\chi_{\rm 0}=0.7$, we find an initial spin of $\chi_{\rm i}=0.695$ after gauge adjustments. The critical coupling associated with a spin magnitude of $|\chi|=0.695$ and mass $m_{\rm \mathfrak{n}}=\mathrm{M}/2$ is $\beta_{\rm c}=-2.75$, and it has an error of $\pm 0.14$.

A further issue is that the measured spin can vary depending on the method used to extract it. 
The data in Ref.~\cite{Herdeiro:2020wei}, upon which Eq.~\eqref{eq:beta_c_spinning} is based, measures spin using the asymptotic behavior of semianalytically computed metric functions for a single \bh{}. We cannot apply such a method to a binary system simulated on a full numerical grid and thus do not know how closely our spin measurements correspond to those of Ref.~\cite{Herdeiro:2020wei}. 
While we use Eq.~\eqref{eq:beta_c_spinning} as an approximate guide for what the critical coupling of a spinning \bh{} in our simulations may be, we must find a way to verify whether initially spinning \bh{s} begin scalarized or unscalarized.

The initial scalar field is constructed assuming a non-spinning configuration; see Sec.~\ref{sec:NRFramework_scalarevolution}. 
Consequently, there is an initial period during which the scalar field adjusts to the geometry, making it difficult to determine whether a \bh{} is scalarizing or descalarizing in the short interval preceding an encounter.
We thus run a series of control simulations in which the incident angle is large, such that the spin-up during the encounter is marginal (see Fig.~\ref{fig:Spinup_Explanation}).
We can be reasonably certain that the behavior of the scalar field after the encounter matches its behavior before the encounter in these systems. 
Therefore, we can use them to verify the initial state of spinning \bh{s} in other systems with the same dimensionless coupling.

\textbf{CtrlBm300Xp07}, listed in Table~\ref{tab:Simulation_Suite}, is set up as a control run to verify scenarios in which \bh{s} with an initial spin parameter magnitude of $|\chi_{\rm 0}|=0.7$ are initially unscalarized. The simulation is chosen to have a large incident angle, such that the change in \bh{} parameters is marginal; the \bh{s} only experience a spin-down of $\Delta \chi=-0.008$. Through trial and error, the dimensionless coupling is chosen to be $\beta=-3.00$. The scalar field ultimately decays, suggesting that \bh{s} with this initial spin magnitude and dimensionless coupling are unscalarized.

\textbf{CtrlBm350Xp07}, listed in Table~\ref{tab:Simulation_Suite}, is set up as a control run to verify scenarios in which \bh{s} with an initial spin parameter magnitude of $|\chi_{\rm 0}|=0.7$ are initially scalarized. The background spacetime is the same as in CtrlBm300Xp07, but here the dimensionless coupling is chosen to be $\beta=-3.50$. Note that the initial scalar field amplitude is set to $c_{\rm 0}=0.1$, as is also the case in the ExptBm350Xm07 run. 
The scalar field ultimately grows exponentially, suggesting that \bh{s} with the same initial spin magnitude and dimensionless coupling are scalarized.
A more detailed analysis of the results from both control simulations can be found in Sec.~\ref{sec:results_controls}.

\subsubsection{Numerical experiments with spin}
\label{sec:Setup_experiments}

Now that we understand the initial conditions for scalarized and unscalarized \bh{} binaries with spin, we proceed with the study of \textit{spin-induced (de)scalarization}. We begin by returning to the matter of \textit{dynamical scalarization}.

\textbf{ExptBm300Xm07}, depicted in Fig.~\ref{fig:Feynmann_SpinEncounter} and listed in Table~\ref{tab:Simulation_Suite}, is set up to achieve the first goal of finding \textit{dynamical scalarization} in systems with initially spinning \bh{s} and a negative coupling constant.
The parameters of ExptBm300Xm07 are chosen such that the \bh{s} scatter and thus undergo a single close encounter before separating. 
The \bh{s} have an initial spin of $\chi_{\rm i}=-0.695$ and experience a spin-up of $\Delta \chi=0.304$ during the encounter, leading to a final spin of $\chi_{\rm f}=-0.391$.
In order to see if we can observe \textit{spin-induced dynamical scalarization}, we want the \bh{s} to begin (and end) unscalarized and thus set the dimensionless coupling to $\beta=-3.00$ in order to match the unscalarized control CtrlBm300Xp07. The results are displayed in Sec.~\ref{sec:results_SpinEncounter}. We ultimately find that the system does temporarily display \textit{spin-induced dynamical scalarization} during the encounter. 
Note that the change in spin is not necessary for this process and is, in fact, likely to hinder it.
We then find that the \bh{s} descalarize after the encounter because the individual \bh{s'} spins are still too low for \textit{spin-induced scalarization}.

\textbf{ExptBm300Xp07}, depicted in Fig.~\ref{fig:Feynmann_SpinScalarize} and listed in Table~\ref{tab:Simulation_Suite}, is set up to achieve the second goal of finding permanent \textit{spin-up scalarization} due to an increase in the \bh{} spin. The initial parameters of ExptBm300Xp07 are chosen such that the \bh{s} scatter and thus undergo a single close encounter, which results in a small spin-up of $\Delta\chi=0.023$. 
The \bh{s} have an initial spin of $\chi_{\rm i}=0.695$ and a final spin of $\chi_{\rm f}=0.718$.
In order to see if we can observe \textit{spin-up scalarization}, we want the \bh{s} to begin unscalarized and thus set the dimensionless coupling to $\beta=-3.00$.
The change in spin may then result in a system where the critical coupling of the final \bh{s} is less negative than the dimensionless coupling, leading the \bh{s} to scalarize. The results are displayed in Sec.~\ref{sec:results_SpinScalarize}. Like ExptBm300Xm07, the system first scalarizes during the encounter. However, after the encounter, the scalar field continues to grow, rather than decay, indicating that the \bh{s} become permanently scalarized. We thus find that the system does display \textit{spin-up scalarization} over the course of the encounter.

\textbf{ExptBm350Xm07}, depicted in Fig.~\ref{fig:Feynmann_SpinDeScalarize} and listed in Table~\ref{tab:Simulation_Suite}, is set up to achieve the second goal of finding permanent \textit{spin-up descalarization} due to changes in the \bh{} spin. The initial parameters of ExptBm350Xm07 are chosen such that the \bh{s} scatter and thus undergo a single close encounter, which results in a spin-up of $\Delta \chi=0.304$. The initial spin of the \bh{s} is negative, $\chi_{\rm i}=-0.695$, so the spin-up results in an overall decrease in the spin magnitude such that the final spin is $\chi_{\rm f}=-0.391$. 
In order to see if we can observe \textit{spin-up descalarization}, we want the \bh{s} to begin scalarized and thus set the dimensionless coupling to $\beta=-3.50$.
As the final spin is below the minimum value for \textit{spin-induced scalarization}, $|\chi_{\rm f}|<0.5$, we expect the final \bh{s} to descalarize.
Since the dimensionless coupling has a particularly large magnitude, we set the initial scalar field amplitude to $c_{\rm 0}=0.1$. As discussed in Sec.~\ref{sec:NRFramework}, this does not change the dynamics in the decoupling limit; however, it helps minimize reflection off the inner refinement boundaries. 
The results are displayed in Sec.~\ref{sec:results_SpinDeScalarize}. During the encounter, scalarization appears to be enhanced and the scalar field grows even faster than it does initially. However, after the  encounter, the scalar field decays, indicating that the \bh{s} become permanently unscalarized. We thus find that the system does display \textit{spin-up descalarization} over the course of the encounter.

Having described the simulation suites and setup, we next describe the results in Sec.~\ref{sec:Results}.

\section{Results} \label{sec:Results}

\begin{figure*}[htbp!]
    \centering
     \includegraphics[width=\textwidth]{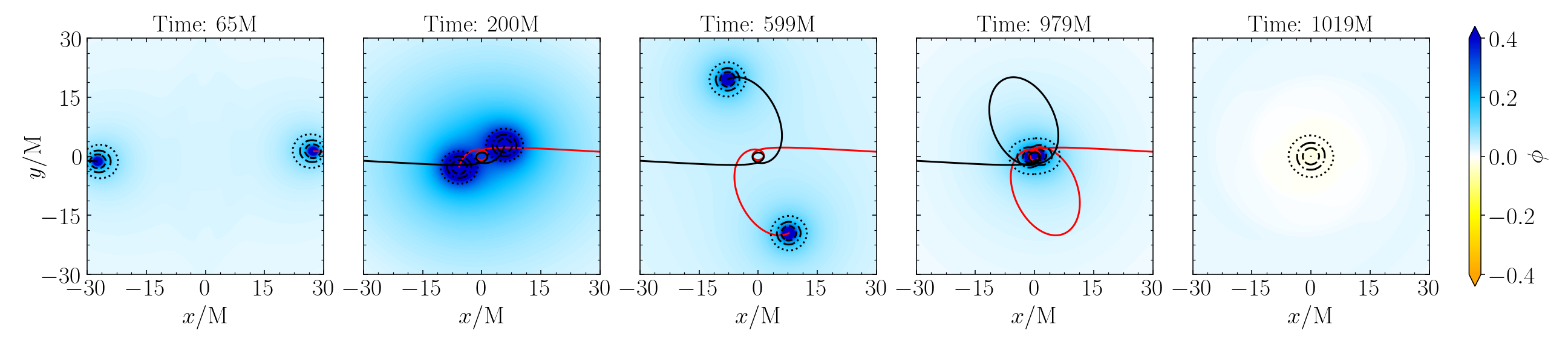}
    \caption{\label{fig:2DScalarPlots_NoSpin} Snapshots of the scalar field, $\phi$, and Gauss-Bonnet invariant, $\mathcal{G}$, in the xy-plane (i.e., the orbital plane) from the ExptBp0355X0 run (see Fig.~\ref{fig:Feynmann_NoSpin} and Table~\ref{tab:Simulation_Suite}). Positive scalar field is shown in blue and negative scalar field is shown in yellow. Positive contours of the Gauss-Bonnet invariant, $\mathcal{G}\mathrm{M}^4=\{10^{-3},10^{-2},10^{-1}\}$, are depicted using dotted, dash-dotted, and dashed black lines, respectively. The trajectories of the \bh{s} are denoted using solid black and red lines.
    }
\end{figure*}

In the following, we analyze the detailed results of the simulations summarized in Sec.~\ref{sec:Setup}. In Sec.~\ref{sec:results_NoSpin} we detail the results of the numerical “experiment” without spin, ExptBp0355X0. In Sec.~\ref{sec:results_controls} we detail the results of the “control” simulations for spinning \bh{s}, CtrlBm300Xp07 and CtrlBm350Xp07.
In Secs.~\ref{sec:results_SpinEncounter},~\ref{sec:results_SpinScalarize}, and~\ref{sec:results_SpinDeScalarize} we detail the results of the numerical “experiments” with spin: ExptBm300Xm07, ExptBm300Xp07, and ExptBm350Xm07. Animations depicting the different simulations can be found at Ref.~\cite{CanudaAnimations}.

\subsection{Dynamical scalarization of two initially non-spinning black holes}
\label{sec:results_NoSpin}

In this section, we analyze the results of the ExptBp0355X0 run listed in Table~\ref{tab:Simulation_Suite} and depicted in Fig.~\ref{fig:Feynmann_NoSpin}. The \bh{s} are initially non spinning and begin unscalarized. 
They undergo a zoom whirl and merger.
In the following, we comment upon the successive \textit{dynamical scalarization}, descalarization, and brief rescalarization of the binary \bh{s} as they perform the zoom whirl, followed by their ultimate descalarization after merger. Animations created from this simulation can be found at Ref.~\cite{CanudaAnimations}.

In order to understand the behavior of the system, we begin by plotting snapshots of the scalar field in the orbital plane at a series of representative times in Fig.~\ref{fig:2DScalarPlots_NoSpin}.
The value of the scalar field is shown using a color scale, where darker blues show more positive values and darker yellows show more negative values. The \bh{} trajectories are plotted over the scalar field. Furthermore, we plot contours of the Gauss-Bonnet invariant, $\mathcal{G}\mathrm{M}^4=\{10^{-3},10^{-2},10^{-1}\}$.

Qualitatively, we organize the evolution of the system into five phases, each of which aligns with one of the panels in Fig.~\ref{fig:2DScalarPlots_NoSpin}.
Note that the time of the first close encounter (i.e., the smallest \bh{} separation prior to merger) and the time at which the \CAH is formed during the final inspiral are $t_{\rm CE}=140\mathrm{M}$ and $t_{\rm CAH}=987\mathrm{M}$, respectively. \textbf{Phase I ($0\mathrm{M}<t<t_{\rm CE}$):} The first panel on the far left shows the \bh{s} as they are approaching one another before the first encounter. While the \bh{s} are seeded with an initial scalar field, they are unscalarized in the sense defined in Sec.~\ref{sec:NRFramework_scalarevolution}. 
\textbf{Phase II ($t_{\rm CE}<t\lesssim 250\mathrm{M}$):} The second panel from the left shows the \bh{s} a short time after their first encounter. 
One can see from the color coding that the scalar field subsequently becomes larger across a wide region around the \bh{s}.
This is likely a manifestation of the finding that the critical coupling of two non-spinning \bh{s} is lower when they are in close proximity than when they are separated~\cite{Julie:2023ncq,Nee:2024bur, Lara:2025kzj}. During this phase, the \bh{s} thus temporarily exhibit \textit{dynamical scalarization}.
\textbf{Phase III ($250 \mathrm{M} \lesssim t \lesssim 900 \mathrm{M}$):} The central panel shows the \bh{s} during the eccentric zoom-whirl orbit following their encounter. One can see that the scalar field is diminished in comparison to the prior panel. This indicates that the \bh{s} start to descalarize due to their separation during this phase.
\textbf{Phase IV ($900 \mathrm{M} \lesssim t \leq t_{\rm CAH}$):} The second panel from the right shows the \bh{s} shortly before their coalescence. Although it is difficult to discern here, we demonstrate in due course that the \bh{s} undergo \textit{dynamical scalarization} for a second time during this phase.
\textbf{Phase V ($t_{\rm CAH} \leq t$):} The panel on the far right shows the remnant \bh{} shortly after merger. 
During this phase, the scalar field temporarily becomes negative as it decays, and the remnant \bh{} undergoes descalarization.

\begin{figure}[htbp!]
    \begin{center}

    \includegraphics[width=1\columnwidth]{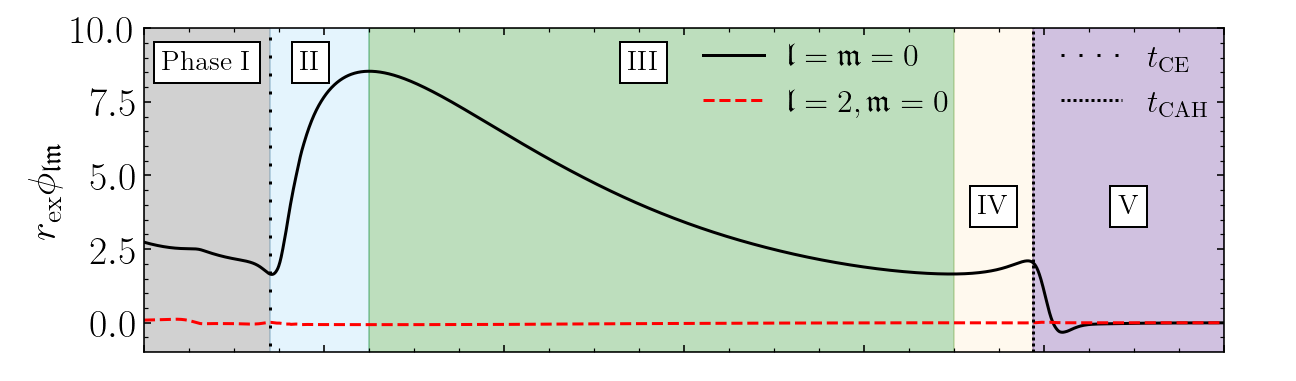}
    \includegraphics[width=1\columnwidth]{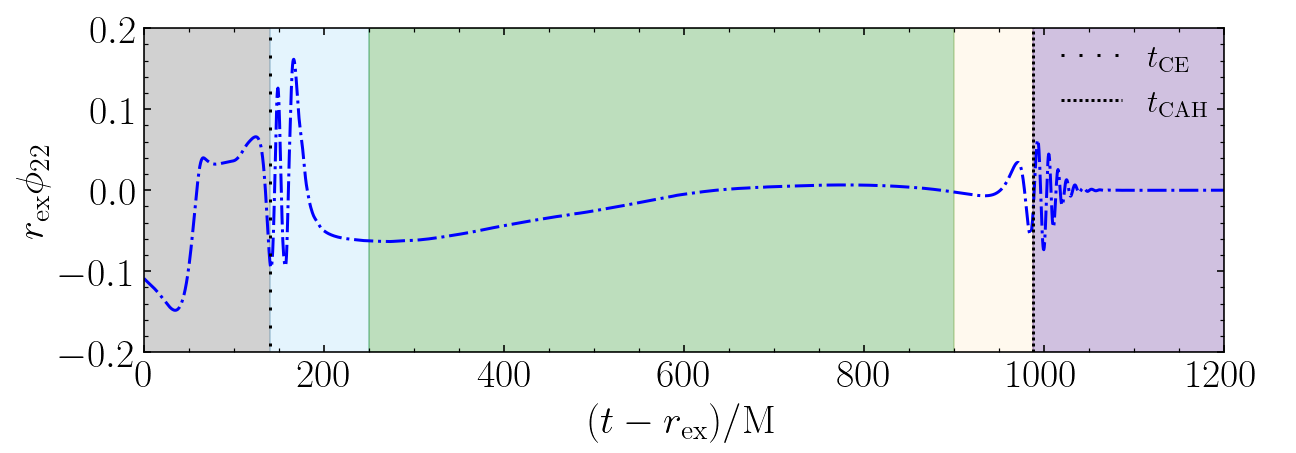}
    \caption{\label{fig:1DScalarvTime_NoSpin} Multipoles of the scalar field, $\phi$, plotted as a function of time for the ExptBp0355X0 run (see Fig.~\ref{fig:Feynmann_NoSpin} and Table~\ref{tab:Simulation_Suite}). The modes ($\mathfrak{l}=\mathfrak{m}=0$) and ($\mathfrak{l}=2,\mathfrak{m}=0$) are shown together in the top panel. The bottom panel shows the $\mathfrak{l}=\mathfrak{m}=2$ mode in isolation. All other modes with $\mathfrak{l}\leq2$ are zero due to the symmetries of the system. We rescale the scalar field by the extraction radius $\rex=100\mathrm{M}$. We shift the time by the extraction radius to account for the propagation time of the scalar field. We denote the time of the first close encounter, $t_{\rm CE}=140\mathrm{M}$, and of the formation of a common apparent horizon, $t_{\rm CAH}=987\mathrm{M}$, via sparsely and densely dotted lines, respectively. Note that these times need not be shifted by the extraction radius.
    } 
    \end{center}
\end{figure}

\begin{figure}[htbp!]
    \begin{center}
    \includegraphics[width=1\columnwidth]{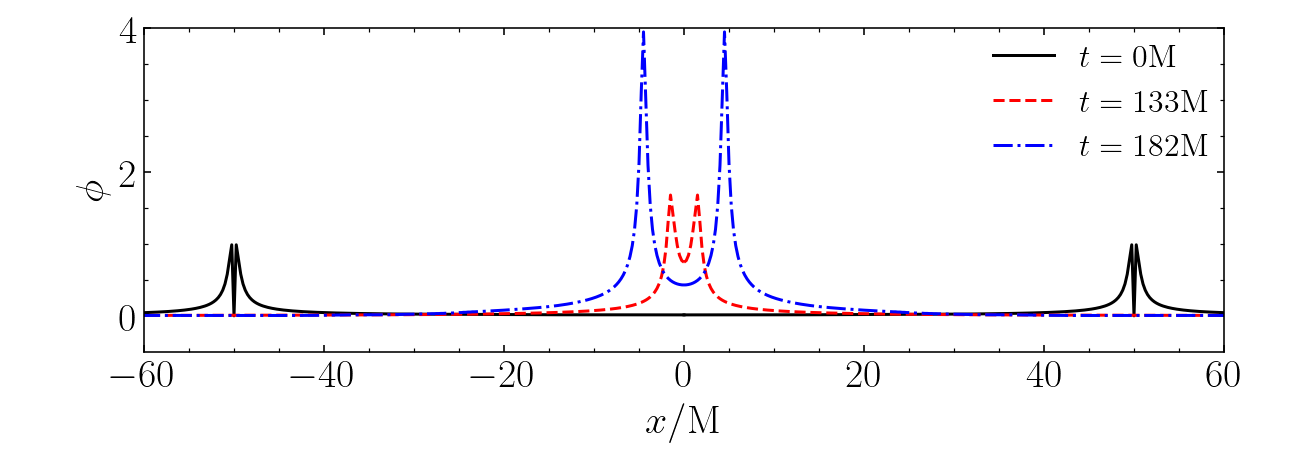}
    \includegraphics[width=1\columnwidth]{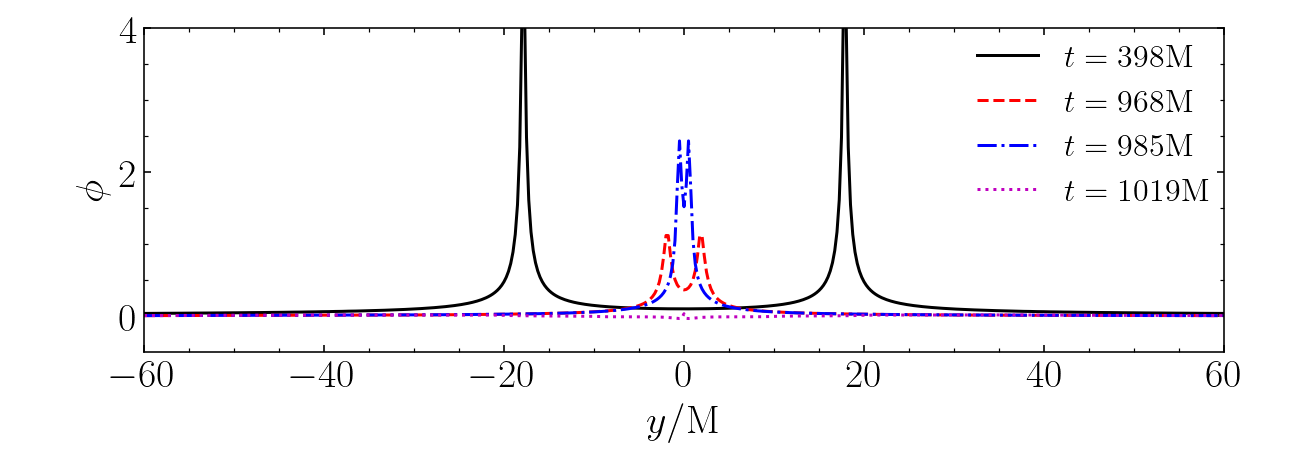}
    \caption{\label{fig:1DScalarvXaxis_NoSpin} Profile of the scalar field, $\phi$, plotted along the x-axis (top panel) and y-axis (bottom panel) for the ExptBp0355X0 run (see Fig.~\ref{fig:Feynmann_NoSpin} and Table~\ref{tab:Simulation_Suite}). The \bh{} trajectories cross either the x-axis or y-axis at each time shown. Data is taken from the orbital plane, where $z=0\mathrm{M}$. The \bh{s} form a common apparent horizon at time $t_{\rm CAH}=987\mathrm{M}$.}
    \end{center}
\end{figure}

To provide a broad overview of the scalar field's evolution, we plot the three lowest non-zero multipoles of the scalar field as a function of time in Fig.~\ref{fig:1DScalarvTime_NoSpin}.
The multipoles are measured at an extraction radius of $\rex=100\mathrm{M}$. In order to account for the propagation time of the scalar field, we subtract the extraction radius from the simulation time. 
The time of the first close encounter, $t_{\rm CE}=140\mathrm{M}$, and the time at which the \CAH forms, $t_{\rm CAH}=987\mathrm{M}$, are denoted by sparsely and densely dotted black lines, respectively.
Note that these times are not shifted by the extraction radius. The monopole, $\rex \phi_{\rm 00}$, corresponds to the scalar charge of the system (see Eq.~\eqref{eq:scalar_charge}), and it is significantly larger than the $\mathfrak{l}=2$ modes plotted for comparison (top panel). We plot the $\mathfrak{l}=\mathfrak{m}=2$ mode, which is related to the scalar radiation, individually so that it is more visible (bottom panel).

During phase I ($0\mathrm{M}<t-\rex<t_{\rm CE}$), the scalar charge gradually declines as the \bh{s} descalarize prior to the first close encounter. 
Following the first close encounter in phase II ($t_{\rm CE}<t-\rex\lesssim 250\mathrm{M}$), the scalar charge increases and the \bh{s} undergo \textit{dynamical scalarization}. One can also see that the \bh{s} emit a burst of scalar radiation.
In phase III ($250 \mathrm{M} \lesssim t-\rex \lesssim 900 \mathrm{M}$), as the \bh{s} separate and undergo an eccentric orbit, the scalar charge gradually declines and the \bh{s} start to descalarize. 
As the \bh{s} reapproach one another and coalesce during phase IV ($900 \mathrm{M} \lesssim t - \rex \leq t_{\rm CAH}$), the scalar charge slightly increases. The \bh{s} thus appear to experience \textit{dynamical scalarization} for a second time. They also emit a second burst of scalar radiation.   
Finally, once the \bh{s} form a \CAH in phase V ($t_{\rm CAH} \leq t-\rex$), the scalar charge decays to zero and the remnant undergoes \textit{“dynamical” descalarization} in the sense of Refs.~\cite{Silva:2020omi,Elley:2022ept,Doneva:2022byd}.

To more clearly observe the quantitative evolution of the scalar field, we plot its profile in Fig.~\ref{fig:1DScalarvXaxis_NoSpin} along the x (top panel) and y (bottom panel) axes for a selection of times at which the \bh{s} cross the said axes.
By comparing the profiles at different times within the panels we can more clearly observe the different phenomena described above. 
The scalar profile at time $t=0\mathrm{M}$ shows two peaks centered on the \bh{s} ($x=\pm50\mathrm{M}$) at the start of the simulation.
Due to a coordinate artifact in the initial data construction, the initial scalar field is zero at the location of the \bh{} punctures. This profile occurs during phase I as described above. 
By comparing the scalar profiles at times $t=133\mathrm{M}$ and $t=182\mathrm{M}$, one can see the behavior of phase II. Namely, the scalar field increases on account of the \textit{dynamical scalarization} during the encounter. 
By comparing the scalar field profiles at times $t=398\mathrm{M}$ and $t=968\mathrm{M}$, we see the effect of phase III. Namely, that once the \bh{s} separate, they descalarize, and the scalar field decays. We can then see in the profiles at times $t=968\mathrm{M}$ and $t=985\mathrm{M}$, when the \bh{s} reapproach one another (i.e., during the inspiral), that the scalar field again increases as the system briefly dynamically rescalarizes. This behavior corresponds to phase IV. Finally, at time $t=1019\mathrm{M}$, after the merger, the scalar field decays to near zero due to the descalarization of the remnant \bh{}. This behavior corresponds to phase V.

Physically, the scalarization or descalarization of a system is determined by the behavior of the effective mass and, ergo, the Gauss-Bonnet invariant (see Eq.~\eqref{eq:mueff}). 
Here, we are considering a system with a positive coupling constant, so a necessary (but not sufficient) condition for scalarization to occur is that the Gauss-Bonnet invariant is positive. Generally, the more positive the Gauss-Bonnet invariant is, the more prone such a system is to scalarize. The magnitude of the Gauss-Bonnet invariant indicates the spacetime curvature; the larger the Gauss-Bonnet invariant, the larger the curvature.

\begin{figure}[htbp!]
    \begin{center}
    \includegraphics[width=1\columnwidth]{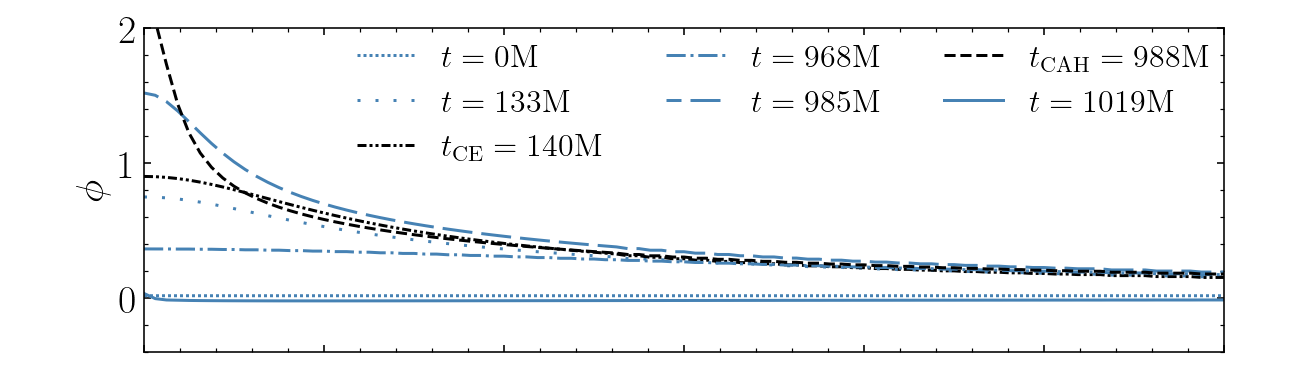}
    \includegraphics[width=1\columnwidth]{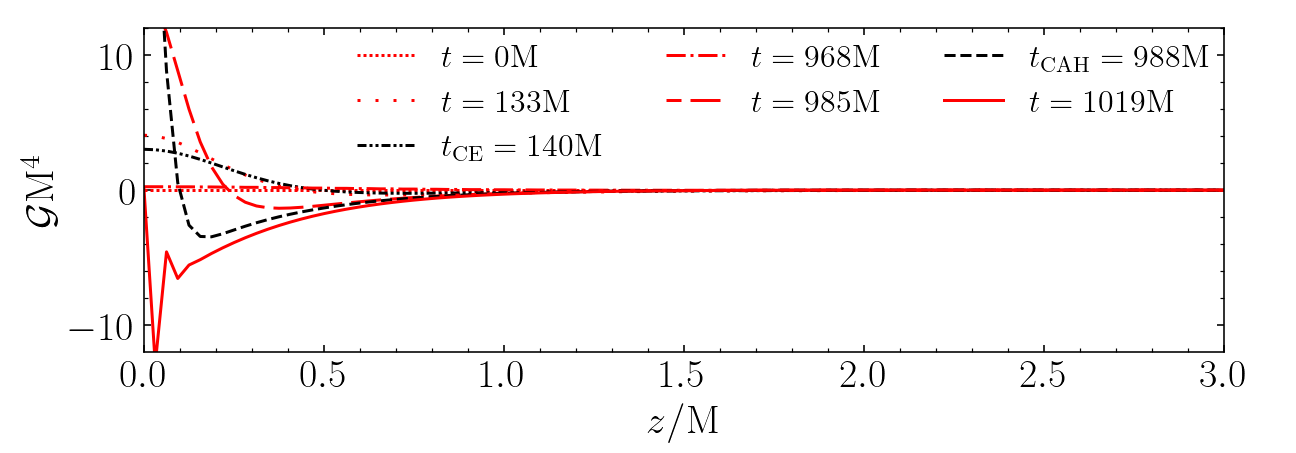}
    \caption{\label{fig:1DScalarvZaxis_NoSpin} The scalar field, $\phi$, (top panel) and Gauss-Bonnet invariant, $\mathcal{G}$, (bottom panel) plotted along the z-axis for the ExptBp0355X0 run (see Fig.~\ref{fig:Feynmann_NoSpin} and Table~\ref{tab:Simulation_Suite}).  
    We mark the time of the close encounter, $t_{\rm CE}$, and of the formation of a common apparent horizon, $t_{\rm CAH}$, via black lines. We note that the Gauss-Bonnet invariant is negative outside the horizon of the final \bh{}.} 
    \end{center}
\end{figure}

In Fig.~\ref{fig:1DScalarvZaxis_NoSpin}, we plot the scalar field (top panel) and Gauss-Bonnet invariant (bottom panel) along the z-axis at different times. 
At time $t=0\mathrm{M}$, the \bh{s} are far from the z-axis and thus (along the axis) the scalar field and Gauss-Bonnet invariant are both near zero. This behavior corresponds to phase I. 
At times $t=133\mathrm{M}$ and $t_{\rm CE}=140\mathrm{M}$, one can see that the Gauss-Bonnet invariant increases near $z=0\mathrm{M}$ (i.e., the orbital plane) and so does the scalar field. This behavior corresponds to phase II, where the \bh{s} approach one another and undergo \textit{dynamical scalarization} during their close encounter. Afterwards, at time $t=968\mathrm{M}$, the Gauss-Bonnet invariant nears zero again on account of the \bh{s} being widely separated during their eccentric orbit. 
Consequently, the scalar field also decays along the z-axis. This behavior corresponds to phase III. 
At times $t=985\mathrm{M}$ and $t_{\rm CAH}=988\mathrm{M}$, the Gauss-Bonnet invariant and scalar field increase again near the orbital plane.  
This behavior corresponds to phase IV, where the \bh{s} undergo \textit{dynamical scalarization} for a second time during their inspiral. 
One can also see that the Gauss-Bonnet invariant becomes negative above the orbital plane ($0.25\mathrm{M}\lesssim z\lesssim1\mathrm{M}$) during the inspiral.
Finally, at time $t=1019\mathrm{M}$, one can see that the Gauss-Bonnet invariant is negative across the z-axis, including outside of the \CAH ($0.5\mathrm{M} \lesssim z$). This is because the remnant acquires a spin of $\chi=0.733$. Consequently, the remnant \bh{} descalarizes with the scalar field approaching zero. This behavior corresponds to phase V.

The fact that the Gauss-Bonnet invariant becomes negative above the orbital plane during the final inspiral (i.e., phase IV) in Fig.~\ref{fig:1DScalarvZaxis_NoSpin} is intriguing because it suggests that \bh{} interactions could enhance \textit{spin-induced scalarization} in systems where the coupling constant is negative.
In fact, this was observed during the inspiral of spinning \bh{s} in Ref.~\cite{Elley:2022ept}. We further consider the implications of such phenomena on the close encounters of spinning \bh{s} in the following sections.

\subsection{Control simulations for spinning black holes} 
\label{sec:results_controls}

In this section, we analyze the control runs CtrlBm300Xp07 and CtrlBm350Xp07 listed in Table~\ref{tab:Simulation_Suite} and described in Sec.~\ref{sec:Setup_SimulationSuite}. 
The \bh{s} in these systems are designed to scatter with a large impact parameter, such that the resulting spin-down, $\Delta\chi=-0.008$, is very small. Depending on the dimensionless coupling, the \bh{s} ultimately scalarize or descalarize after scattering. Because the change in spin is so small, we can use these systems as benchmarks to verify the initial conditions of \bh{s} with the same initial spin parameter magnitude, $|\chi_{\rm 0}|=0.7$, and dimensionless coupling. Animations created from these simulations can be found at Ref.~\cite{CanudaAnimations}.

\subsubsection{Unscalarized black holes: $\beta=-3.00$}

Here we present the CtrlBm300Xp07 run. In this case, the dimensionless coupling, $\beta=-3.00$, is chosen to be small so that the \bh{s} are unscalarized.

\begin{figure}[tbp!]
    \centering
     \includegraphics[width=1\columnwidth]{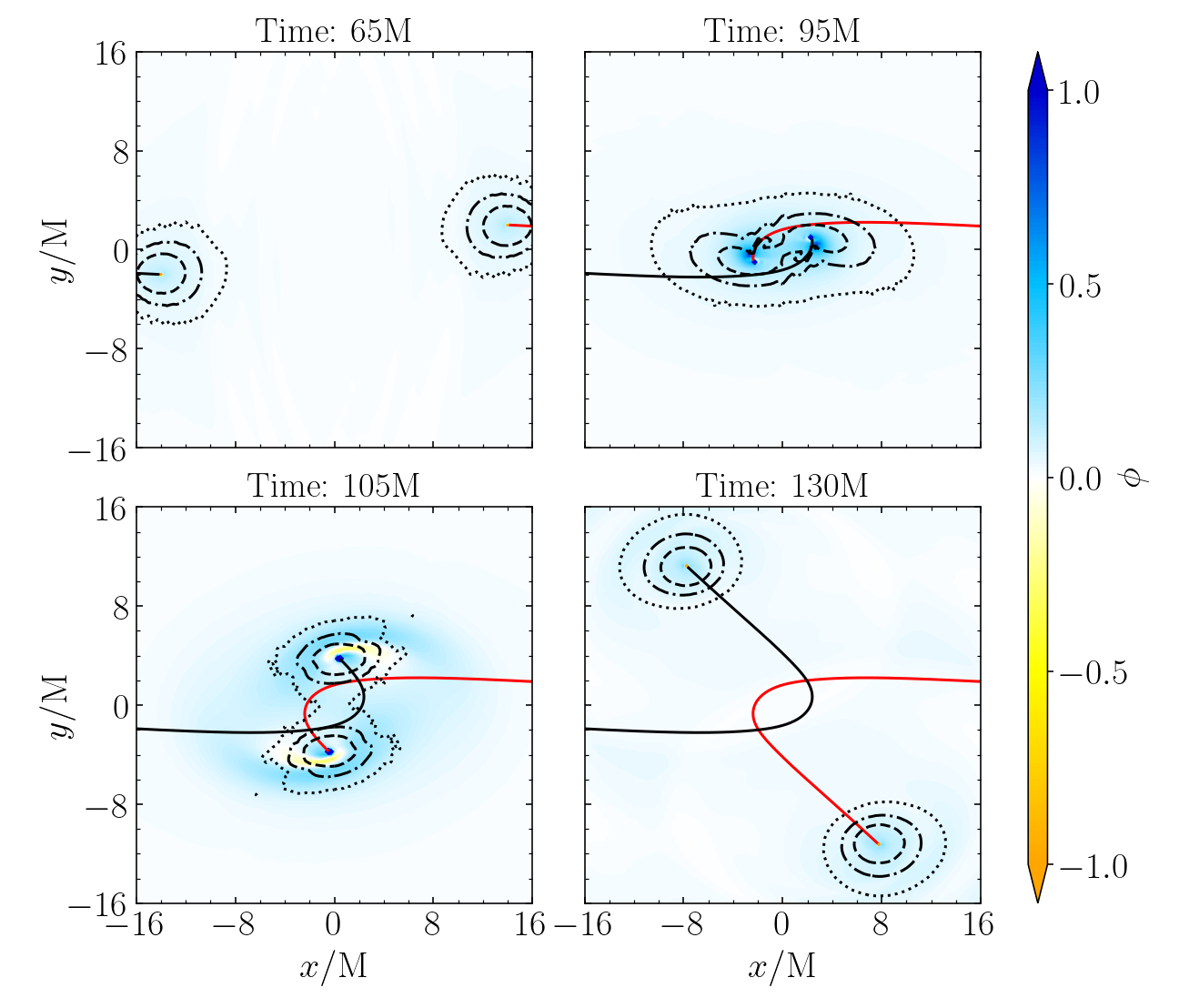}
    \caption{\label{fig:2DScalarPlots_Control_Betam3} Snapshots of the scalar field, $\phi$, and Gauss-Bonnet invariant, $\mathcal{G}$, in the xy-plane (i.e., the orbital plane) from the CtrlBm300Xp07 run (see Table~\ref{tab:Simulation_Suite}). Positive scalar field is shown in blue and negative scalar field is shown in yellow. Positive contours of the Gauss-Bonnet invariant, $\mathcal{G}\mathrm{M}^4=\{10^{-3},10^{-2},10^{-1}\}$, are depicted using dotted, dash-dotted, and dashed black lines, respectively. The trajectories of the \bh{s} are denoted using solid black and red lines. 
    }
\end{figure}

In Fig.~\ref{fig:2DScalarPlots_Control_Betam3}, we plot snapshots of the scalar field in the orbital plane at four different times. 
The value of the scalar field is shown using a color scale, where darker blues show more positive values and darker yellows show more negative values. The \bh{} trajectories are plotted over the scalar field. Furthermore, we plot contours of the Gauss-Bonnet invariant, $\mathcal{G}\mathrm{M}^4=\{10^{-3},10^{-2},10^{-1}\}$. We do not find substantial regions in which the Gauss-Bonnet invariant is negative.

Here we are primarily interested in the late time behavior of the scalar field after the \bh{s'} encounter. However, some non-trivial behavior is visible prior to this point. We thus break the system's evolution into three phases. 
Note that the time of the close encounter (i.e., the smallest \bh{} separation) is $t_{\rm CE}=87\mathrm{M}$.
\textbf{Phase I ($0\mathrm{M}<t\lesssim70\mathrm{M}$):} The top left panel of Fig.~\ref{fig:2DScalarPlots_Control_Betam3} shows the \bh{s} as they approach one another prior to the close encounter. The \bh{s} have some scalar field remaining from the initial seed provided to all \bh{s}; see Eq.~\eqref{eq:initialscalar}.
\textbf{Phase II ($70\mathrm{M}\lesssim t \lesssim 130\mathrm{M}$):} The top right and bottom left panels show the \bh{s} during the second phase, around the time of the close encounter. 
There is a slight increase in the scalar field between the first and second panels. This is not merely a consequence of superposition, but due to curvature.
In the third panel, some portions of the scalar field become negative as it begins to decay. We discuss these effects further in the context of the experiment runs. 
\textbf{Phase III ($130\mathrm{M}\lesssim t$):} The bottom right panel marks the beginning of the third phase. Here the \bh{s} become widely separated again, and the scalar field notably decreases compared to phase II as the \bh{s} begin to descalarize.

\begin{figure}[htbp!]
    \begin{center}
    
    \includegraphics[width=1\columnwidth]{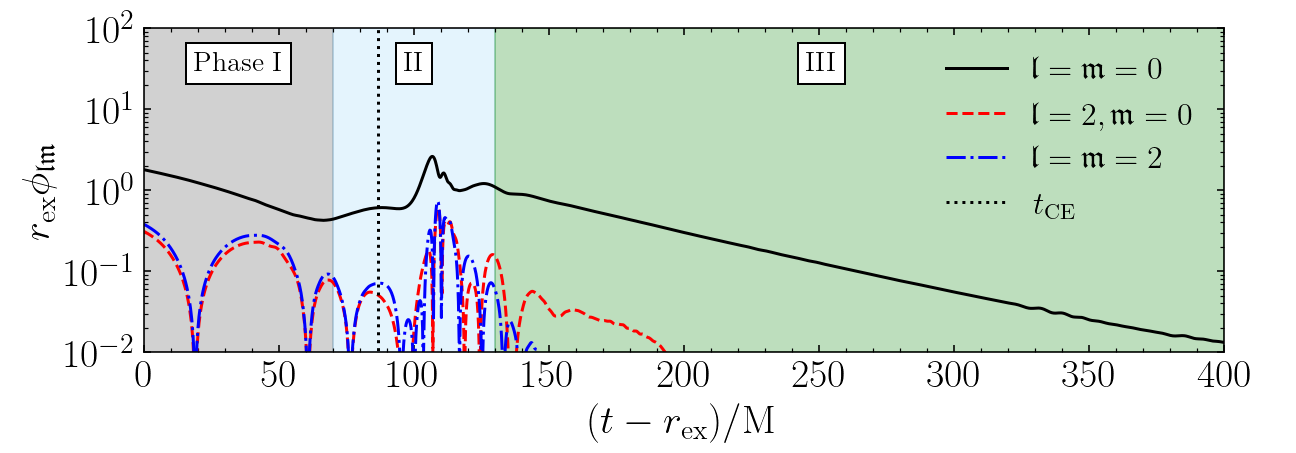}
    \caption{\label{fig:1DScalarvTime_Control_Betam3} Multipoles of the scalar field, $\phi$, plotted as a function of time for the CtrlBm300Xp07 run (see Table~\ref{tab:Simulation_Suite}). The three lowest modes ($\mathfrak{l}=\mathfrak{m}=0$), ($\mathfrak{l}=2,\mathfrak{m}=0$), and ($\mathfrak{l}=\mathfrak{m}=2$) are shown. All other modes with $\mathfrak{l}\leq2$ are zero due to the symmetries of the system. We rescale the scalar field by the extraction radius $\rex=100\mathrm{M}$. We shift the time by the extraction radius to account for the propagation time of the scalar field. We denote the time of the close encounter, $t_{\rm CE}=87\mathrm{M}$, via a dotted line. Note that this time is not shifted by the extraction radius.} 
    \end{center}
\end{figure}

To provide a broad overview of the scalar field's evolution, we plot the three lowest non-zero multipoles as a function of time in Fig.~\ref{fig:1DScalarvTime_Control_Betam3}.
The multipoles are measured at an extraction radius of $\rex=100\mathrm{M}$. In order to account for the propagation time of the scalar field, we subtract the extraction radius from the simulation time. The time of the close encounter, $t_{\rm CE}=87\mathrm{M}$, is denoted by a dotted black line. Note that this time is not shifted by the extraction radius. The monopole, $\rex \phi_{\rm 00}$, corresponds to the scalar charge of the system (see Eq.~\eqref{eq:scalar_charge}), and it is significantly larger than the $\mathfrak{l}=2$ modes plotted for comparison. 

During phase I ($0\mathrm{M}<t-\rex\lesssim70\mathrm{M}$), the scalar charge begins to decline. However, as the initial scalar field is not well adapted to \bh{s} with initial spin (see Sec.~\ref{sec:NRFramework_scalarevolution}), it is difficult to make any definitive statements about this stage of the scalar field's evolution.
Leading up to and following the encounter in phase II ($70\mathrm{M}\lesssim t-\rex \lesssim 130\mathrm{M}$), there is still some non-trivial behavior and growth in the scalar charge. We remark on this behavior in more detail in Sec.~\ref{sec:results_SpinEncounter}, where it is more prominent. 
Once the \bh{s} separate following the encounter in phase III ($130\mathrm{M}\lesssim t-\rex$), we can see that the scalar charge undergoes an exponential decay characterized by a linear slope in the logarithmic scale of Fig.~\ref{fig:1DScalarvTime_Control_Betam3}. This demonstrates that, in isolation, \bh{s} with initial spin parameters of magnitude $|\chi_{\rm 0}|=0.7$ and dimensionless coupling $\beta=-3.00$ cannot sustain a scalar hair.

\subsubsection{Scalarized black holes: $\beta=-3.50$}

Here we present the CtrlBm350Xp07 run. In this case, the dimensionless coupling, $\beta=-3.50$, is chosen to be large so that the \bh{s} scalarize. Recall that the initial scalar field amplitude is $c_{\rm 0}=0.1$; see Sec~\ref{sec:Setup}. 

\begin{figure}[tbp!]
    \centering
     \includegraphics[width=1\columnwidth]{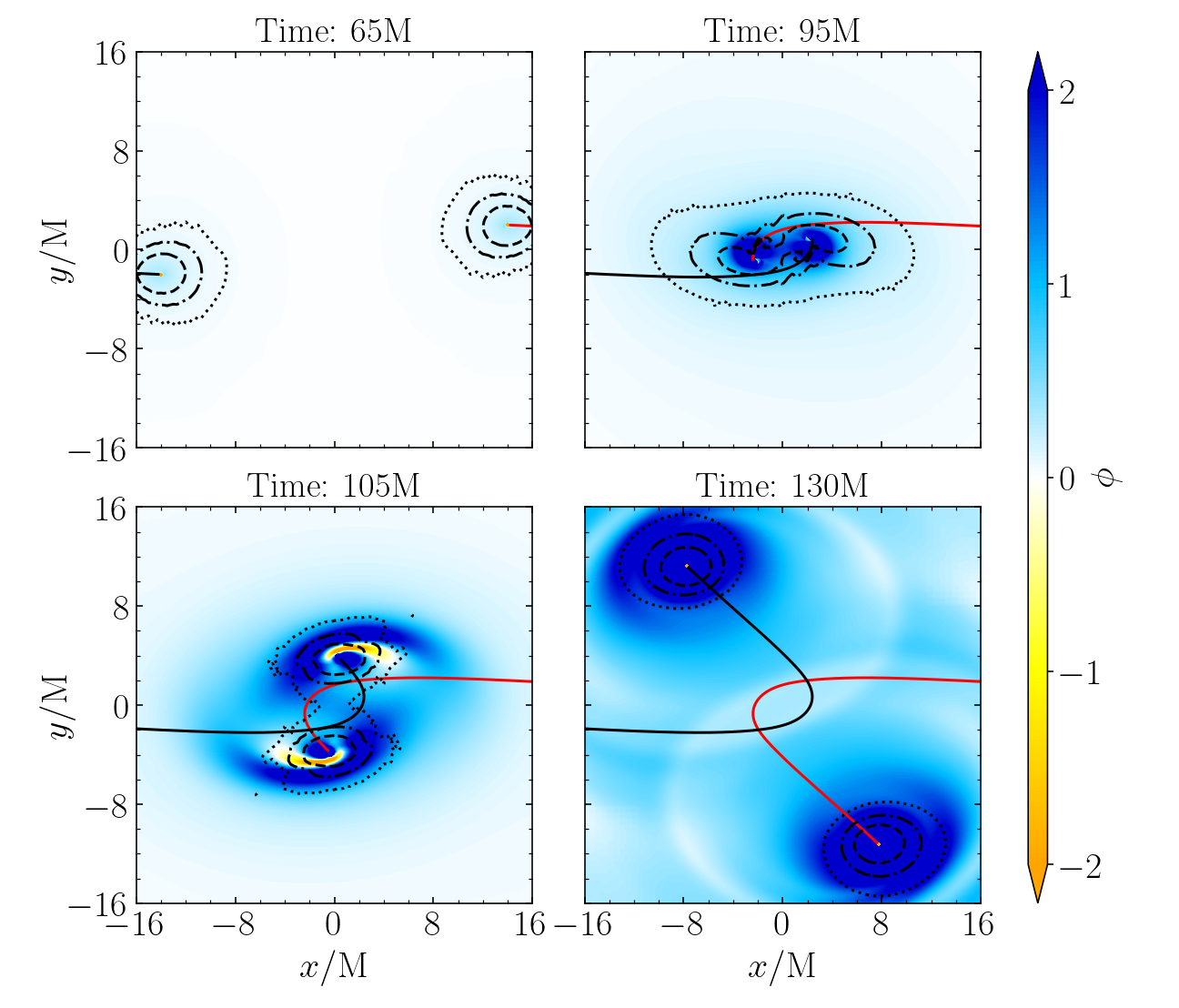}
    \caption{\label{fig:2DScalarPlots_Control_Betam3.5} Snapshots of the scalar field, $\phi$, and Gauss-Bonnet invariant, $\mathcal{G}$, in the xy-plane (i.e., the orbital plane) from the CtrlBm350Xp07 run (see Table~\ref{tab:Simulation_Suite}). Positive scalar field is shown in blue and negative scalar field is shown in yellow. Positive contours of the Gauss-Bonnet invariant, $\mathcal{G}\mathrm{M}^4=\{10^{-3},10^{-2},10^{-1}\}$, are depicted using dotted, dash-dotted, and dashed black lines, respectively. The trajectories of the \bh{s} are denoted using solid black and red lines. 
    }
\end{figure}

In Fig.~\ref{fig:2DScalarPlots_Control_Betam3.5}, we plot snapshots of the scalar field in the orbital plane at four different times. 
The value of the scalar field is shown using a color scale, where darker blues show more positive values and darker yellows show more negative values. The \bh{} trajectories are plotted over the scalar field. Furthermore, we plot contours of the Gauss-Bonnet invariant, $\mathcal{G}\mathrm{M}^4=\{10^{-3},10^{-2},10^{-1}\}$. We do not find substantial regions in which the Gauss-Bonnet invariant is negative.

Here we are primarily interested in the late time behavior of the scalar field after the encounter. However, some non-trivial behavior is visible prior to this point. We thus break the system's evolution into three phases. 
Note that the time of the close encounter is $t_{\rm CE}=87\mathrm{M}$.
\textbf{Phase I ($0\mathrm{M<t\lesssim70\mathrm{M}}$):} The top left panel of Fig.~\ref{fig:2DScalarPlots_Control_Betam3.5} shows the \bh{s} as they approach one another prior to the close encounter.
As in all simulations, the \bh{s} are seeded with an initial scalar field, and one can see that there is a light blue around the \bh{s} denoting the scalar field in the panel.
\textbf{Phase II ($70\mathrm{M}\lesssim t \lesssim 120\mathrm{M}$):} The top right and bottom left panels show the \bh{s} during the second phase, around the time of the close encounter. 
There is already some growth of the scalar field in the second and third panels as compared to the first. There are also regions where the scalar field oscillates and temporarily turns negative.
We discuss the dynamical effects of the close encounter later in the context of the experiment runs. 
\textbf{Phase III ($120\mathrm{M}\lesssim t$):} The bottom right panel is illustrative of the third phase. Here the \bh{s} become widely separated again and are surrounded with an increasing scalar field as the \bh{s} scalarize. 

\begin{figure}[htbp!]
    \begin{center}

    \includegraphics[width=1\columnwidth]{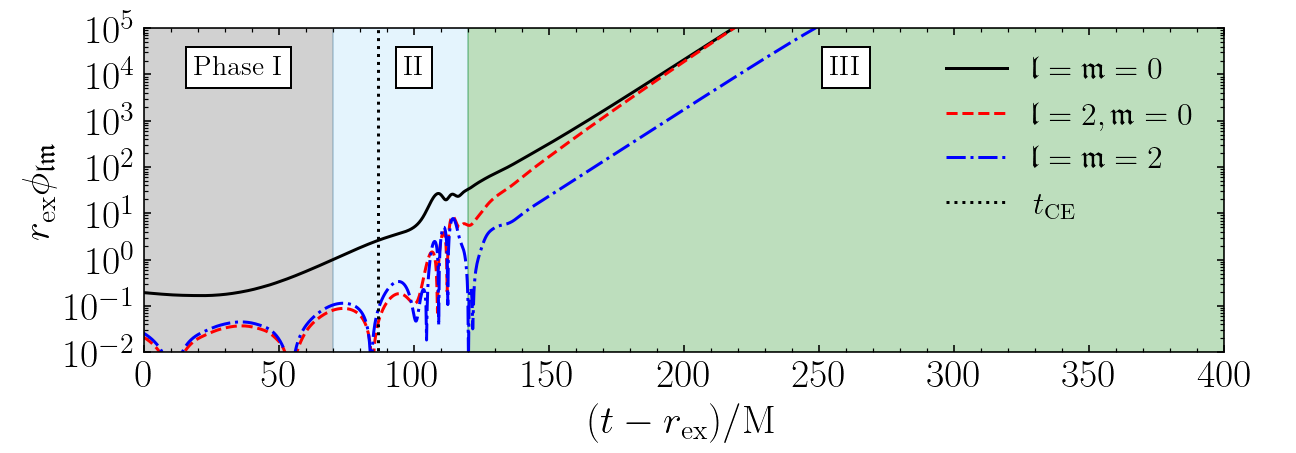}
    \caption{\label{fig:1DScalarvTime_Control_Betam3.5} Multipoles of the scalar field, $\phi$, plotted as a function of time for the CtrlBm350Xp07 run (see Table~\ref{tab:Simulation_Suite}). The three lowest modes ($\mathfrak{l}=\mathfrak{m}=0$), ($\mathfrak{l}=2,\mathfrak{m}=0$), and ($\mathfrak{l}=\mathfrak{m}=2$) are shown. All other modes with $\mathfrak{l}\leq2$ are zero due to the symmetries of the system. We rescale the scalar field by the extraction radius $\rex=200\mathrm{M}$. We shift the time by the extraction radius to account for the propagation time of the scalar field. We denote the time of the close encounter, $t_{\rm CE}=87\mathrm{M}$, via a dashed line. Note that this time is not shifted by the extraction radius.}
    \end{center}
\end{figure}

To provide a broad overview of the scalar field's evolution, we plot the three lowest non-zero multipoles as a function of time in Fig.~\ref{fig:1DScalarvTime_Control_Betam3.5}.
The multipoles are measured at an extraction radius of $\rex=200\mathrm{M}$. In order to account for the propagation time of the scalar field, we subtract the extraction radius from the simulation time. The time of the close encounter, $t_{\rm CE}=87\mathrm{M}$, is denoted by a dotted black line. Note that this time is not shifted by the extraction radius. The monopole, $\rex \phi_{\rm 00}$, corresponds to the scalar charge of the system (see Eq.~\eqref{eq:scalar_charge}), and it is typically larger than the $\mathfrak{l}=2$ modes plotted for comparison. 

During phase I ($0\mathrm{M<t-\rex\lesssim70\mathrm{M}}$), the scalar charge initially appears constant and then gradually increases. However, as the initial scalar field is not well adapted to \bh{s} with initial spin (see Sec.~\ref{sec:NRFramework_scalarevolution}), it is difficult to make definitive statements about this stage of the scalar field's evolution. 
Leading up to and following the encounter in phase II ($70\mathrm{M}\lesssim t- \rex \lesssim 120\mathrm{M}$), there is some non-trivial behavior and growth in the scalar charge. We remark upon this behavior in more detail in Sec.~\ref{sec:results_SpinEncounter}, where it is more prominent.
Once the \bh{s} separate following the encounter in phase III ($120\mathrm{M}\lesssim t - \rex$), we can see that the scalar modes each undergo an exponential growth characterized by a linear increase in the logarithmic scale of Fig.~\ref{fig:2DScalarPlots_Control_Betam3.5}. This demonstrates that, in isolation, \bh{s} with an initial spin parameter of magnitude $|\chi_{\rm 0}|=0.7$ and dimensionless coupling $\beta=-3.50$ can sustain a scalar hair through \textit{spin-induced scalarization}.

With control runs established for both scalarized and unscalarized initial conditions, we proceed to consider the remaining experiment runs.

\subsection{Spin-induced dynamical scalarization of scattering black holes} \label{sec:results_SpinEncounter}

\begin{figure*}[htbp!]
    \centering
    \subfloat[xy-plane (i.e., the orbital plane)\label{fig:2DScalarPlots_XY_SpinEncounter}]{%
        \includegraphics[width=2\columnwidth]{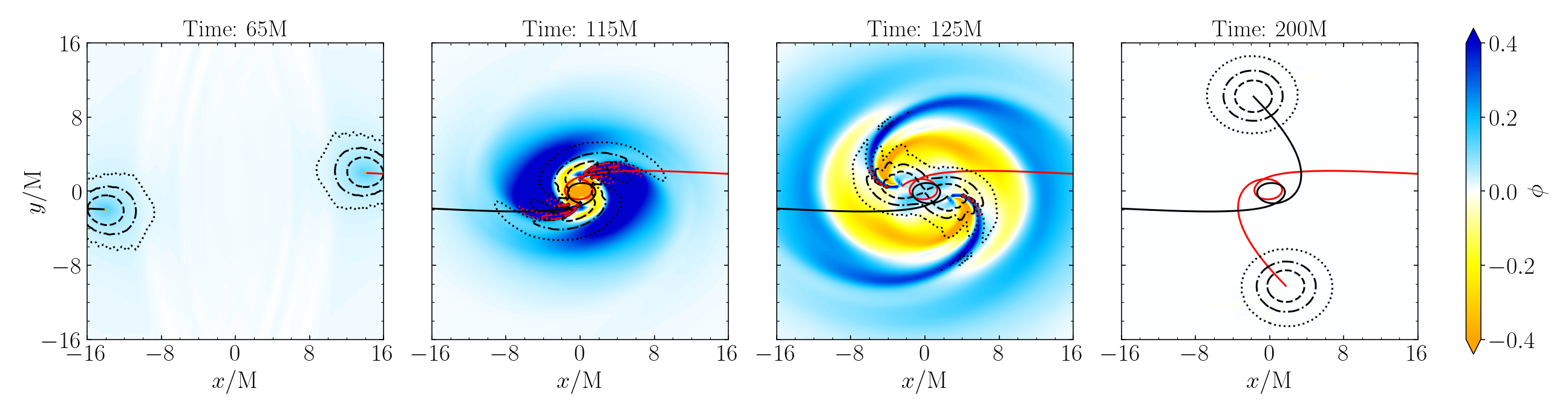}%
    }\par
    \subfloat[xz-plane \label{fig:2DScalarPlots_XZ_SpinEncounter}]{%
        \includegraphics[width=2\columnwidth]{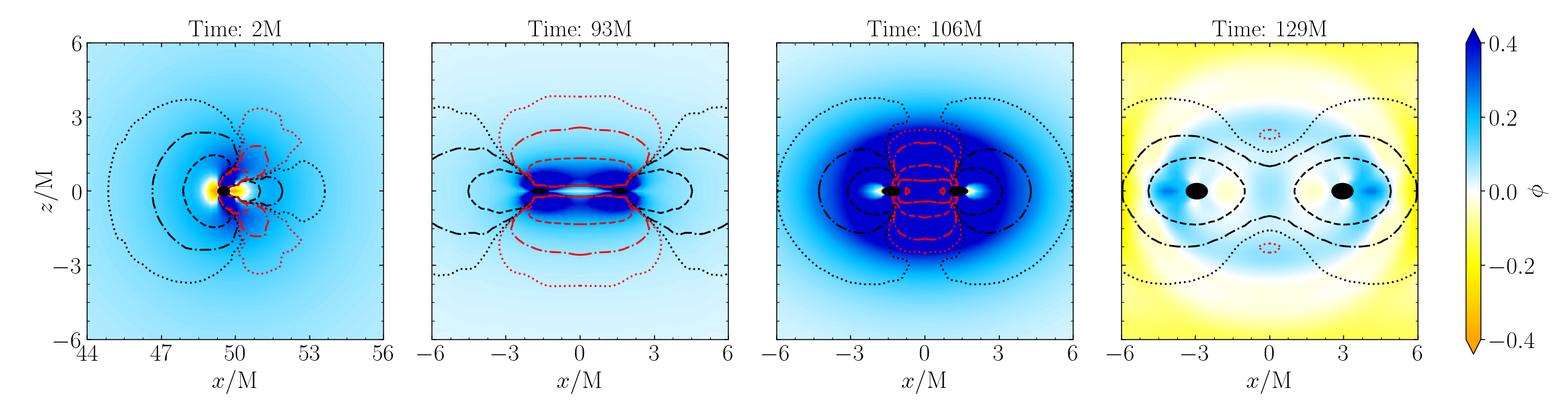}%
    }
    \caption{\label{fig:2DScalarPlots_SpinEncounter} Snapshots of the scalar field, $\phi$, and Gauss-Bonnet invariant, $\mathcal{G}$, from the ExptBm300Xm07 run (see Fig.~\ref{fig:Feynmann_SpinEncounter} and Table~\ref{tab:Simulation_Suite}). The top panel shows the xy-plane (i.e., the orbital plane). The bottom panel shows the xz-plane at times when the \bh{s} cross the x-axis. Positive scalar field is shown in blue, and negative scalar field is shown in yellow. Positive (negative) contours of the Gauss-Bonnet invariant, $|\mathcal{G}\mathrm{M}^4|=\{10^{-3},10^{-2},10^{-1}\}$, are depicted using dotted, dash-dotted, and dashed black (red) lines, respectively. We note that the negative contours may be difficult to discern in grayscale. The trajectories of the \bh{s} are denoted using solid black and red lines.}
\end{figure*}

In this section, we analyze the results of the ExptBm300Xm07 run listed in Table~\ref{tab:Simulation_Suite} and depicted in Fig.~\ref{fig:Feynmann_SpinEncounter}. The \bh{s} have initial spin parameter $\chi_{\rm 0}=-0.7$ and begin unscalarized. The dimensionless coupling is set to $\beta=-3.00$ to match the unscalarized control. In the following, we comment upon the temporary \textit{spin-induced dynamical scalarization} and subsequent descalarization of the binary \bh{s} as they undergo a close encounter and scatter. We note that the \bh{s} experience a spin-up of $\Delta\chi=0.304$ and have a final spin of $\chi_{\rm f}=-0.391$.
The individual \bh{s} should always be below the threshold for scalarization;
therefore, the change in spin is not particularly important to the dynamics of this system. 
Animations of this simulation can be found at Ref.~\cite{CanudaAnimations}.

As discussed in Sec.~\ref{sec:sGB_Theory_coupling}, the \textit{spin-induced scalarization} of an individual \bh{} is driven by negative regions of the Gauss-Bonnet invariant along its spin axis.
Since the \bh{s} in this system have spins oriented along the z-axis, it is important to understand how the scalar field and Gauss-Bonnet invariant evolve perpendicular to the orbital plane. 

In Fig.~\ref{fig:2DScalarPlots_SpinEncounter}, we thus plot snapshots of the scalar field in both the orbital plane (top panels) and the xz-plane (bottom panels) at different instances in time. The snapshots of the orbital plane are selected at times representative of the system's evolution. The snapshots of the xz-plane correspond to times at which the \bh{s} cross the x-axis. The value of the scalar field is shown using a color scale, where darker blues show more positive values and darker yellows show more negative values. We plot positive (negative) contours of the Gauss-Bonnet invariant, $|\mathcal{G}\mathrm{M}^4|=\{10^{-3},10^{-2},10^{-1}\}$, using black (red) lines. In the orbital plane, the \bh{} trajectories are plotted over the scalar field. In the xz-plane, we plot the approximate shape of the \bh{s'} apparent horizons as ellipses using the maximum and minimum horizon radii from the \verb|AHFinderDirect| thorn's output.

Qualitatively, we organize the evolution of the system into four phases, each of which aligns with one of the snapshots of the orbital plane shown in Fig.~\ref{fig:2DScalarPlots_XY_SpinEncounter}.
Note that the time of the close encounter (i.e., the smallest \bh{} separation) is $t_{\rm CE}=100\mathrm{M}$. 
\textbf{Phase I ($0\mathrm{M}<t<t_{\rm CE}$):} The first panel on the left shows the \bh{s} as they approach one another prior to the close encounter. While the \bh{s} are seeded with an initial scalar field, they are unscalarized in the sense defined in Sec.~\ref{sec:NRFramework_scalarevolution}. 
There is also some junk scalar radiation passing each \bh{}; this artifact of the initial conditions is best seen in the wide view animations~\cite{CanudaAnimations}. 

\textbf{Phase II ($t_{\rm CE}<t\lesssim 120\mathrm{M}$):} The second panel from the left shows the \bh{s} shortly after the close encounter. A dark blue region forms around the \bh{s}, indicating an increase in the scalar field. During this phase, we can thus see that the \bh{s} exibit \textit{spin-induced dynamical scalarization}. We can understand this growth of the scalar field in part by observing that there are regions in which the Gauss-Bonnet invariant is negative trailing behind each \bh{}. Notably, the scalar field appears to lag behind these regions as the \bh{s} move. Conversely, one can see that there are regions in which the Gauss-Bonnet invariant is positive and the scalar field is negative.

\textbf{Phase III ($120\mathrm{M}\lesssim t\lesssim 150\mathrm{M}$):} We can further understand the development of these two regions by looking at the second panel from the right, which shows the \bh{s} as they complete their orbit and separate again.
As is best seen in the animations~\cite{CanudaAnimations}, the contours of the Gauss-Bonnet invariant rotate around the origin with the \bh{s}.
Consequently, the positive contours pass over regions with a large scalar field. This causes the scalar field to rapidly oscillate and decay, temporarily turning it negative. 
One can see that a wake of negative scalar field forms behind the \bh{s}. 

\textbf{Phase IV ($150\mathrm{M}\lesssim t$):} The panel on the far right shows the \bh{s} long after they separate. During this phase, the scalar field decays monotonically; i.e., the \bh{s} descalarize after the encounter.

We can gain additional insight into the system by considering the snapshots of the xz-plane shown in Fig.~\ref{fig:2DScalarPlots_XZ_SpinEncounter}. The first panel on the left shows one of the \bh{s} shortly after the start of the simulation (i.e., at the beginning of phase I). One can see that the Gauss-Bonnet invariant is negative roughly along the \bh{'s} spin-axis (i.e., the z-axis) in a “dumbbell” shape. Elsewhere, the Gauss-Bonnet invariant is positive, particularly along the orbital plane (i.e., $z=0$). Because the \bh{} is moving in the negative x-direction, and the curvature takes time to adjust to its changing position, the contours are not symmetric.
Although the scalar field is initialized in a spherically symmetric configuration around each \bh{} (see Eq.~\eqref{eq:initialscalar}), it is already adapting to this “dumbbell” geometry.  
One can see that the scalar field is largest within the inner negative contours of the Gauss-Bonnet invariant, where the curvature is largest. However, within the inner positive contours of the Gauss-Bonnet invariant, the scalar field becomes negative as it oscillates and decays.

The second panel from the left shows the \bh{s} as they approach one another shortly before the close encounter. 
One can see that the regions of negative Gauss-Bonnet invariant above and below the \bh{s} have merged, presumably on account of the frame dragging caused by the \bh{s} orbital angular momentum.
The innermost contour coincides with a dark blue region of scalar field, suggesting the possibility of \textit{spin-induced dynamical scalarization} prior to the encounter. 

The second panel from the right shows the \bh{s} shortly after the close encounter. 
There continues to be a joint region of negative Gauss-Bonnet invariant between the \bh{s}, and we can see that the scalar field substantially increases with respect to the previous panel. 
This behavior corresponds to the \textit{spin-induced dynamical scalarization} observed in phase II and suggests that it is caused by the orbital angular momentum of the \bh{s}.
Simultaneously, one can see that the scalar field is decreased and negative within the regions of positive Gauss-Bonnet invariant along the orbital plane, which coincides with the phenomena observed in Fig.~\ref{fig:2DScalarPlots_XY_SpinEncounter}. 

The panel on the far right of Fig.~\ref{fig:2DScalarPlots_XZ_SpinEncounter} shows the \bh{s} as they begin to separate following the encounter. One can see that the Gauss-Bonnet invariant is predominantly positive due to the decrease of the \bh{s'} spin magnitudes.
Consequently, the scalar field oscillates and decays turning negative in some regions.

\begin{figure}[htbp!]
    \begin{center}

    \includegraphics[width=1\columnwidth]{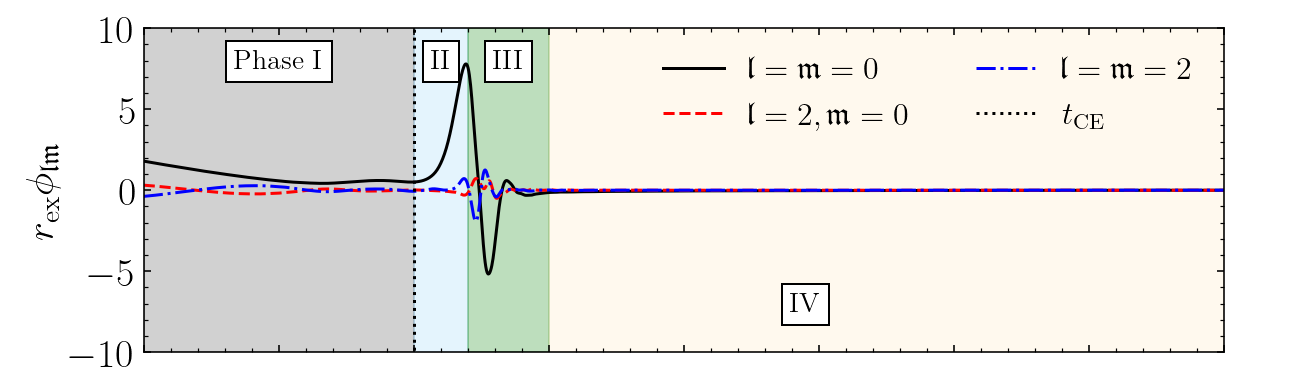}
    \includegraphics[width=1\columnwidth]{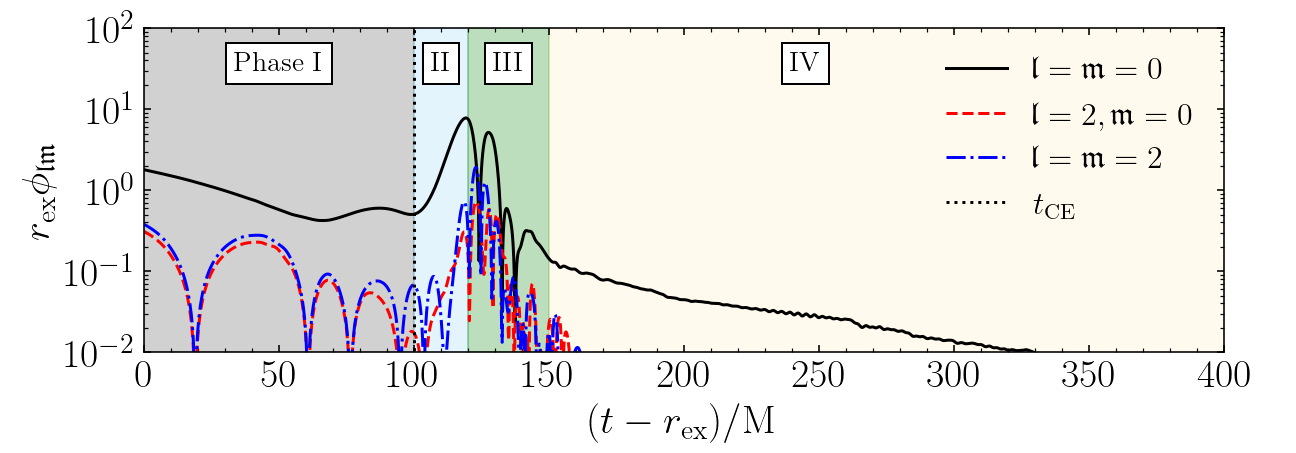}
    \caption{\label{fig:1DScalarvTime_SpinEncounter} Multipoles of the scalar field, $\phi$, plotted as a function of time for the ExptBm300Xm07 run (see Fig.~\ref{fig:Feynmann_SpinEncounter} and Table~\ref{tab:Simulation_Suite}). The top (bottom) panel uses a linear (logarithmic) scale. The three lowest modes ($\mathfrak{l}=\mathfrak{m}=0$), ($\mathfrak{l}=2,\mathfrak{m}=0$), and ($\mathfrak{l}=\mathfrak{m}=2$) are shown. All other modes with $\mathfrak{l}\leq2$ are zero due to the symmetries of the system. We rescale the scalar field by the extraction radius $\rex=100\mathrm{M}$. We shift the time by the extraction radius to account for the propagation time of the scalar field. We denote the time of the close encounter, $t_{\rm CE}=100\mathrm{M}$, via a dotted line. Note that this time is not shifted by the extraction radius.} 
    \end{center}
\end{figure}

To provide a broad overview of the scalar field's evolution, we plot the three lowest non-zero multipoles as a function of time in Fig.~\ref{fig:1DScalarvTime_SpinEncounter} using both a linear (top panel) and logarithmic scale (bottom panel).
The multipoles are measured at an extraction radius of $\rex=100\mathrm{M}$. In order to account for the propagation time of the scalar field, we subtract the extraction radius from the simulation time. The time of the close encounter, $t_{\rm CE}=100\mathrm{M}$, is denoted by a dotted black line. Note that this time is not shifted by the extraction radius. The monopole, $\rex \phi_{\rm 00}$, corresponds to the scalar charge of the system (see Eq.~\eqref{eq:scalar_charge}), and it is significantly larger than the $\mathfrak{l}=2$ modes plotted for comparison. 

During phase I ($0\mathrm{M}<t-\rex<t_{\rm CE}$), the scalar field seed adjusts to the spinning \bh{s}, and the scalar charge decreases; i.e., the \bh{s} are unscalarized in the sense of Sec.~\ref{sec:NRFramework_scalarevolution}.
As the \bh{s} approach one another  ($60\mathrm{M}\lesssim t-\rex < t_{\rm CE}$), we find a subtle increase in the scalar charge. 
We suspect this may either be due to junk scalar radiation produced by the initial conditions, or to the early scalarization observed in the second panel from the left in Fig.~\ref{fig:2DScalarPlots_XZ_SpinEncounter}.
Following the close encounter in phase II ($t_{\rm CE}<t-\rex\lesssim 120\mathrm{M}$), the scalar charge increases and the \bh{s} undergo \textit{spin-induced dynamical scalarization}. 
In phase III ($120\mathrm{M}\lesssim t-\rex \lesssim 150\mathrm{M}$), as the \bh{s} separate following the close encounter, the scalar charge undergoes a period of damped oscillation, and the \bh{s} begin to descalarize.
As the \bh{s} continue to separate and descalarize during phase IV ($150\mathrm{M}\lesssim t-\rex$), the scalar charge ceases to oscillate and decays exponentially.

\begin{figure}[tbp!]
    \begin{center}
    \includegraphics[width=1\columnwidth]{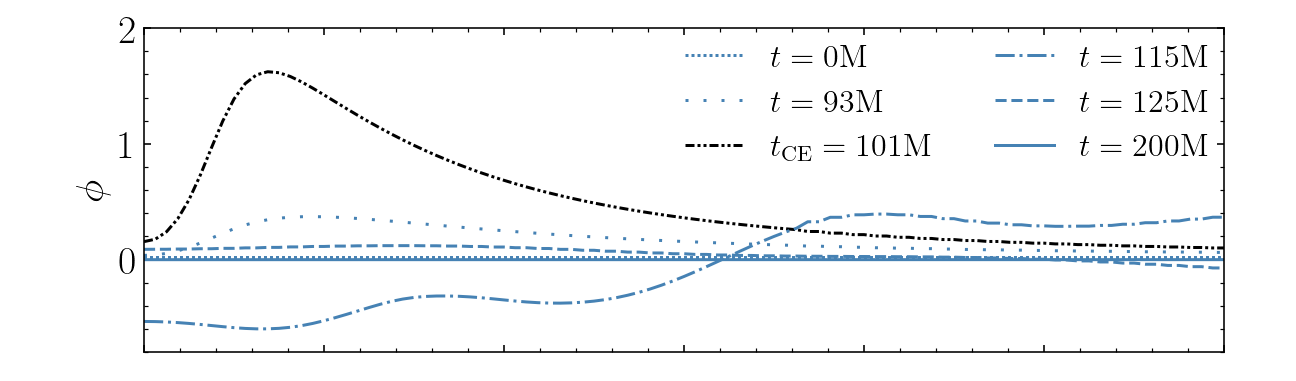}
    \includegraphics[width=1\columnwidth]{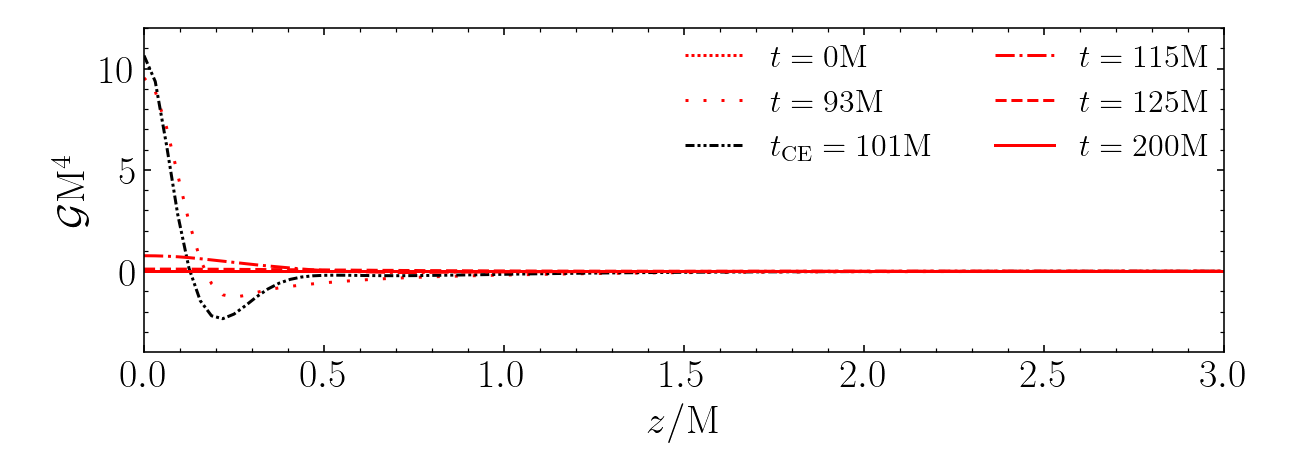}
    \caption{\label{fig:1DScalarvZaxis_SpinEncounter}  The scalar field, $\phi$, (top panel) and Gauss-Bonnet invariant, $\mathcal{G}$, (bottom panel) plotted along the z-axis for the ExptBm300Xm07 run (see Fig.~\ref{fig:Feynmann_SpinEncounter} and Table~\ref{tab:Simulation_Suite}). 
    We mark the time of the close encounter, $t_{\rm CE}$, via a black line.} 
    \end{center}
\end{figure}

We can further our understanding of how the scalar field is produced by analyzing its profile along the z-axis, which runs through the center of the joint “dumbbell” region observed in Fig.~\ref{fig:2DScalarPlots_XZ_SpinEncounter}. In Fig.~\ref{fig:1DScalarvZaxis_SpinEncounter}, we plot the scalar field (top panel) and Gauss-Bonnet invariant (bottom panel) along the z-axis at different times. Initially, at time $t=0\mathrm{M}$, the \bh{s} are far from the z-axis and thus the scalar field and Gauss-Bonnet invariant are both near zero. This corresponds with phase I. Later, at times $t=93\mathrm{M}$ and $t_{\rm CE}=101\mathrm{M}$, one can see that the Gauss-Bonnet invariant becomes negative at intermediate distances from the orbital plane ($0.1 \mathrm{M} \lesssim z \lesssim 0.5 \mathrm{M}$), which coincides with a large increase in the scalar field centered around the same region. This behavior corresponds to phase II, where the \bh{s} undergo \textit{spin-induced dynamical scalarization} during their encounter. 
Afterwards, at times $t=115\mathrm{M}$ and $t=125\mathrm{M}$, the Gauss-Bonnet invariant decays to zero and the scalar field oscillates both in time and space as the field begins to decay. This behavior corresponds to phase III. 
We note that we observe phase II and phase III behavior earlier here than in Fig.~\ref{fig:1DScalarvTime_SpinEncounter}. This is not entirely surprising as Fig.~\ref{fig:1DScalarvTime_SpinEncounter} depicts an averaged measure of the scalar field at large radius, assuming propagation from the origin, whereas Fig.~\ref{fig:1DScalarvZaxis_SpinEncounter} shows local dynamics along the z-axis. Finally, at time $t=200\mathrm{M}$, the \bh{s} move far from the z-axis, and both the Gauss-Bonnet invariant and scalar field decay to near zero. This behavior corresponds to phase IV.

Here we find that \bh{s} can exhibit a spin-induced form of \textit{dynamical scalarization} during a close encounter when the dimensionless coupling is negative. This is a temporary effect, and the \bh{s} descalarize when they move apart.
This phenomena appears to be driven by the orbital angular momentum of the \bh{s}. Note that the spin-up of the \bh{s} is largely irrelevant to this process as we would expect the individual \bh{s} to be unscalarized based on both their initial and final spins.

Next, we consider the possibility of systems where the spin-up has a non-trivial effect on the dynamics of the system and can lead to a permanent change in the state of the \bh{s}.

\subsection{Scalarization of black holes due to spin-up} \label{sec:results_SpinScalarize}

In this section, we analyze the results of the ExptBm300Xp07 run listed in Table~\ref{tab:Simulation_Suite} and depicted in Fig.~\ref{fig:Feynmann_SpinScalarize}. 
The \bh{s} have an initial spin parameter of $\chi_{\rm 0}=0.7$ and scatter, undergoing a lengthy close encounter. This encounter results in a spin-up of $\Delta \chi=0.023$ and a final spin of $\chi_{\rm f}=0.718$.
The dimensionless coupling is tuned to $\beta=-3.00$ such that the \bh{s} are unscalarized before the encounter, but scalarize due to the spin-up during the encounter. 
In the following, we briefly discuss \textit{dynamical scalarization}, which remains active during the encounter, but primarily focus on the permanent \textit{spin-up scalarization} of the \bh{s}, which results due to their change in spin. Animations of this simulation can be found at Ref.~\cite{CanudaAnimations}.

\begin{figure*}[htbp!]
    \centering
    \subfloat[xy-plane (i.e., the orbital plane)\label{fig:2DScalarPlots_XY_SpinScalarize}]{%
        \includegraphics[width=2\columnwidth]{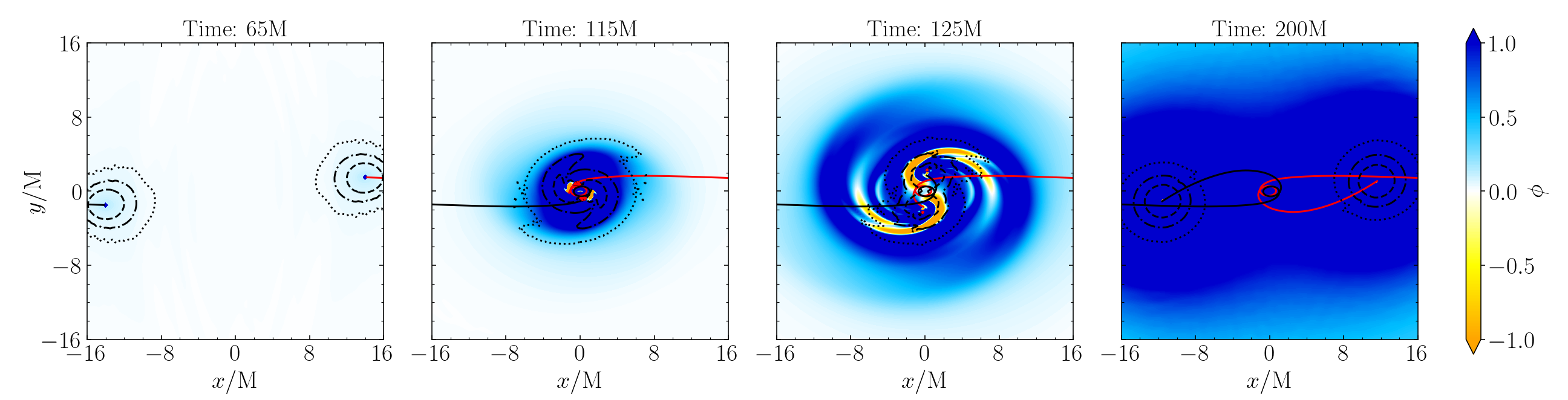}%
    }\par
    \subfloat[xz-plane \label{fig:2DScalarPlots_XZ_SpinScalarize}]{%
        \includegraphics[width=2\columnwidth]{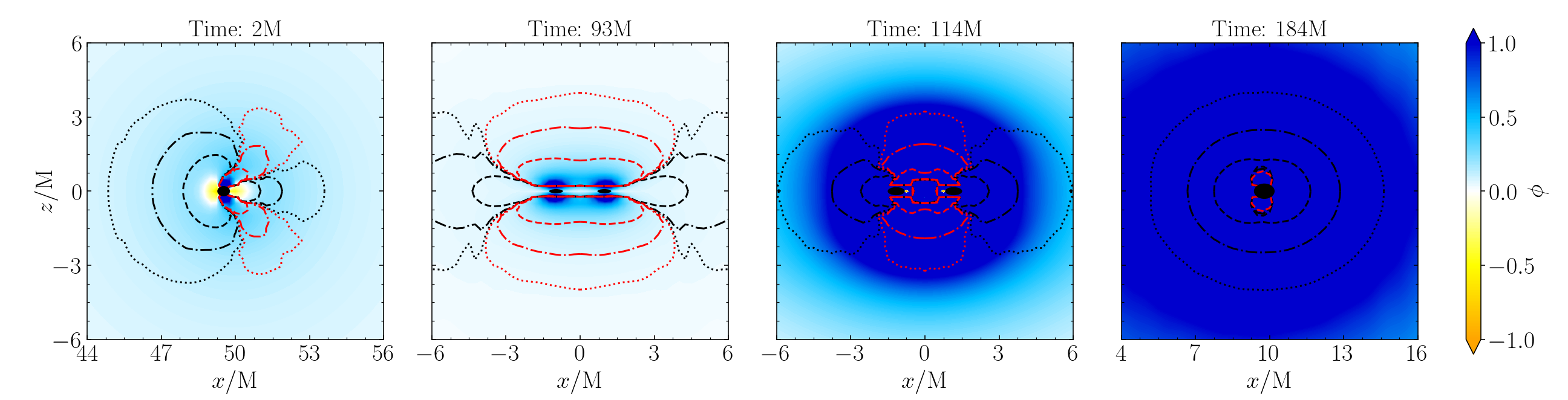}%
    }
    \caption{\label{fig:2DScalarPlots_SpinScalarize} Snapshots of the scalar field, $\phi$, and Gauss-Bonnet invariant, $\mathcal{G}$, from the ExptBm300Xp07 run (see Fig.~\ref{fig:Feynmann_SpinScalarize} and Table~\ref{tab:Simulation_Suite}). The top panel shows the xy-plane (i.e., the orbital plane). The bottom panel shows the xz-plane at times when the \bh{s} cross the x-axis. Positive scalar field is shown in blue and negative scalar field is shown in yellow. Positive (negative) contours of the Gauss-Bonnet invariant, $|\mathcal{G}\mathrm{M}^4|=\{10^{-3},10^{-2},10^{-1}\}$, are depicted using dotted, dash-dotted, and dashed black (red) lines, respectively. We note that the negative contours can be difficult to discern in grayscale. The trajectories of the \bh{s} are denoted using solid black and red lines.}
\end{figure*}

In Fig.~\ref{fig:2DScalarPlots_SpinScalarize}, we plot snapshots of the scalar field in both the orbital plane (top panels) and the xz-plane (bottom panels) at different instances in time. The snapshots of the orbital plane are selected at times representative of the system's evolution. The snapshots of the xz-plane correspond to times at which the \bh{s} cross the x-axis. The value of the scalar field is shown using a color scale, where darker blues show more positive values and darker yellows show more negative values. We plot positive (negative) contours of the Gauss-Bonnet invariant, $|\mathcal{G}\mathrm{M}^4|=\{10^{-3},10^{-2},10^{-1}\}$, using black (red) lines.
In the orbital plane, the \bh{} trajectories are plotted over the scalar field. 
In the xz-plane, we plot the approximate shape of the \bh{s'} apparent horizons as ellipses using the maximum and minimum horizon radii from the \verb|AHFinderDirect| thorn's output.

Qualitatively, we organize the evolution of the system into four phases, each of which aligns with one of the snapshots of the orbital plane shown in Fig.~\ref{fig:2DScalarPlots_XY_SpinScalarize}. 
Note that the time of the close encounter (i.e., the smallest \bh{} separation) is $t_{\rm CE}=98\mathrm{M}$.
\textbf{Phase I ($0\mathrm{M}<t< t_{\rm CE}$):} From left to right, the first panel shows the \bh{s} as they approach one another prior to the close encounter.
The \bh{s} are unscalarized in the sense described in Sec.~\ref{sec:NRFramework_scalarevolution}, but still have scalar field remaining from the initial seed.  

The two central panels show the \bh{s} as they orbit around each other and separate following the close encounter. 
\textbf{Phase II ($t_{\rm CE} < t \lesssim  125\mathrm{M}$):} As in Sec.~\ref{sec:results_SpinEncounter}, one can see that the scalar field is much greater in these central panels than in the first panel. During this phase, we thus observe the presence of \textit{spin-induced dynamical scalarization}.
\textbf{Phase III ($125\mathrm{M}\lesssim t \lesssim 150\mathrm{M}$):} Furthermore, in the third panel, the scalar field oscillates in regions where the Gauss-Bonnet invariant is positive. During this phase, we thus observe that the scalar field temporarily decays and turns negative in some regions. 
However, unlike in Sec.~\ref{sec:results_SpinEncounter}, this does not result in the descalarization of the \bh{s}. 

\textbf{Phase IV ($150\mathrm{M} \lesssim t$):} The panel on the far right shows the \bh{s} well after they have separated. One can see that they are now independently scalarized as the scalar field continues to increase, indicating the presence of \textit{spin-up scalarization} during this phase. 

We now consider the snapshots of the xz-plane shown in Fig.~\ref{fig:2DScalarPlots_XZ_SpinScalarize}. From left to right, the first panel shows one of the \bh{s} shortly after the start of the simulation (i.e., at the beginning of phase I). Here the negative contours of the Gauss-Bonnet invariant have the same characteristic “dumbbell” shape as remarked upon in Sec.~\ref{sec:results_SpinEncounter}.
The scalar field distribution is also similar, but the colors appear fainter here because we use a broader color scale.

The two central panels show the \bh{s} around the time of the close encounter. As in Sec.~\ref{sec:results_SpinEncounter}, one can see that the \bh{s'} “dumbbells” join into a larger region leading to \textit{spin-induced dynamical scalarization} and the growth of the scalar field, which increases across the two panels. This behavior corresponds to phase II. 

The final panel on the far right shows one of the \bh{s} long after the close encounter (i.e., during phase IV). 
Again we find a “dumbbell” shaped region in which the Gauss-Bonnet invariant is negative.
To improve clarity, we show only the most negative red contour of the Gauss-Bonnet invariant, as the contours otherwise lie on top of one another.
Although the “dumbbell” appears smaller than in the far left panel, the Gauss-Bonnet invariant within it is more negative on account of the \bh{'s} increased spin. 
Consequently, the \bh{} scalarizes individually, as indicated by the saturation of the panel with a dark blue. We thus see that the strengthening of the “dumbbell” region via spin-up leads to \textit{spin-up scalarization}.

\begin{figure}[tbp!]
    \begin{center}

    \includegraphics[width=1\columnwidth]{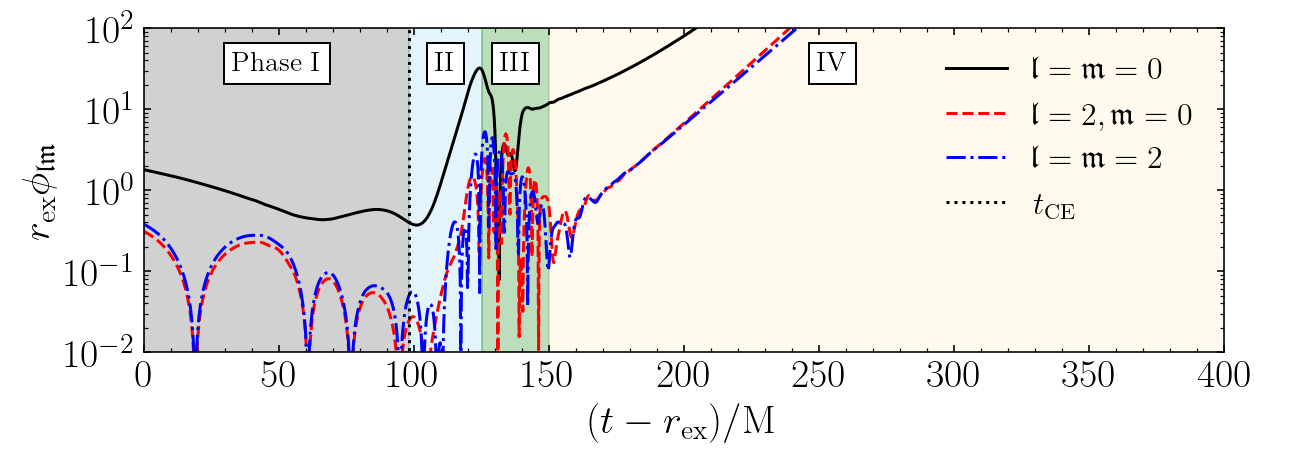}
    \includegraphics[width=1\columnwidth]{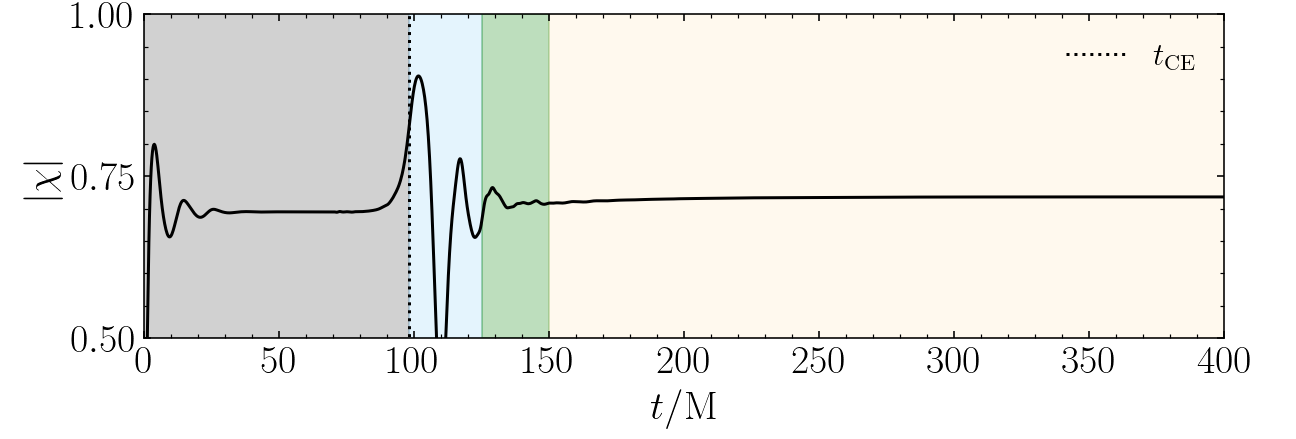}
    \caption{\label{fig:1DScalarvTime_SpinScalarize} Top panel: Multipoles of the scalar field, $\phi$, plotted as a function of time for the ExptBm300Xp07 run (see Fig.~\ref{fig:Feynmann_SpinScalarize} and Table~\ref{tab:Simulation_Suite}). The three lowest modes ($\mathfrak{l}=\mathfrak{m}=0$), ($\mathfrak{l}=2,\mathfrak{m}=0$), and ($\mathfrak{l}=\mathfrak{m}=2$) are shown. All other modes with $\mathfrak{l}\leq2$ are zero due to the symmetries of the system. We rescale the scalar field by the extraction radius $\rex=100\mathrm{M}$. We shift the time by the extraction radius to account for the propagation time of the scalar field. Bottom panel: The spin magnitude of one of the \bh{s} is plotted as a function of time. We denote the time of the close encounter, $t_{\rm CE}=98\mathrm{M}$, via a dotted line. Note that this time is not shifted by the extraction radius.
    } 
    \end{center}
\end{figure}

To provide a broad overview of the scalar field's evolution, we plot the three lowest non-zero multipoles as a function of time in the top panel of Fig.~\ref{fig:1DScalarvTime_SpinScalarize}. 
The multipoles are measured at an extraction radius of $\rex=100\mathrm{M}$. In order to account for the propagation time of the scalar field, we subtract the extraction radius from the simulation time. The time of the close encounter, $t_{\rm CE}=98\mathrm{M}$, is denoted by a dotted black line. Note that this time is not shifted by the extraction radius. The monopole, $\rex \phi_{\rm 00}$, corresponds to the scalar charge of the system (see Eq.~\eqref{eq:scalar_charge}), and it is significantly larger than the $\mathfrak{l}=2$ modes plotted for comparison. In the bottom panel of Fig.~\ref{fig:1DScalarvTime_SpinScalarize}, we plot the magnitude of one of the \bh{s'} spins as a function of time using Eq.~\eqref{eq:chi_extract}. Note that there is no propagation time associated with the spin.

During phase I ($0\mathrm{M}<t-\rex< t_{\rm CE}$), the scalar field both decays and adjusts to the spinning \bh{s} ($\chi_{\rm i}=0.695$), which descalarize prior to the close encounter, as indicated by the decreasing scalar charge. However, as the \bh{s} approach one another ($60\mathrm{M}\lesssim t-\rex< t_{\rm CE}$), the logarithmic scale reveals a subtle increase in the scalar charge similar to the one discussed in Sec.~\ref{sec:results_SpinEncounter}.
During the close encounter in phase II ($t_{\rm CE} < t-\rex \lesssim  125\mathrm{M}$), the scalar charge increases, and the \bh{s} undergo \textit{spin-induced dynamical scalarization}.  
In phase III ($125\mathrm{M}\lesssim t-\rex \lesssim 150\mathrm{M}$), as the \bh{s} begin to separate following the close encounter, they partially descalarize, and for a brief period, the scalar charge declines. During this time, the \bh{} spins also stabilize to larger values ($\chi_{\rm f}=0.718$).
As the \bh{s} continue to separate during phase IV ($150\mathrm{M} \lesssim t-\rex$), the scalar charge exponentially increases. 
The \bh{s} thus appear to undergo permanent \textit{spin-up scalarization} due to the increase in their spin.

\begin{figure}[htbp!]
    \begin{center}
    \includegraphics[width=1\columnwidth]{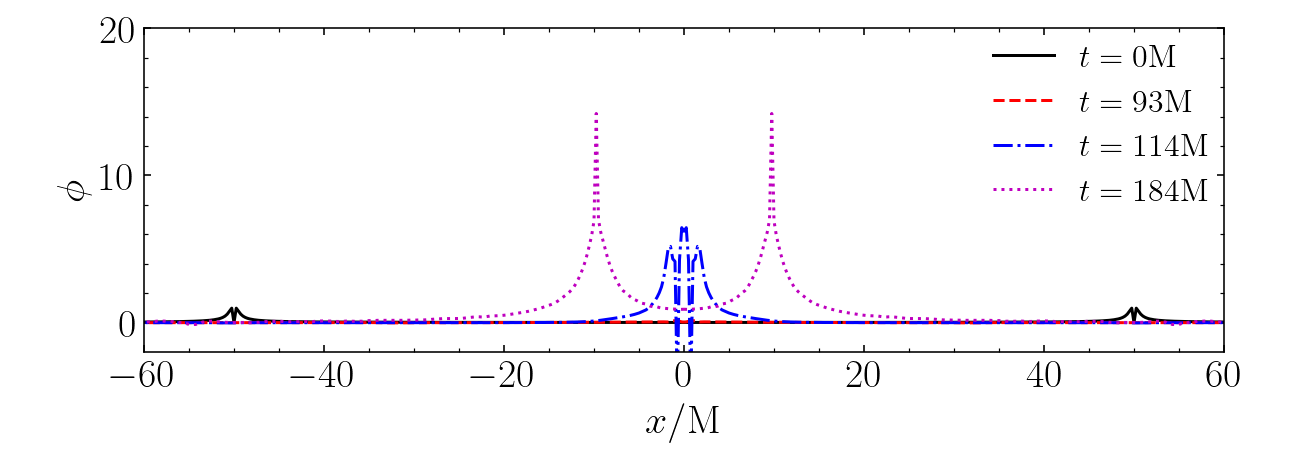}
    \caption{\label{fig:1DScalarvXaxis_SpinScalarize} Profile of the scalar field, $\phi$, plotted along the x-axis for the ExptBm300Xp07 run (see Fig.~\ref{fig:Feynmann_SpinScalarize} and Table~\ref{tab:Simulation_Suite}). The \bh{s'} trajectories cross the x-axis at each time shown. Data is taken from the orbital plane, where $z=0\mathrm{M}$.} 
    \end{center}
\end{figure}

To more clearly observe the quantitative evolution of the scalar field, we plot its profile in Fig.~\ref{fig:1DScalarvXaxis_SpinScalarize} along the x-axis for a selection of times at which the \bh{s} cross that axis.
The initial scalar profile at time $t=0 \mathrm{M}$ consists of two peaks centered on the \bh{s} at $x=\pm50\mathrm{M}$.
By comparing the scalar profiles at times $t=0 \mathrm{M}$ and $t=93 \mathrm{M}$, one can see that the scalar field decays as the \bh{s} individually descalarize prior to the encounter. 
Note that the scalar profile at time $t=93 \mathrm{M}$ consists of two peaks located at $x\sim\pm1\mathrm{M}$ which are small and difficult to discern. 
This behavior corresponds to phase I. 
By comparing the scalar profiles at times $t=93 \mathrm{M}$ and $t=114 \mathrm{M}$, we can see an increase in the scalar field. This behavior corresponds to the \textit{spin-induced dynamical scalarization} found during the \bh{'s} encounter in phase II. 
Furthermore, we can see the beginning of the formation of a negative scalar field in the scalar profile at time $t=114 \mathrm{M}$.
This behavior preempts phase III. At time $t=184 \mathrm{M}$, when the \bh{s} have moved apart to positions of $x\sim \pm 10\mathrm{M}$, one can see that the scalar field grows and forms peaks around them.
The peaks' radial falloff behaves as $\phi\propto1/r$, consistent with scalar hair due to \textit{spin-induced scalarization}. 
This behavior corresponds to phase IV, where the \bh{s} exhibit \textit{spin-up scalarization} and individually scalarize. 

Here we observe that initially unscalarized \bh{s} can become permanently scalarized after being spun up during a close encounter. We call this process \textit{spin-up scalarization}.
This phenomenon principally concerns the change in the state of the individual \bh{s} before and after the encounter, rather than the dynamical effects of their encounter.

\subsection{Descalarization of black holes due to spin-up} 
\label{sec:results_SpinDeScalarize}

\begin{figure*}[htbp!]
    \centering
    \subfloat[xy-plane (i.e., the orbital plane)\label{fig:2DScalarPlots_XY_SpinDeScalarize}]{%
        \includegraphics[width=2\columnwidth]{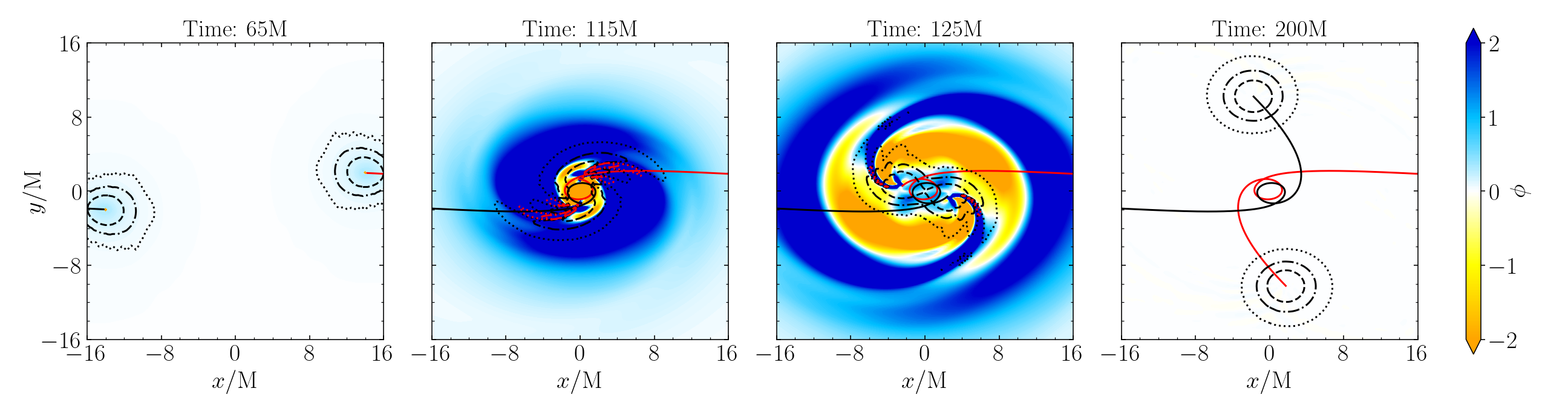}%
    }\par
    \subfloat[xz-plane\label{fig:2DScalarPlots_XZ_SpinDeScalarize}]{%
        \includegraphics[width=2\columnwidth]{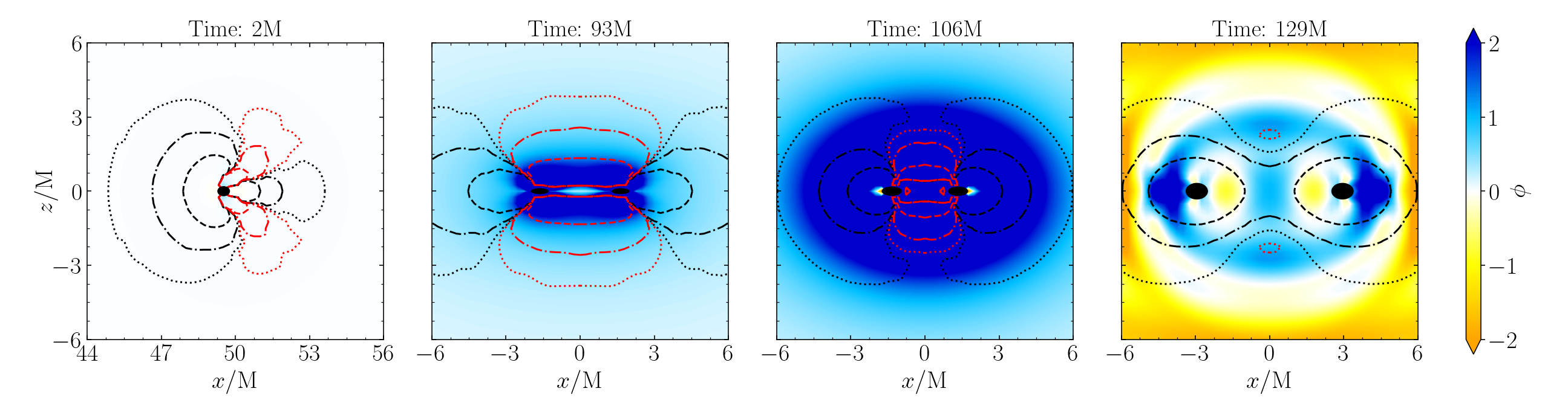}%
    }
    \caption{\label{fig:2DScalarPlots_SpinDeScalarize} Snapshots of the scalar field, $\phi$, and Gauss-Bonnet invariant, $\mathcal{G}$, from the ExptBm350Xm07 run (see Fig.~\ref{fig:Feynmann_SpinDeScalarize} and Table~\ref{tab:Simulation_Suite}). The top panel shows the xy-plane (i.e., the orbital plane). The bottom panel shows the xz-plane at times when the \bh{s} cross the x-axis. Positive scalar field is shown in blue and negative scalar field is shown in yellow. Positive (negative) contours of the Gauss-Bonnet invariant, $|\mathcal{G}\mathrm{M}^4|=\{10^{-3},10^{-2},10^{-1}\}$, are depicted using dotted, dash-dotted, and dashed black (red) lines, respectively. We note that the negative contours can be difficult to discern in grayscale. The trajectories of the \bh{s} are denoted using solid black and red lines.}
\end{figure*}

In this section, we analyze the results of the ExptBm350Xm07 run listed in Table~\ref{tab:Simulation_Suite} and depicted in Fig.~\ref{fig:Feynmann_SpinDeScalarize}. The \bh{s} have initial spin parameter $\chi_{\rm 0}=-0.7$ and begin scalarized. The \bh{s} scatter and undergo a close encounter, resulting in a large spin-up of $\Delta \chi= 0.304$, such that the \bh{s} have a final spin of $\chi_{\rm f}=-0.391$. 
Consequently, the \bh{s} should be unable to remain individually scalarized after the encounter; see Eq.~\eqref{eq:beta_c_spinning}.
The \gr background is the same as in the ExptBm300Xm07 run discussed in Sec.~\ref{sec:results_SpinEncounter}. Here, however, the dimensionless coupling between the scalar field and Gauss-Bonnet invariant is larger, namely $\beta=-3.50$. Furthermore, the initial scalar field amplitude is smaller $c_{\rm 0}=0.1$; see Sec~\ref{sec:Setup}.
In the following, we comment briefly upon the dynamical enhancement of scalarization, which occurs during the encounter, but primarily focus on the permanent \textit{spin-up descalarization} of the \bh{s}, which results from the change in spin during the close encounter. Animations created from this simulation can be found at Ref.~\cite{CanudaAnimations}.

In Fig.~\ref{fig:2DScalarPlots_SpinDeScalarize}, we plot snapshots of the scalar field in both the orbital plane (top panels) and the xz-plane (bottom panels) at different instances in time. The snapshots of the orbital plane are selected at times representative of the system's evolution. The snapshots of the xz-plane correspond to times at which the \bh{} punctures cross the x-axis. The value of the scalar field is shown using a color scale, where darker blues show more positive values and darker yellows show more negative values. We plot positive (negative) contours of the Gauss-Bonnet invariant, $|\mathcal{G}\mathrm{M}^4|=\{10^{-3},10^{-2},10^{-1}\}$, using black (red) lines.
In the orbital plane, the \bh{} trajectories are plotted over the scalar field.
In the xz-plane, we plot the approximate shape of the \bh{s'} apparent horizons as ellipses using the maximum and minimum horizon radii from the \verb|AHFinderDirect| thorn's output.

Qualitatively, we organize the evolution of the system into four phases, each of which aligns with one of the snapshots of the orbital plane shown in Fig.~\ref{fig:2DScalarPlots_XY_SpinDeScalarize}.
Note that the time of the close encounter (i.e., the smallest \bh{} separation) is $t_{\rm CE}=100\mathrm{M}$.
\textbf{Phase I ($0\mathrm{M}<t<t_{\rm CE}$):} From left to right, the first panel shows the \bh{s} as they approach one another prior to the encounter. Here, the \bh{s} are scalarizing. 

The two central panels show the \bh{s} as they orbit around each other and separate following the close encounter. Qualitatively, the evolution of the scalar field is similar to that of Sec.~\ref{sec:results_SpinEncounter}.  
\textbf{Phase II ($t_{\rm CE} < t \lesssim 120\mathrm{M}$):} First, one notices the formation of a dark blue region around the \bh{s}. This is due both to the initial \textit{spin-induced scalarization} of the \bh{s} prior to the encounter and to the dynamical enhancement of scalarization during this phase. 
\textbf{Phase III ($120\mathrm{M} \lesssim t \lesssim 150\mathrm{M}$):} In the third panel, regions where the Gauss-Bonnet invariant is positive pass over the scalar field causing it to oscillate.
During this phase, the scalar field temporarily turns negative as it begins to decay.
We note that the magnitude of the scalar field is much larger in these panels than in Sec.~\ref{sec:results_SpinEncounter} (in spite of the smaller initial scalar field amplitude) due to the greater magnitude of the dimensionless coupling and scalarization prior to the encounter. 

\textbf{Phase IV ($150\mathrm{M} \lesssim t$):} The panel on the far right shows the \bh{s} well after they have separated. Recall that the \bh{s} have a final spin magnitude of $|\chi_{\rm f}|=0.391$, which is below the threshold for \textit{spin-induced scalarization}. Indeed, we see that the scalar field vanishes, consistent with \textit{spin-up descalarization}.

We now consider the snapshots of the xz-plane shown in Fig.~\ref{fig:2DScalarPlots_XZ_SpinDeScalarize}. From left to right, the first panel shows one of the \bh{s} shortly after the start of the simulation, i.e., at the beginning of phase I. Here the negative Gauss-Bonnet invariant contours have the same characteristic “dumbbell” shape remarked upon in Secs.~\ref{sec:results_SpinEncounter} and~\ref{sec:results_SpinScalarize}. The scalar field appears fainter due to the smaller initial scalar field amplitude and broader color scale.

The two central panels show the \bh{s} shortly before and after the close encounter. As in Secs.~\ref{sec:results_SpinEncounter} and~\ref{sec:results_SpinScalarize}, the \bh{s'} “dumbbells” join into a larger region, which dynamically enhances the scalarization of the \bh{s}. 
In the short time span between the central panels, one can see that the scalar field increases sufficiently to encompass both \bh{s}.
This behavior corresponds to phase II. 

The final panel on the far right shows the \bh{s} as they begin to separate following the encounter. At this stage, the \bh{s'} spins have a magnitude $|\chi|<0.5$. One can see that the “dumbbell” regions around the \bh{s} disappear. There is only a small region between the \bh{s} where the Gauss-Bonnet invariant is negative due to the systems' orbital angular momentum. 
As a result, the scalar field undergoes a damped oscillation and decays.
We can thus see that the removal of the “dumbbell” region via the reduction of the \bh{s'} spin magnitudes leads to \textit{spin-up descalarization}.

\begin{figure}[tbp!]
    \begin{center}

    \includegraphics[width=1\columnwidth]{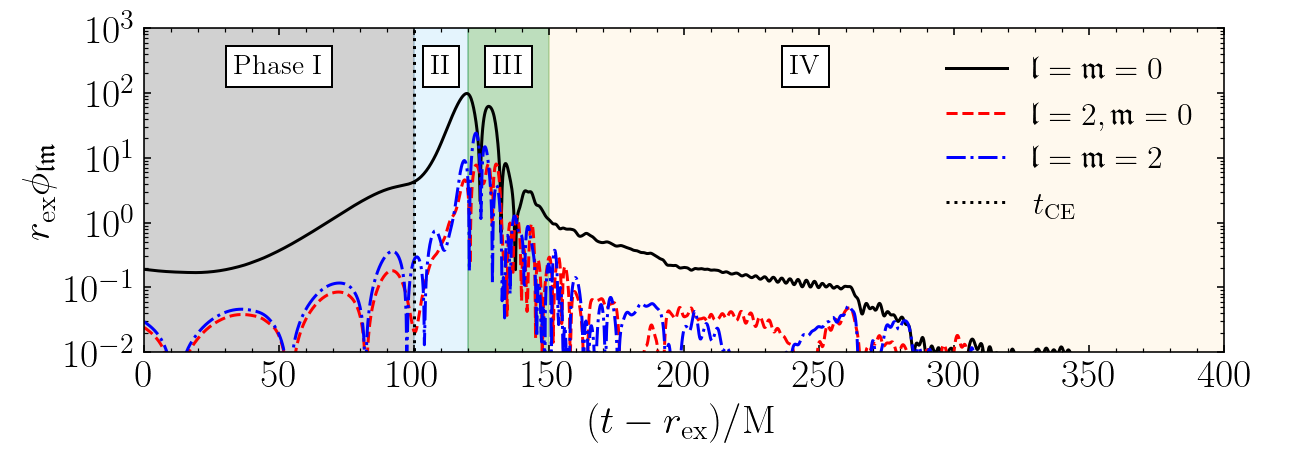}
    \includegraphics[width=1\columnwidth]{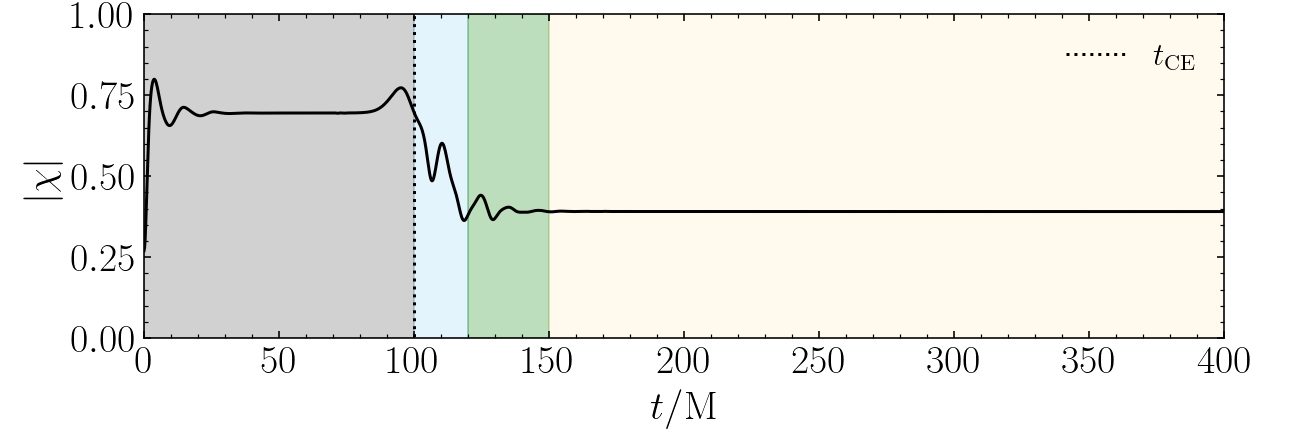}
    \caption{\label{fig:1DScalarvTime_SpinDeScalarize} Top panel: Multipoles of the scalar field, $\phi$, plotted as a function of time for the ExptBm350Xm07 run (see Fig.~\ref{fig:Feynmann_SpinDeScalarize} and Table~\ref{tab:Simulation_Suite}). The three lowest modes ($\mathfrak{l}=\mathfrak{m}=0$), ($\mathfrak{l}=2,\mathfrak{m}=0$), and ($\mathfrak{l}=\mathfrak{m}=2$) are shown. All other modes with $\mathfrak{l}\leq2$ are zero due to the symmetries of the system. We rescale the scalar field by the extraction radius $\rex=100\mathrm{M}$. We shift the time by the extraction radius to account for the propagation time of the scalar field. Bottom panel: The spin magnitude of one of the \bh{s} is plotted as a function of time. We denote the time of the close encounter, $t_{\rm CE}=100\mathrm{M}$, via a dotted line. Note that this time is not shifted by the extraction radius. 
    } 
    \end{center}
\end{figure}

To provide a broad overview of the scalar field's evolution, we plot the three lowest non-zero multipoles as a function of time in the top panel of Fig.~\ref{fig:1DScalarvTime_SpinDeScalarize}. 
The multipoles are measured at an extraction radius of $\rex=100\mathrm{M}$. In order to account for the propagation time of the scalar field, we subtract the extraction radius from the simulation time. The time of the close encounter, $t_{\rm CE}=100\mathrm{M}$, is denoted by a dotted black line. Note that this time is not shifted by the extraction radius. The monopole, $\rex \phi_{\rm 00}$, corresponds to the scalar charge of the system (see Eq.~\eqref{eq:scalar_charge}), and it is significantly larger than the $\mathfrak{l}=2$ modes plotted for comparison. In the bottom panel of Fig.~\ref{fig:1DScalarvTime_SpinDeScalarize}, we plot the magnitude of one of the \bh{s'} spins as a function of time using Eq.~\eqref{eq:chi_extract}. Note that there is no propagation time associated with the spin.

During phase I ($0\mathrm{M}<t-\rex<t_{\rm CE}$), the scalar charge notably increases as the \bh{s} begin to scalarize prior to the close encounter. 
However, the scalar charge initially displays a brief plateau, presumably due to the adjustment of the scalar field seed to the spinning \bh{s} ($\chi_{\rm i}=-0.695$).
During the close encounter in phase II ($t_{\rm CE} < t-\rex \lesssim 120\mathrm{M}$), the scalarization of the \bh{s} is dynamically enhanced and the scalar charge increases more rapidly. 
During this time, the spin magnitude also rapidly declines.
In phase III ($120\mathrm{M} \lesssim t-\rex \lesssim 150\mathrm{M}$), as the \bh{s} separate following the close encounter, their spin magnitudes stabilize to values below the threshold for the \textit{spin-induced scalarization} of isolated \bh{s} ($\chi_{\rm f}=-0.391$). We observe that the scalar charge initially undergoes a damped oscillation as the \bh{s} begin to descalarize.
As the \bh{s} continue to separate during phase IV ($150\mathrm{M} \lesssim t-\rex$), the scalar charge decays exponentially. 
The \bh{s} thus appear to undergo permanent \textit{spin-up descalarization} due to their decrease in spin magnitude.
We note that the scalar charge is dominated by noise beyond time $t-\rex\sim250\mathrm{M}$ due to reflection off of the refinement boundaries.

\begin{figure}[tbp!]
    \begin{center}
    \includegraphics[width=1\columnwidth]{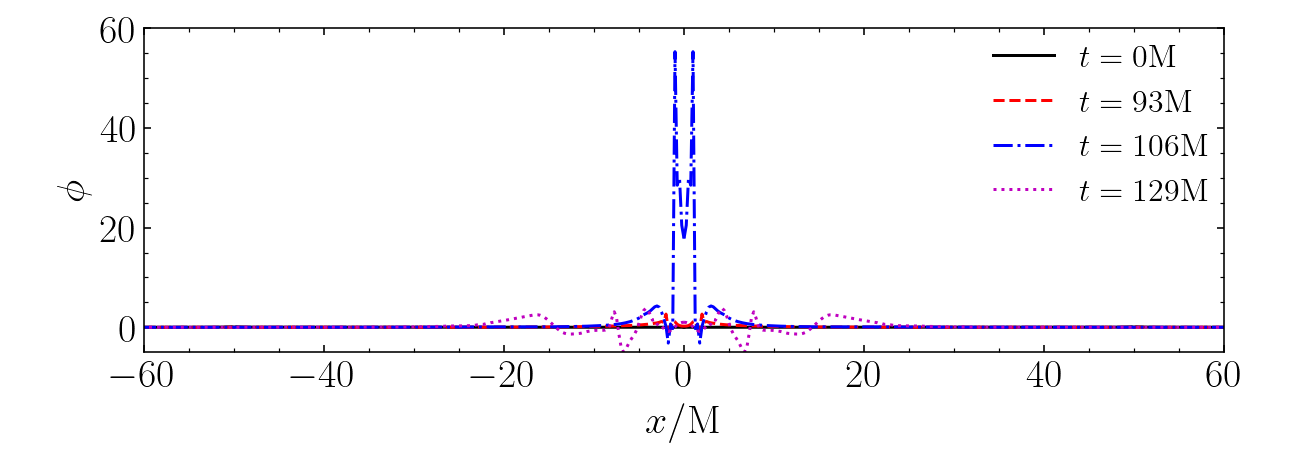}
    \includegraphics[width=1\columnwidth]{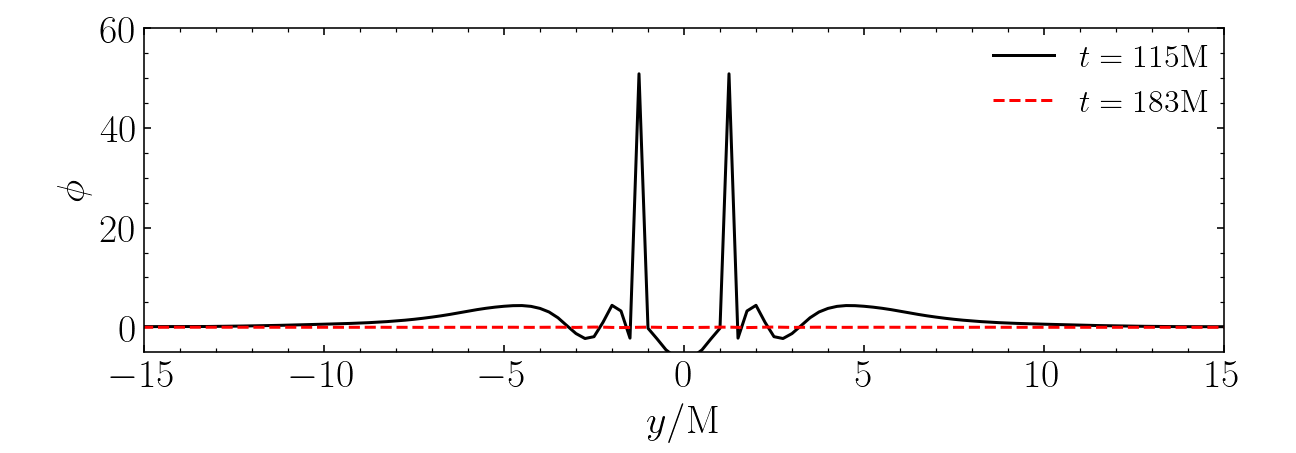}
    \caption{\label{fig:1DScalarvXaxis_SpinDeScalarize} Profile of the scalar field, $\phi$, plotted along the x-axis (top panel) and y-axis (bottom panel) for the ExptBm350Xm07 run (see Fig.~\ref{fig:Feynmann_SpinDeScalarize} and Table~\ref{tab:Simulation_Suite}). The \bh{} trajectories cross either the x-axis or y-axis at each time shown. Data is taken from the orbital plane, where $z=0\mathrm{M}$. 
    } 
    \end{center}
\end{figure}

To more clearly observe the quantitative evolution of the scalar field, we plot its profile in Fig.~\ref{fig:1DScalarvXaxis_SpinDeScalarize} along the x (top panel) and y (bottom panel) axes for a selection of times at which the \bh{s} cross those axes.
The initial scalar profile at time $t=0\mathrm{M}$ consists of two peaks centered on the \bh{s} at $x=\pm 50\mathrm{M}$.
However, the initial peaks are so small compared to scalar profiles at later times that they are not visible. In contrast, the scalar profile at time $t=93\mathrm{M}$ is visible due to the \textit{spin-induced scalarization} of the \bh{s} in phase I. By comparing the profiles at times $t=93\mathrm{M}$ and $t=106\mathrm{M}$, one can observe the dynamically enhanced scalarization that occurs during the encounter as the scalar field rapidly increases. This behavior corresponds to phase II. At times $t=115\mathrm{M}$ and $t=129\mathrm{M}$, the scalar field develops negative regions as it decays immediately following the encounter. This behavior corresponds to phase III. At time $t=183\mathrm{M}$, when the \bh{s} move apart to positions of $x\sim \pm9\mathrm{M}$, one can see that the scalar field rapidly decays to the point where it is no longer distinguishable from zero. This behavior corresponds to phase IV, where the \bh{s} exhibit \textit{spin-up descalarization} and individually descalarize.

Here, we observe that initially scalarized \bh{s} can become permanently unscalarized after experiencing a decline in spin magnitude caused by a close encounter. We call this process \textit{spin-up descalarization}. This phenomenon principally concerns the change in the state of the individual \bh{s} before and after the encounter, rather than the dynamical effects of their encounter. We thus conclude the presentation of our simulation suite and the demonstration of the phenomena illustrated in Fig.~\ref{fig:Feynmann_Diagrams}.

\section{Conclusions And Outlook} \label{sec:Conclusion}

In this work, we have explored how hyperbolic encounters of binary \bh{s} can cause scalarization or descalarization within quadratic \sGB gravity. 
To this end, we performed a series of simulations in the decoupling approximation, in which the \bh{s} either undergo a zoom-whirl or scatter according to \gr. The \bh{s} are seeded with an initial scalar field, which evolves on top of the space time.
In many systems, the \bh{s} are initially unable to sustain this seed and the scalar field decays. We call these \bh{s} unscalarized.
We consider cases with both positive and negative coupling constants between the scalar field and Gauss-Bonnet invariant. In the former case, non-spinning \bh{s} may be individually scalarized. 
In the latter case, \bh{s} must have a spin magnitude of at least $|\chi|>0.5$ to undergo \textit{spin-induced scalarization}. 

In the case where the coupling constant is positive, we consider the behavior of initially unscalarized, non-spinning \bh{s}, which undergo a single zoom whirl and then merge. We observe that the \bh{s} temporarily undergo \textit{dynamical scalarization} during the close encounter. 
The \bh{s} descalarize following the first encounter, but then undergo a second round of \textit{dynamical scalarization} during the final inspiral and briefly rescalarize. After the merger, the remnant \bh{} descalarizes due to its larger mass (and smaller curvature). 

In the case where the coupling constant is negative, we consider the behavior of \bh{s} with initial spin parameters of magnitude $|\chi_{\rm 0}|=0.7$, which undergo a close encounter and scatter. When the \bh{s} begin unscalarized, we find that the \bh{s} can temporarily undergo \textit{spin-induced dynamical scalarization}. This is because the regions of negative Gauss-Bonnet invariant along the \bh{s'} spin axes are temporarily enhanced and combine during the close encounter. This enhancement is caused by the orbital angular momentum of the \bh{s}.

We have further considered how the spin-up of \bh{s} during scattering can affect the scalar field when the coupling constant is negative.
We find that when the \bh{s} begin unscalarized and the \bh{s'} spin magnitudes increase (here from $|\chi_{\rm i}|=0.695$ to $|\chi_{\rm f}|=0.718$) due to the encounter, the \bh{s} can become permanently scalarized. Furthermore, we find that when the \bh{s} begin scalarized and the \bh{s'} spin magnitudes decrease (here from $|\chi_{\rm i}|=0.695$ to $|\chi_{\rm f}|=0.391$) due to the encounter, the \bh{s} can become permanently unscalarized. We refer to these processes as \textit{spin-up scalarization} and \textit{spin-up descalarization}, respectively. We thus find that close encounters have the power to bestow scalar hair upon a \bh{} or to rescind it. 

While we have primarily considered the impact of hyperbolic encounters on the scalarization of \bh{s}, we can also make inferences about how that scalarization might influence the evolution of the encounters if backreaction were taken into account. 
Properties such as the scattering angle or the number of the zoom-whirl orbits undertaken prior to merger can be very sensitive to the initial conditions of a binary; see e.g. Refs.~\cite{Gold:2012tk,Rettegno:2023ghr}. In \gr, these quantities depend on the amount of kinetic energy lost to gravitational radiation. When more energy is emitted, it results in larger scattering angles. If sufficient energy is lost such that the system becomes bound, then the \bh{s} undergo a series of zoom-whirls until they loose enough energy to merge.
In quadratic \sGB gravity, scalar radiation provides an additional mechanism through which kinetic energy can be lost. Consequently, one might expect to find larger scattering angles and fewer zoom-whirl orbits than one would in \gr. Similarly, one might expect to find the transition between scattering, zoom-whirl, and direct merger morphologies at larger incident angles.

Furthermore, close encounters result in a brief emission of \gw{s}~\cite{Garcia-Bellido:2017qal,Teuscher:2024xft,Caldarola:2023ipo,Roskill:2023bmd,Bae:2023sww,Fontbute:2024amb}, potentially including uniquely identifiable quasi-normal modes~\cite{Bae:2023sww}, which may be detectable in the future~\cite{Garcia-Bellido:2021jlq,Kerachian:2023gsa, Kocsis:2006hq,Mukherjee:2020hnm,Morras:2021atg,Bini:2023gaj}. 
The temporary or permanent scalarization of \bh{s} during close encounters could leave imprints on these emissions. For example, one might expect to see a change in frequency, or a phase shift, as is sometimes observed during \bh{} mergers in theories beyond \gr and in scalar dark matter environments; see e.g. Refs.~\cite{Aurrekoetxea:2023jwk, Corman:2024cdr,AresteSalo:2025sxc,Lara:2025kzj,Cheng:2025wac,Corman:2025wun}.
These phenomena could be interesting to explore in systems with full backreaction.

In the future, there are numerous other initial conditions and phenomena which may be interesting to explore within quadratic \sGB gravity and other varieties of \sGB gravity that allow for scalarization. Many of these theories (including quadratic \sGB gravity) are invariant under a change in the sign of the scalar field. The finding that the scalar charge oscillates (and changes sign) in the aftermath of \textit{spin-induced dynamical scalarization}, suggests the possibility that close encounters could result in permanent sign flips, where the scalar charge becomes negative. Such behaviors have already been found during inspirals~\cite{Lara:2025kzj}. 
One could further consider systems in which the two \bh{s} are seeded by scalar fields with opposite signs. It would be interesting to see whether this enhances or suppresses \textit{dynamical scalarization}. 

It may also be interesting to study systems where one \bh{} begins scalarized, but the other does not. This could potentially lead to a permanent form of \textit{induced scalarization}, where one \bh{} seeds another with scalar field thus allowing it to scalarize. Furthermore, one could consider the scattering of \bh{s} with unequal mass. The findings of Refs.~\cite{Nee:2024bur} demonstrate that \bh{s} too massive to scalarize can experience \textit{induced scalarization} from a smaller \bh{}.
This suggests that a temporary dynamical form of \textit{induced scalarization} may occur in close encounters.

Finally, \sGB gravity is far from the only theory beyond \gr that posits a scalar field coupled to higher order curvature terms. Among these theories, there are others which exhibit spin-dependence or scalarization; see e.g. Refs.~\cite{Doneva:2022ewd,Richards:2025ows}.
It would be interesting to see whether the equivalents of the phenomena presented here can be found in the close encounters of other such theories.

\section{Acknowledgments}

We thank
C.~-~H.~Cheng,
H.~Kogan,
N.~Ghadiri,
and
H.~O.~da~Silva,
for insightful discussions and comments.
The authors acknowledge support provided by the National Science Foundation under NSF Award No.
No.~OAC-2411068 and No.~PHY-2409726.
We acknowledge the Texas Advanced Computing Center (TACC) at the University of Texas at Austin for providing HPC resources on Frontera via allocation PHY22041.
This work used Delta at NCSA through allocation PHY260102 from the Advanced Cyberinfrastructure Coordination Ecosystem: Services and Support (ACCESS) program, which is supported by U.S. National Science Foundation grants No. 2138259, 2138286, 2138307, 2137603, and 2138296~\cite{ACCESS}. 
This research used resources provided by the Delta research computing project, which is supported by the NSF Award No. OAC-2005572 and the State of Illinois. 
This research was supported in part by the Illinois Computes project which is supported by the University of Illinois Urbana-Champaign and the University of Illinois System.
This work made use of the Illinois Campus Cluster, a computing resource that is operated by the Illinois Campus Cluster Program (ICCP) in conjunction with the National Center for Supercomputing Applications (NCSA) and which is supported by funds from the University of Illinois Urbana-Champaign. 
This work used the open-source softwares \textsc{xTensor}~\cite{xAct:web,Brizuela:2008ra}, the \ETK~\cite{maxwell_rizzo_2025_15520463,Loffler:2011ay, Zilhao:2013hia},
\canuda~\cite{witek_2023_7791842}, and kuibit~\cite{kuibit}.



\appendix

\section{Convergence Tests} \label{sec:Convergence}

We perform a series of convergence tests to assess the validity and numerical accuracy of our simulations. In this aim, we rerun a subset of the simulations (ExptBp0355X0, CtrlBm350Xp07, and ExptBm300Xm07) with step sizes $dx_{\rm low}=1\mathrm{M}$, $dx_{\rm med}=0.95\mathrm{M}$, and $dx_{\rm high}=0.85\mathrm{M}$ on the outermost refinement level (see Table~\ref{tab:Convergence_qsGB}). Within successive (inner) refinement levels, the step sizes are halved. Note that in the ExptBp0355X0 test we use a dimensionless coupling of $\beta=0.36281$, rather than $\beta=0.355$ as in the production run. This is because the test was conducted as a preliminary check before the production runs. 
We expect the ExptBp0355X0 production run to have better convergence than the test as its dimensionless coupling is below the critical value for the scalarization of individual \bh{s}.
In the following, we show convergence plots and make error estimates for $\phi_{\rm 00}$, which is related to the scalar charge, and $\phi_{\rm 22}$, which is related to the scalar radiation. Both quantities are evaluated at an extraction radius of $\rex=100\mathrm{M}$.
We use the same simulations to produce convergence plots and error estimates for the spin and other properties of the \gr background in our previous work Ref.~\cite{Kogan:2025vml}, which we do not reproduce here. 

\begin{table}[htbp!]
    \centering
    \begin{tabular}{|c|c|c|c|c|c|}
    \hline
     Run & $\beta$ & $c_{\rm 0}$ & $\chi_{\rm 0}$  & $|\vec{P}_{\rm i}|/\mathrm{M}$ & $\theta$ \\
     \hline
     ExptBp0355X0 & $0.36281$ & $1.0$ & $0.0$ & $0.245$ & $0.0580$ \\
     CtrlBm350Xp07 & $-3.50$ & $0.1$ & $0.7$ & $0.490$ & $0.05685$ \\
     ExptBm300Xm07 & $-3.00$ & $1.0$ & $-0.7$ & $0.490$ & $0.05685$ \\
     \hline
    \end{tabular}
    \caption{\label{tab:Convergence_qsGB} Parameters for the suite of convergence tests. In each case, we run three simulations with step sizes $dx_{\rm low}=1.0\mathrm{M}$, $dx_{\rm med}=0.95 \mathrm{M}$, and $dx_{\rm high}=0.85\mathrm{M}$ at the outermost refinement level. The $dx_{\rm low}=1.0\mathrm{M}$ simulations correspond to those in Table~\ref{tab:Simulation_Suite}. The dimensionless coupling of the ExptBp0355X0 test is larger than that of the production run. Here we list the run name, the dimensionless coupling, $\beta$, the initial scalar field amplitude, $c_{\rm 0}$, the \bh{s'} initial spin parameter, $\chi_{\rm 0}$, and the \bh{s'} initial momentum, $|\vec{P}_{\rm i}|$. Each system has a separation of $d=100\mathrm{M}$.}
\end{table}

As we increase the resolution of the simulations, we expect them to converge to the true solution at a rate determined by the convergence order, $n$. For a given quantity, $q$, the difference between the low and medium resolutions, $q_{\rm low}-q_{\rm med}$, should be proportional to the difference between the medium and high resolutions, $q_{\rm med}-q_{\rm high}$, times the convergence factor, 
\begin{equation}
    Q_{n}(dx_{\rm low},dx_{\rm med},dx_{\rm high})=\frac{dx_{\rm low}^n - dx_{\rm med}^n}{dx_{\rm med}^n - dx_{\rm high}^n} \, .
\end{equation}
In the simulations, we use fourth order finite differencing to evaluate spatial derivatives and a fourth order Runge-Kutta scheme to evolve in time. Therefore, we expect fourth order convergence and use the convergence factor $Q_{4}(1.0,0.95,0.85)=0.634$. However, at the refinement boundaries, we use second order and fifth order interpolations in time and space, respectively. Therefore, the convergence order may be mixed. For comparison, the third and second order convergence factors are $Q_{3}=0.586$ and $Q_{2}=0.542$, respectively.

We compute the approximate percent error of a quantity by using the highest resolution simulation as a reference,
\begin{equation}
    \%\mathrm{Error}=100 \, \left| \frac{q_{\rm low}-q_{\rm high}}{q_{\rm high}} \right| \, .
\end{equation}
Whenever the high resolution quantity crosses zero, the percent error diverges. Neither the scalar charge, nor the scalar radiation are positive definite, so we must take this into consideration in both cases. The scalar charge changes sign on rare occasions in the midst of encounters and mergers. We thus comment on its typical error outside of these isolated incidents. The scalar radiation, however, undergoes frequent oscillations. We thus report the error found at the maxima of the radiation.

\subsection{Zoom-Whirl}

\begin{figure}[tbp!]
    \begin{center}
    \includegraphics[width=1\columnwidth]{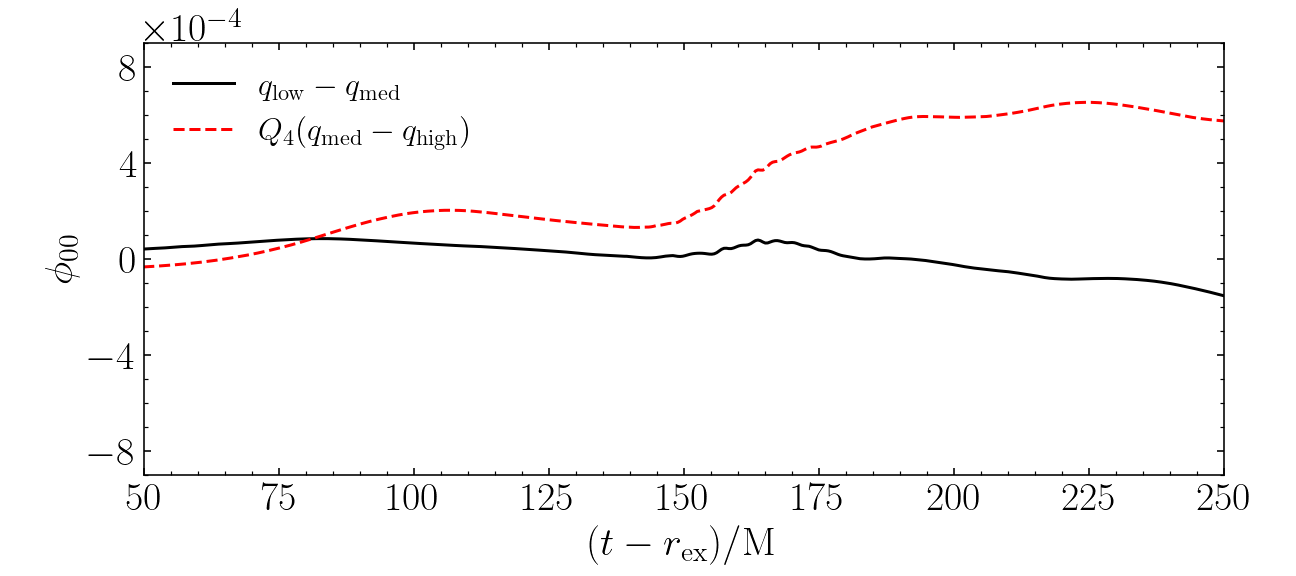}
    \includegraphics[width=1\columnwidth]{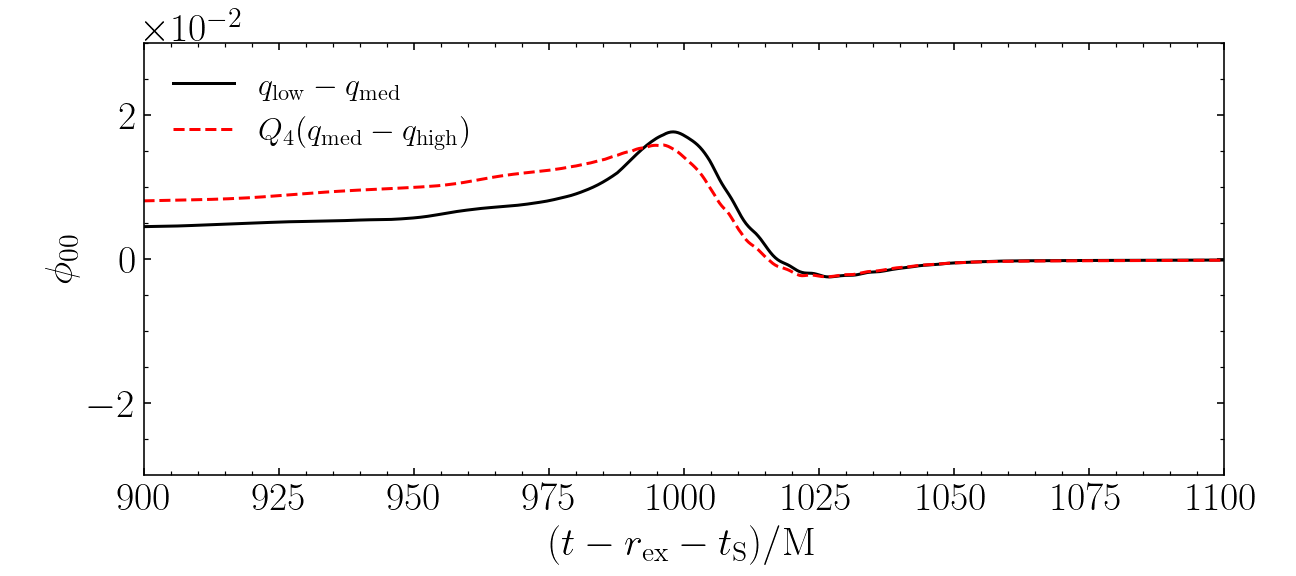}
    \caption{\label{fig:ZWConvergence_phi,00} Convergence plot of the scalar charge in the ExptBp0355X0 run, which corresponds to a zoom-whirl and merger. Top: Convergence centered on the first encounter between the \bh{s}. Bottom: Convergence centered on the \bh{} merger. The simulation times are shifted by $t_{\rm S}$, which is defined such that their times of merger coincide (see text).}
    \end{center}
\end{figure}

\begin{figure}[htbp!]
    \begin{center}
    \includegraphics[width=1\columnwidth]{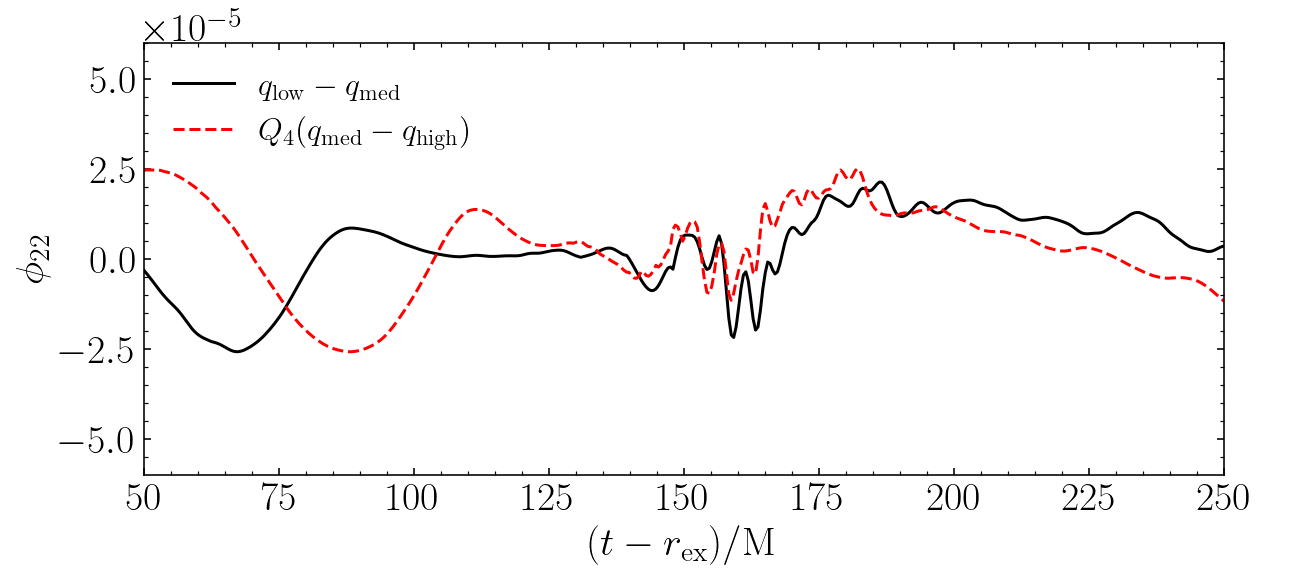}
    \includegraphics[width=1\columnwidth]{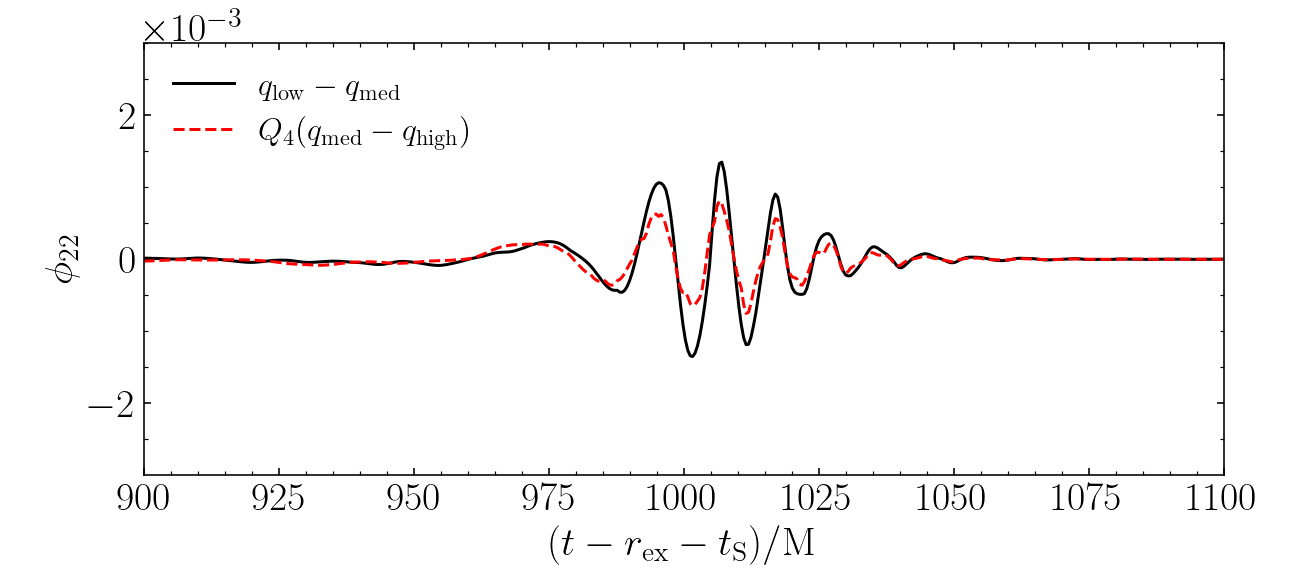}
    \caption{\label{fig:ZWConvergence_phi,22} Convergence plot of the scalar radiation in the ExptBp0355X0 run, which corresponds to a zoom-whirl and merger. Top: Convergence centered on the first encounter between the \bh{s}. Bottom: Convergence centered on the \bh{} merger. The simulation times are shifted by $t_{\rm S}$, which is defined such that their times of merger coincide (see text).}
    \end{center}
\end{figure}

\begin{figure}[tbp!]
    \begin{center}
    \includegraphics[width=1\columnwidth]{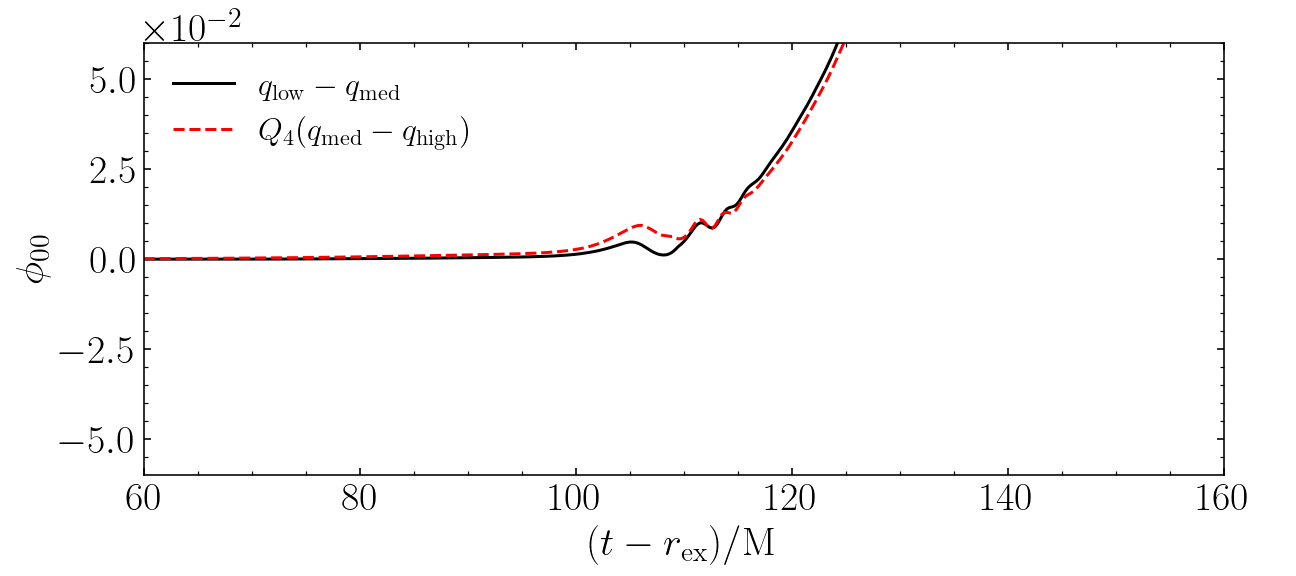}
    \includegraphics[width=1\columnwidth]{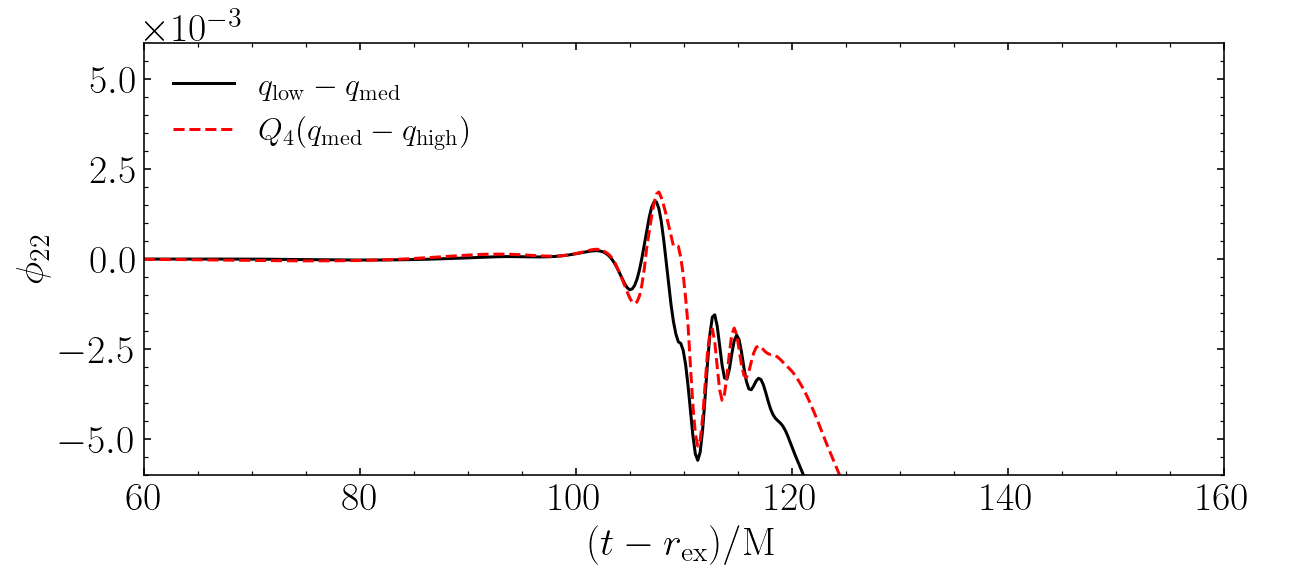}
    \caption{\label{fig:SpinConvergence_phi_The005685_spinp07} Convergence plot of the scalar field in the CtrlBm350Xp07 run, which corresponds to a scattering encounter between the \bh{s}. Both panels are centered on the time of the close encounter. Top: Convergence of the scalar charge. Bottom: Convergence of the scalar radiation.}
    \end{center}
\end{figure}

We first present the convergence test for the ExptBp0355X0 run. In this simulation, the \bh{s} undergo a single zoom-whirl before merging. 
Depending on the resolution, the separation of the \bh{s} during this intermediate phase varies. The \bh{} trajectories realign as they approach merger. However, this variation leads to a shift in the merger time. For this reason, we perform convergence tests in two blocks, where the first is focused on the close encounter, and the second is focused on the merger. In the second block, we shift the time by the value, $t_{\rm S}=t_{\mathrm{ merge,}\, dx}-t_{\mathrm{ merge, \,} dx_{\rm low}}$, in order to align the data with the time of merger in the $dx_{\rm low}=1\mathrm{M}$ simulation.

The convergence of the scalar charge ($\sim \phi_{\rm 00}$) is depicted in Fig.~\ref{fig:ZWConvergence_phi,00}. The convergence during the first encounter is displayed in the top panel. While the trends are somewhat different, the difference between the scalar charges is of order $10^{-4}$. The convergence during the merger is displayed in the bottom panel. We find fourth order convergence. 
We find percent errors of $<3\%$ across both intervals. The only exception being a brief period following the merger when the scalar charge changes sign.

The convergence of the scalar radiation ($\sim \phi_{\rm 22}$) is displayed in Fig.~\ref{fig:ZWConvergence_phi,22}. The convergence during the first encounter is displayed in the top panel. The convergence during the merger is displayed in the bottom panel. We find fourth order convergence. The percent error evaluated at the waveform peak is approximately $0.35 \%$ during the first encounter and about $2.34\%$  during the merger.

\begin{figure}[tbp!]
    \begin{center}
    \includegraphics[width=1\columnwidth]{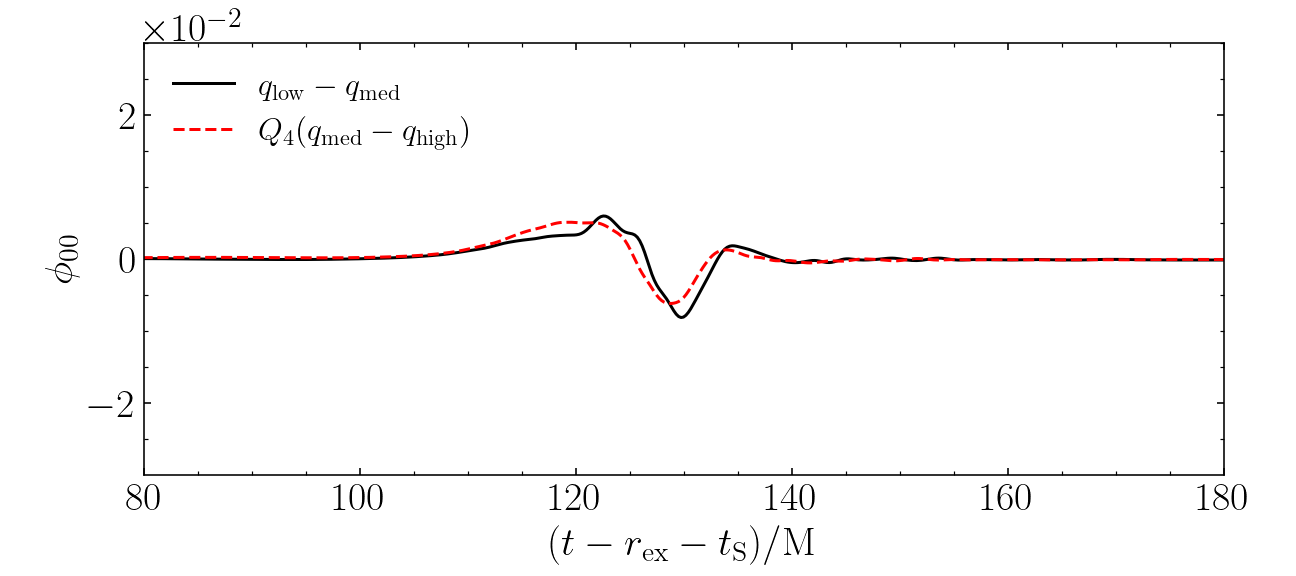}
    \includegraphics[width=1\columnwidth]{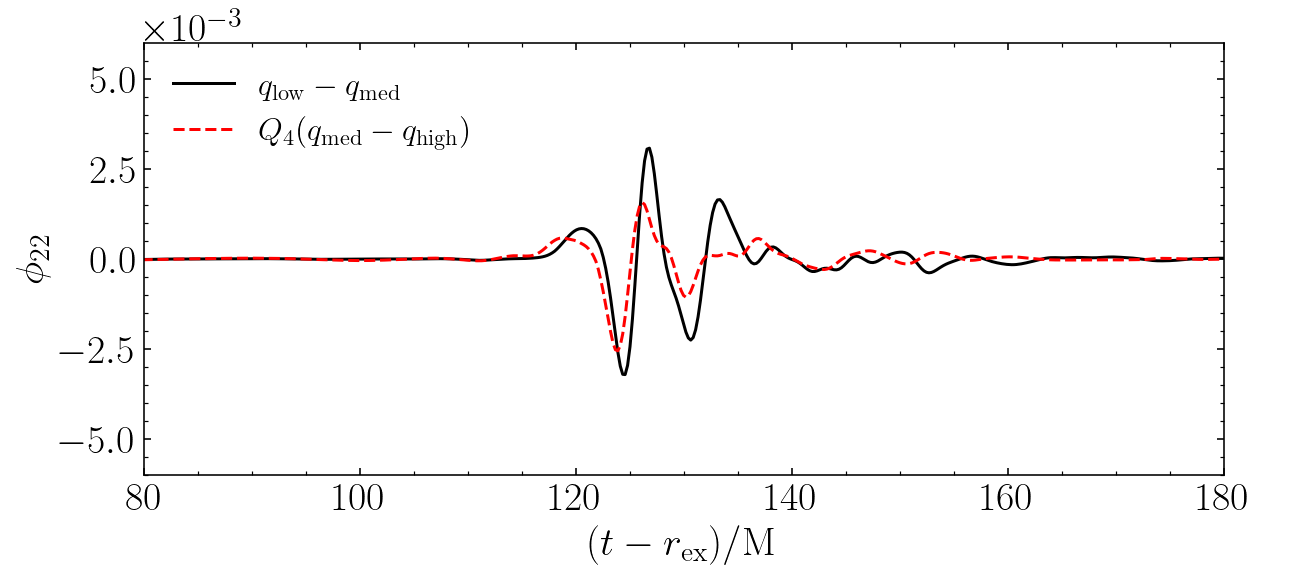}
    \caption{\label{fig:SpinConvergence_phi_The005685_spinm07} Convergence plot of the scalar field in the ExptBm300Xm07 run, which corresponds to a scattering encounter between the \bh{s}. Both panels are centered on the time of the close encounter. Top: Convergence of the scalar charge. The simulation times are shifted by $t_{\rm S}$, which is defined such that the maxima of their scalar charges coincide (see text). Bottom: Convergence of the scalar radiation.}
    \end{center}
\end{figure}

\subsection{Scattering black holes with spin}

Here we present the convergence tests for the CtrlBm350Xp07 and ExptBm350Xm07 runs, which each result in scattering. Therefore, we center the convergence tests on the close encounter between the \bh{s}. 

The convergence test for the CtrlBm350Xp07 run is shown in Fig.~\ref{fig:SpinConvergence_phi_The005685_spinp07}, with the scalar charge ($\sim \phi_{\rm 00}$) in the top panel and the scalar radiation ($\sim \phi_{\rm 22}$) in the bottom panel.
We find fourth order convergence. Prior to and during the encounter we find a percent error of $<20\%$ for the scalar charge and a percent error of $<30\%$ for the scalar radiation. After the encounter, both quantities grow exponentially as the system scalarizes. At different resolutions the scalar field grows with somewhat differing time constants, which causes the percent error to increase with time.
However, we are principally concerned with the qualitative, rather than the quantitative, behavior.

The convergence test for the ExptBm350Xm07 run is shown in Fig.~\ref{fig:SpinConvergence_phi_The005685_spinm07}, with the scalar charge $\sim(\phi_{\rm 00})$ in the top panel and the scalar radiation $\sim(\phi_{\rm 22})$ in the bottom panel. 
We find a slight phase shift in the scalar charge and adjust the simulation time by the shift time, $t_{\rm S}=t_{\mathrm{peak,}\, dx}-t_{\mathrm{peak, \,} dx_{\rm low}}$, to align the data with the time of the scalar charge maximum in the $dx_{\rm low}=1\mathrm{M}$ simulation. 
We find fourth order convergence. The percent error of the scalar charge is $<10\%$ prior to the encounter and $<40\%$ after the encounter. The percent error evaluated at the waveform peak of the scalar radiation is about $20.2\%$. As we are primarily interested in the qualitative behavior of the scalar field, this is within reason.

\bibliography{Refs_qsGB.bib}
\end{document}